\documentclass[manuscript,iop,twocolappendix,numberedappendix]{emulateapj}
\usepackage{colordvi}
\usepackage{epsfig}
\usepackage{graphicx}
\usepackage{amssymb}
\usepackage{threeparttable}
\usepackage{tabularx}
\usepackage{adjustbox}

\newcommand{\mic}{~$\mu$m}

\newcommand{\spitzer}{{\it Spitzer}}

\newcommand{\Ha}{\hbox{{\rm H}\kern 0.1em$\alpha$}}
\newcommand{\Hb}{\hbox{{\rm H}\kern 0.1em$\beta$}}
\newcommand{\MgII}{\hbox{{\rm Mg}\kern 0.1em{\sc ii}}}
\newcommand{\CIV}{\hbox{{\rm C}\kern 0.1em{\sc iv}}}
\newcommand{\NeV}{\hbox{[{\rm Ne}\kern 0.1em{\sc v}]}}
\newcommand{\OII}{\hbox{[{\rm O}\kern 0.1em{\sc ii}]}}
\newcommand{\NeIII}{\hbox{[{\rm Ne}\kern 0.1em{\sc iii}]}}
\newcommand{\OIII}{\hbox{[{\rm O}\kern 0.1em{\sc iii}]}}
\newcommand{\NII}{\hbox{[{\rm N}\kern 0.1em{\sc ii}]}}
\newcommand{\SII}{\hbox{[{\rm S}\kern 0.1em{\sc ii}]}}

\newcommand{\lmass}{log(M/M$_{\odot}$)}

\def\hst{{\it HST}}
\def\spitzer{{\it Spitzer}}
\def\chandra{{\it Chandra}}

\usepackage{color}
\definecolor{citeRGB}{rgb}{0,0.1,0.7}
\usepackage[hyperfootnotes=true,naturalnames=true,letterpaper,pdfstartview=FitH,pdfpagemode=UseNone,colorlinks=true,citecolor=citeRGB]{hyperref}

\shorttitle{The CANDELS/SHARDS GOODS-N Catalog}
\shortauthors{Barro, Perez-Gonzalez et al.}
\slugcomment{To be submitted to The Astrophysical Journal Supplement Series}

\begin{document}

\title{The CANDELS/SHARDS multi-wavelength catalog in GOODS-N:
  Photometry, Photometric Redshifts, Stellar Masses, Emission line
  fluxes and Star Formation Rates}


\author{Guillermo Barro\altaffilmark{1,2}, Pablo G. P\'{e}rez-Gonz\'{a}lez\altaffilmark{3,4},
  Antonio Cava\altaffilmark{5},
  Gabriel Brammer\altaffilmark{6},
  Viraj Pandya\altaffilmark{7},
  Carmen Eliche Moral\altaffilmark{8},
  Pilar Esquej\altaffilmark{9},
  Helena Dom\'{i}nguez-S\'{a}nchez\altaffilmark{10},
  Belen Alcalde Pampliega\altaffilmark{4},  
  Yicheng Guo\altaffilmark{11}
  Anton M. Koekemoer\altaffilmark{6},
  Jonathan R. Trump\altaffilmark{12},
  Matthew L. N. Ashby\altaffilmark{13},
  Nicolas Cardiel\altaffilmark{4},
  Marco Castellano\altaffilmark{14},
  Christopher J. Conselice\altaffilmark{15},
  Mark E. Dickinson\altaffilmark{16},
  Timothy Dolch\altaffilmark{36},
  Jennifer L. Donley\altaffilmark{17},
  N\'{e}stor Espino Briones\altaffilmark{4},
  Sandra M. Faber\altaffilmark{7},
  Giovanni G. Fazio\altaffilmark{13},
  Henry Ferguson\altaffilmark{6},
  Steve Finkelstein\altaffilmark{18},
  Adriano Fontana\altaffilmark{14},
  Audrey Galametz\altaffilmark{19},  
  Jonathan P. Gardner\altaffilmark{20},
  Eric Gawiser\altaffilmark{21},
  Mauro Giavalisco\altaffilmark{22},
  Andrea Grazian\altaffilmark{14},
  Norman A. Grogin\altaffilmark{6},
  Nimish P. Hathi\altaffilmark{6}, 
  Shoubaneh Hemmati\altaffilmark{23},
  Antonio Hern\'{a}n-Caballero\altaffilmark{4},
  Dale Kocevski\altaffilmark{24},
  David C. Koo\altaffilmark{7},
  Dritan Kodra\altaffilmark{25}, 
  Kyoung-Soo Lee\altaffilmark{26},
  Lihwai Lin\altaffilmark{27},
  Ray A. Lucas\altaffilmark{6},   
  Bahram Mobasher\altaffilmark{28},
  Elizabeth J. McGrath\altaffilmark{24},
  Kirpal Nandra\altaffilmark{19},
  Hooshang Nayyeri\altaffilmark{29},
  Jeffrey A. Newman\altaffilmark{25},
  Janine Pforr\altaffilmark{30},
  Michael Peth\altaffilmark{6},
  Marc Rafelski\altaffilmark{6},
  Lucia Rodr\'{i}guez-Munoz\altaffilmark{31},
  Mara Salvato\altaffilmark{19},
  Mauro Stefanon\altaffilmark{32},
  Arjen van der Wel\altaffilmark{33},
  Steven P. Willner\altaffilmark{13}, 
  Tommy Wiklind\altaffilmark{34},
  Stijn Wuyts\altaffilmark{35}}

\altaffiltext{1}{University of the Pacific, Stockton, CA 90340 USA}
\altaffiltext{2}{Department of Astronomy, University of California at Berkeley, Berkeley, CA 94720-3411, USA}
\altaffiltext{3}{Centro de Astrobiolog\'{\i}a (CAB, CSIC-INTA), Carretera deAjalvir km 4, E-28850 Torrej\'on de Ardoz, Madrid, Spain}
\altaffiltext{4}{Universidad Complutense de Madrid, F. CC. F\'{\i}sicas, 28040 Madrid, Spain}
\altaffiltext{5}{Department of Astronomy, University of Geneva, 51 Ch. des Maillettes, 1290, Versoix, Switzerland}
\altaffiltext{6}{Space Telescope Science Institute, 3700 San Martin Drive, Baltimore, MD 21218, USA}
\altaffiltext{7}{UCO/Lick Observatory, Department of Astronomy and Astrophysics, University of California, Santa Cruz, CA 95064, USA}
\altaffiltext{8}{Instituto de Astrof\'{i}sica de Canarias, E-38200 La Laguna, Tenerife, Spain}
\altaffiltext{9}{European Space Astronomy Centre, Villanueva de la Canada E-28692 Madrid, Spain}
\altaffiltext{10}{Department of Physics and Astronomy, University of Pennsylvania, Philadelphia, PA 19104, USA}
\altaffiltext{11}{Department of Physics and Astronomy, University of Missouri, Columbia, MO 65211, USA}
\altaffiltext{12}{Department of Physics, University of Connecticut, 2152 Hillside Road, U-3046 Storrs, CT 06269, USA}
\altaffiltext{13}{Center for Astrophysics | Harvard \& Smithsonian, 60 Garden Street, MS-66, Cambridge, MA 02138-1516, USA}
\altaffiltext{14}{INAF - Osservatorio Astronomico di Roma, via Frascati 33, I-00040 Monte Porzio Catone (RM), Italy}
\altaffiltext{15}{School of Physics and Astronomy, University of Nottingham, Nottingham, UK}
\altaffiltext{16}{National Optical Astronomy Observatories, 950 N Cherry Avenue, Tucson, AZ 85719, USA}
\altaffiltext{17}{Los Alamos National Laboratory, Los Alamos, NM 87544 USA}
\altaffiltext{18}{Department of Astronomy, The University of Texas at Austin, Austin, TX 78712, USA}
\altaffiltext{19}{Max-Planck-Institut f\"{u}r extraterrestrische Physik, Postfach 1312, Giessenbachstr., D-85741 Garching, Germany}
\altaffiltext{20}{NASA's Goddard Space Flight Center, Greenbelt , MD 20771, USA}
\altaffiltext{21}{Department of Physics and Astronomy, Rutgers, The State University of New Jersey, 136 Frelinghuysen Road, Piscataway, NJ 08854, USA}
\altaffiltext{22}{Department of Astronomy, University of Massachusetts, 710 North Plesant Street, Amherst, MA 01003, USA}
\altaffiltext{23}{California Institute of Technology, MS 100-22, Pasadena, CA 91125}
\altaffiltext{24}{Department of Physics and Astronomy, Colby College, Waterville, Maine 04901, USA}
\altaffiltext{25}{Department of Physics and Astronomy and PITT PACC, University of Pittsburgh, Pittsburgh, PA 15260, USA}
\altaffiltext{26}{Department of Physics, Purdue University, 525 Northwestern Avenue, West Lafayette, USA}
\altaffiltext{27}{Institute of Astronomy \& Astrophysics, Academia Sinica, Taipei 10617, Taiwan}
\altaffiltext{28}{Department of Physics and Astronomy, University of California, Riverside, CA, USA}
\altaffiltext{29}{Department of Physics and Astronomy, University of California, Irvine, CA 92697, USA}
\altaffiltext{30}{European Space Research and Technology Centre, Keplerlaan 1, 2201 AZ Noordwijk, Netherlands} 
\altaffiltext{31}{Dipartimento di Fisica e Astronomia, Universita degli Studi di Padova, Vicolo dell Osservatorio 3, Italy}
\altaffiltext{32}{Leiden Observatory, Leiden University, NL-2300 RA Leiden, Netherlands}
\altaffiltext{33}{Max-Planck-Institut f\"{u}r Astronomie, K\"{o}nigstuhl 17, D-69117 Heidelberg, Germany}
\altaffiltext{34}{Catholic University of America, Department of Physics, 620 Michigan Ave NE, Washington, DC 20064, USA}
\altaffiltext{35}{Department of Physics, University of Bath, Claverton Down, Bath, BA2 7AY, UK}
\altaffiltext{36}{Department of Physics, Hillsdale College, 33 E. College Street, Hillsdale, Michigan 49242, USA}


\slugcomment{To be submitted to the Astrophysical Journal Suplements} 
\slugcomment{Last edited: \today}
\label{firstpage}
\begin{abstract} 

  We present a WFC3 F160W ($H$-band) selected catalog in the
  CANDELS/GOODS-N field containing photometry from the ultraviolet
  (UV) to the far-infrared (IR), photometric redshifts and stellar
  parameters derived from the analysis of the multi-wavelength data.
  The catalog contains 35,445 sources over the 171 arcmin$^{2}$ of the
  CANDELS F160W mosaic. The 5$\sigma$ detection limits (within an
  aperture of radius 0\farcs17) of the mosaic range between $H=27.8$,
  28.2 and 28.7 in the wide, intermediate and deep regions, that span
  approximately 50\%, 15\% and 35\% of the total area. The
  multi-wavelength photometry includes broad-band data from UV (U band
  from KPNO and LBC), optical (HST/ACS F435W, F606W, F775W, F814W, and
  F850LP), near-to-mid IR (HST/WFC3 F105W, F125W, F140W and F160W,
  Subaru/MOIRCS Ks, CFHT/Megacam K, and \spitzer/IRAC 3.6, 4.5, 5.8,
  8.0 $\mu$m) and far IR (\spitzer/MIPS 24$\mu$m, HERSCHEL/PACS 100
  and 160$\mu$m, SPIRE 250, 350 and 500$\mu$m) observations. In
  addition, the catalog also includes, optical medium-band data
  (R$\sim50$) in 25 consecutive bands, $\lambda=500$ to 950~nm, from
  the SHARDS survey and WFC3 IR spectroscopic observations with the
  G102 and G141 grisms (R$\sim210$ and 130). The use of higher
  spectral resolution data to estimate photometric redshifts provides
  very high, and nearly uniform, precision from $z=0-2.5$. The
  comparison to 1,485 good quality spectroscopic redshifts up to
  $z\sim3$ yields $\Delta z$/(1+$z_{\rm spec}$)$=$0.0032 and an
  outlier fraction of $\eta=$4.3\%. In addition to the multi-band
  photometry, we release added-value catalogs with emission line
  fluxes, stellar masses, dust attenuations, UV- and IR- based star
  formation rates and rest-frame colors.

\end{abstract}

\section{Introduction}
\label{s:intro}

Large multi-wavelength photometric surveys have made it possible to
study galaxy populations over most of cosmic history. Near-infrared
selected samples have been used to trace the evolution of the stellar
mass function \citep[e.g.,][]{pg08,marchesini09,muzzin13smf}, the star
formation--mass relation (e.g., \citealt{whitaker12}), and the
structural evolution of galaxies \citep[e.g.,][]{franx08, bell12,
  wuyts12, vdw12}. Until recently most of these surveys relied on
deep, wide-field imaging from ground-based telescopes (e.g.,
\citealt{muzzin13smf, williams09}). The WFC3 camera on the
\textit{Hubble Space Telescope} (\textit{HST}) has opened up the
possibility to select and study galaxies at near-infrared wavelengths
with excellent sensitivity and spatial resolution.

The Cosmic Assembly Near-infrared Deep Extragalactic Legacy Survey
\cite[CANDELS,][]{candelsgro,candelskoe} is a 902-orbit legacy program
designed to study galaxy formation and evolution over a wide redshift
range using the near-infrared {\it HST}/WFC3 camera to obtain deep
imaging of faint and distant objects. So far, CANDELS has imaged over
250,000 distant galaxies within five strategic regions: GOODS-S,
GOODS-N, UDS, EGS, and COSMOS over a combined area of
$\sim$0.22~deg$^{2}$. The extremely deep, high spatial resolution
observations have enabled a broad array of science such as: the
characterization of the UV luminosity functions up to $z=10$ (e.g.,
\citealt{finkelstein15}; \citealt{bouwens16}), the stellar mass
functions and the star formation rate (SFR) sequence at $z=4-6$
(\citealt{duncan14}; \citealt{grazian15}; \citealt{mortlock15};
\citealt{salmon15}), or detailed studies of the structural and stellar
mass growth in star-forming and quiescent galaxies since cosmic noon,
$z\sim2$ (e.g.,\citealt{wuyts13}; \citealt{vdw14}; \citealt{barro14a};
\citealt{guo15}; \citealt{papovich15}).



\begin{figure*}[t]
\centering
\includegraphics[width=12cm,angle=0.]{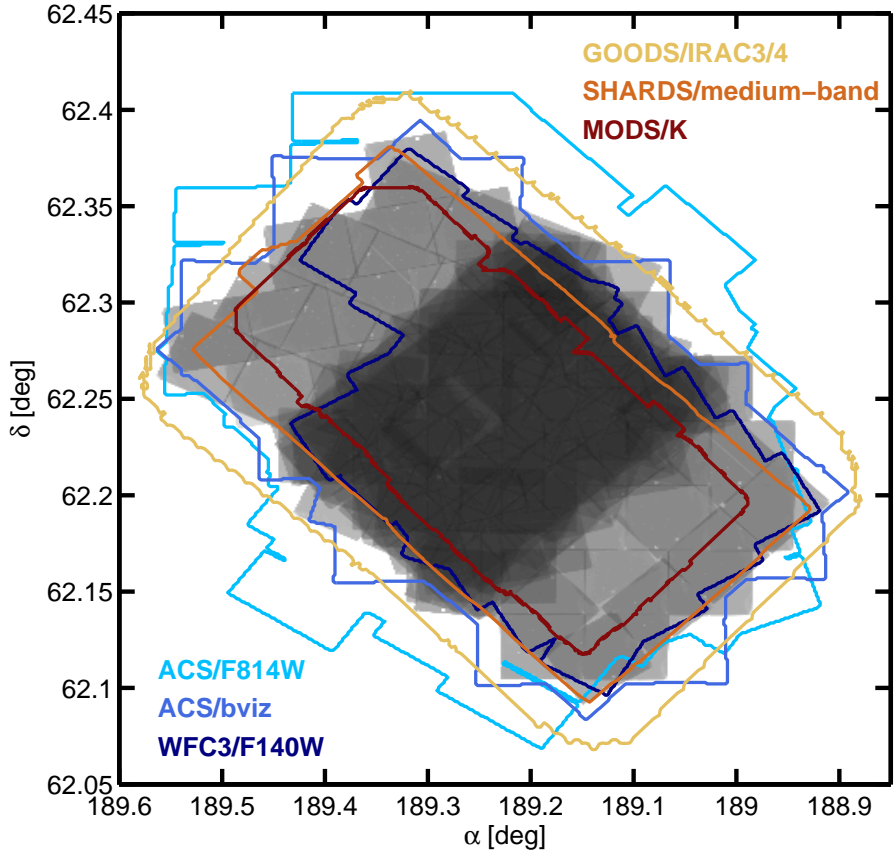}
\caption{\label{fig:footprint} Sky coverage of the multi-wavelength data
  sets used in the GOODS-N F160W catalog. The gray scale image shows
  the exposure time of the F160W mosaic, which includes the CANDELS
  wide and deep region. Coverage of ancillary data from UV to MIR are
  also shown: GOODS \hst/ACS (blue), MOIRCS/MODS Ks (red), GTC/SHARDS
  optical medium-bands (orange), and GOODS \spitzer/IRAC (yellow). The
  entire field is covered by both SEDS \spitzer/IRAC and CFHT/K.}
\end{figure*}

The CANDELS multi-wavelength photometric catalogs for the GOODS-S,
UDS, COSMOS and EGS fields have been presented in \citet{guo13},
\citet{galametz13}, \citet{nayyeri17} and \citet{stefanon17},
respectively; photometric redshifts and stellar population parameters
for the first two fields are presented separately in \citet{dahlen13}
and \citet{santini15}. This paper presents the multi-wavelength
catalog in GOODS-N, based on a CANDELS WFC3/F160W detection and making
use of all the available ancillary data spanning from the UV to FIR
wavelengths. Most notably, this catalog includes photometry in 25
medium-bands from the SHARDS survey \citep{shards}, which follows a
similar observational strategy as previous optical surveys, such as
COMBO17 \citep{wolf01,wolf03} and the COSMOS medium-band survey
\citep{ilbert09}, but provides higher spectral resolution ($R\sim50$)
and deeper photometry (4$\sigma$, H$\sim27$~mag) with an average sub
arcsec seeing.

Furthermore, we expand the high spectral resolution coverage to the
NIR by combining new WFC3 G102 grism observations with the publicly
released G141 data from the 3D-HST survey \citep{3dhstgrism}, which
yields a nearly continuous coverage from $\lambda=0.8-1.7$~$\mu$m with
a resolution better than $R=130$. Lastly, we complement the optical
and NIR photometry with a compilation of all the available FIR data
from {\it Spitzer} and Herschel, spanning from
$\lambda=24-500$~$\mu$m.

This paper is organized as follows: \S~\ref{s:datasets} briefly
summarizes the photometry datasets included in our
catalog. \S~\ref{s:photometry} discusses the detection process in the
CANDELS F160W image and photometry measurements on the {\it HST}\ and
mid-to-low spatial resolution images. \S~\ref{s:quality} presents
several tests to evaluate the quality of the multi-band
photometry. \S~\ref{s:addedvalue} presents the added-value properties
estimated from the fitting of the UV-to-FIR SEDs to stellar population
and dust emission templates.  The summary is given in
\S~\ref{s:summary}. The appendices describe the contents of
photometric and added value the catalogs, released together with this
paper, as well as the methodology to estimate self-consistent SFRs.

The CANDELS GOODS-N multi-wavelength catalog and its associated files
are made publicly available on the Mikulski Archive for Space
Telescopes
(MAST)\footnote[2]{https://archive.stsci.edu/prepds/candels/}. They
are also available in the Rainbow Database\citep{pg08, barro11b},
either through Slicer,\footnote[3]{US:
  http://arcoiris.ucsc.edu/Rainbow\_slicer\_public, and Europe:
  http://rainbowx.fis.ucm.es/Rainbow\_slicer\_public.} which allows a
direct download of images and catalogs, or through
Navigator,\footnote[4]{US:
  http://arcoiris.ucsc.edu/Rainbow\_navigator\_public, and Europe:
  http://rainbowx.fis.ucm.es/Rainbow\_navigator\_public.} which
features a query menu that allows users to search for individual
galaxies, create subsets of the complete sample based on different
criteria, and inspect cutouts of the galaxies in any of the available
bands.

All magnitudes in the paper are on the AB scale \citep{oke74} unless
otherwise noted. We adopt a flat ${\rm \Lambda CDM}$ cosmology with
$\Omega_m=0.3$, $\Omega_{\Lambda}=0.7$ and use the Hubble constant in
terms of $h\equiv H_0/100 {\rm km~s^{-1}~Mpc^{-1}} = 0.70$.

\begin{figure*}[t]
\centering
\includegraphics[width=11.25cm,angle=0.]{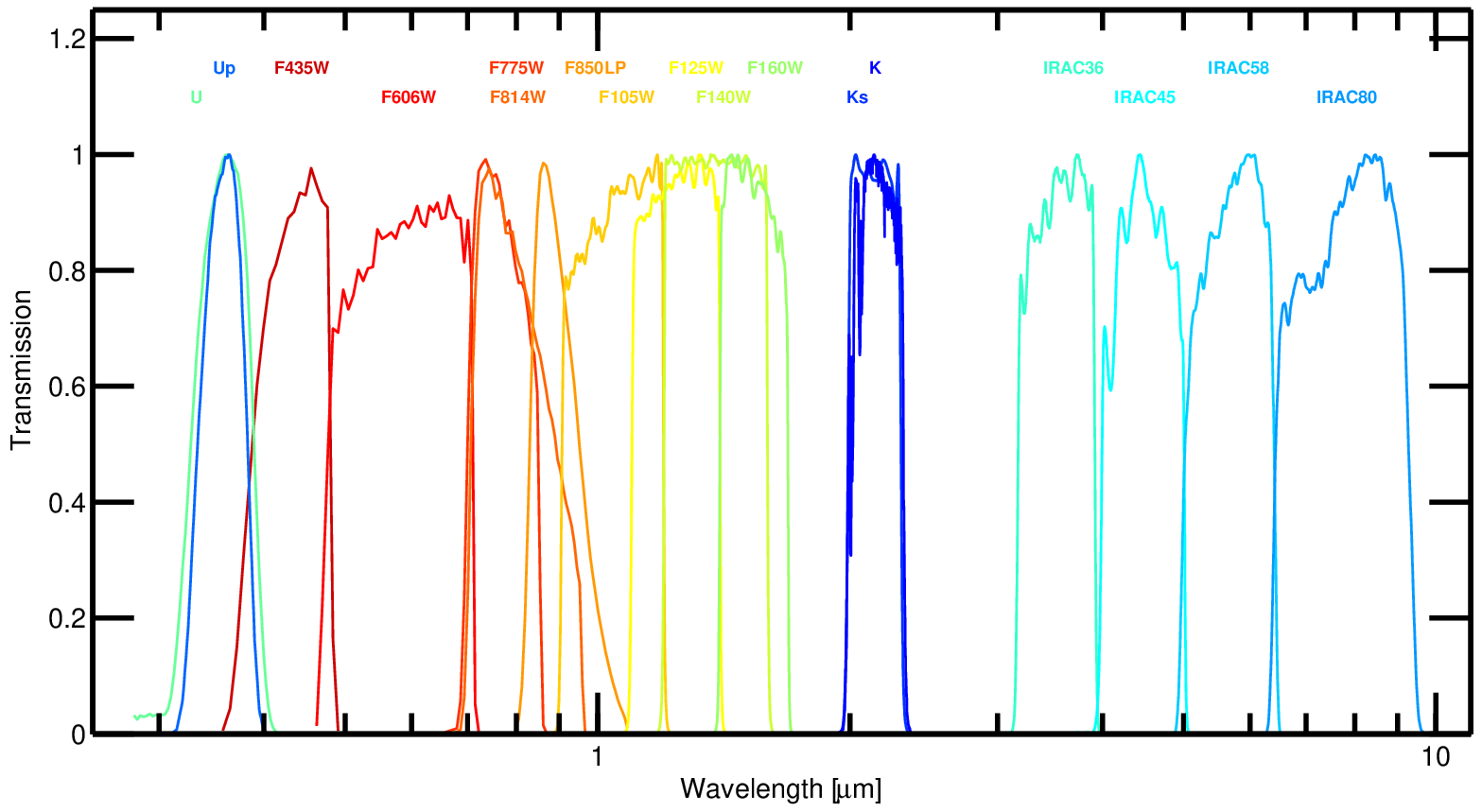}
\includegraphics[width=6cm,angle=0.]{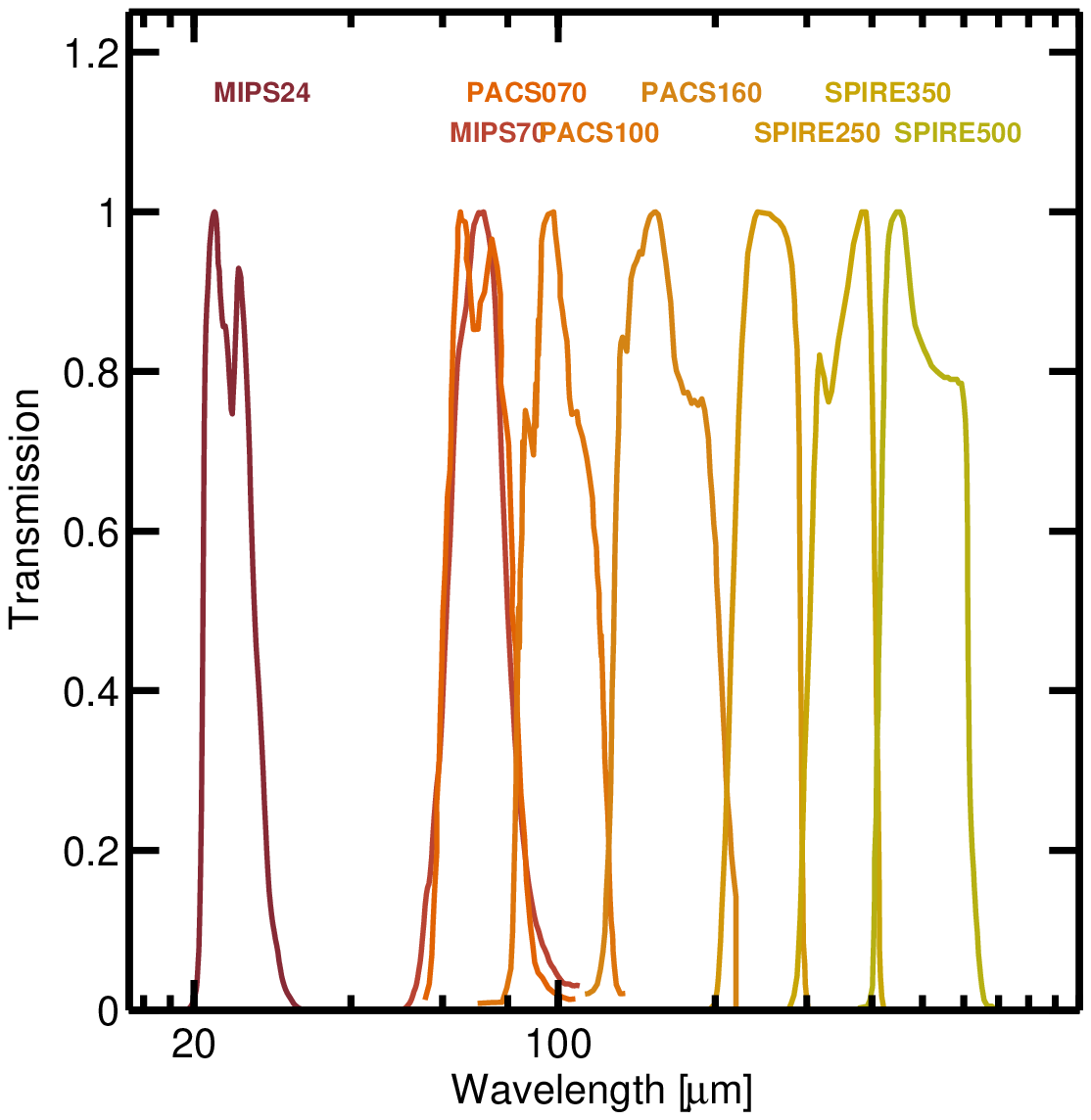}
\hspace*{0.50cm}
\includegraphics[width=13.95cm,angle=0.]{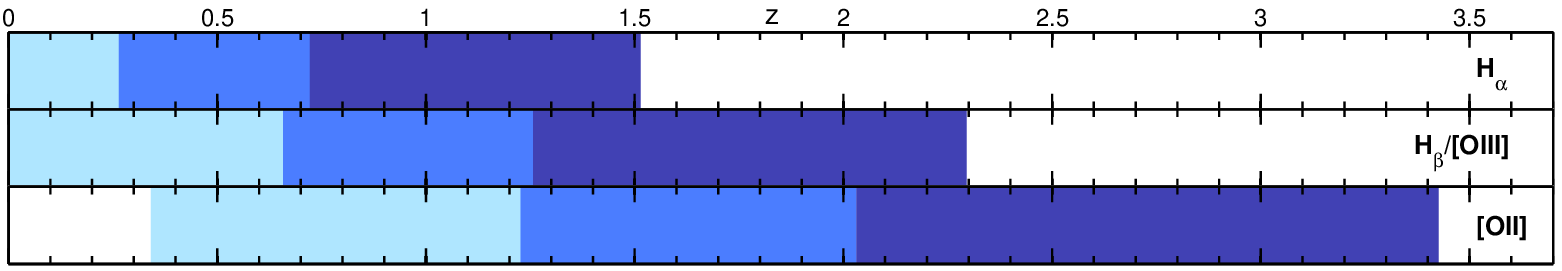}
\includegraphics[width=15cm,angle=0.]{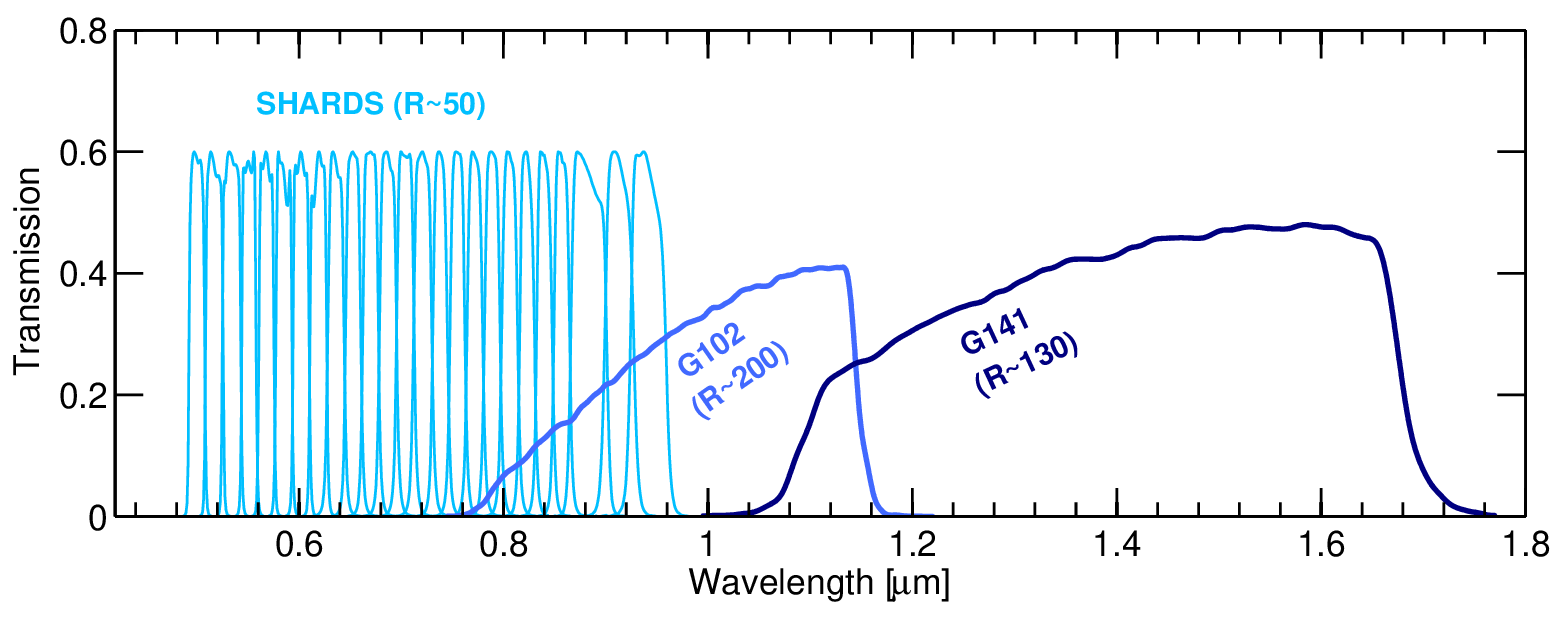}
\caption{\label{fig:filters} {\it Top:} Transmission curves of all the
  broad-band filters used in the CANDELS GOODS-N multi-wavelength
  catalog, from the UV, optical and NIR (left) to the FIR (right).
  {\it Bottom:} Transmission curves of the higher spectral resolution
  data: the optical medium-band survey SHARDS ($R\sim$50), and the two
  HST/WFC3 grisms, G102 ($R\sim200$) and G141 ($R\sim130$). The panel
  above shows the redshifts ranges in which the most prominent
  emission lines (\Ha, \OIII,~\Hb~and \OII) can be detected in each of
  these higher resolution dataset.}
\end{figure*} 

\section{Imaging Datasets}
\label{s:datasets}

The GOODS-N field \citep{goods}, centered around the Hubble Deep Field
North \citep[HDFN][]{williams96} at $\alpha$(J2000) = 12$^{\rm
  h}$36$^{\rm m}$55$^{\rm s}$ and $\delta$(J2000)
=+62$^{\circ}$14$^{\rm m}$11$^{\rm s}$, is a sky region of about 171
arcmin${\rm ^2}$ which has been targeted for some of the deepest
observations ever taken by NASA's Great Observatories: \hst, \spitzer,
and \chandra\ as well as by other world-class telescopes (see
Figure~\ref{fig:footprint}).

The multi-wavelength coverage of GOODS-N spans from X-ray, UV to far
IR and radio data: UV data from GALEX (PI C. Martin), ground-based
optical data from $U$ to $z$ bands taken by the Kitt Peak 4-m
telescope and from Suprime-Cam on the Subaru 8.2-m as a part of the
Hawaii Hubble Deep Field North project \citep{capak04}, 25
medium-bands from the GTC SHARDS \citep{shards} survey, near infrared
(NIR) J, H and Ks imaging from the Subaru MOIRCS deep survey
\citep{kajisawa09} and CFHT/WIRCam Ks photometry (Hsu et al. 2019);
IRAC maps from \spitzer~GOODS (Dickinson et al. 2003), SEDS
\citep{ashby13} and SCANDELS \citep{ashby15}; MIPS data from
GOODS-FIDEL (PI: M. Dickinson); Herschel from the GOODS-Herschel
(\citealt{elbaz11}) and PEP (\citealt{magnelli13}) surveys.

In the following we provide more details about the datasets included
in the multi-band catalog. The telescope/instrument as well as the
reference for the survey are given in
Table~\ref{table:ancil_data}. Table~\ref{table:gn_image_data} lists
the central wavelength of the filters, dust attenuation from Galactic
extinction, image zero point and the average FWHM for each of the
mosaics. Transmission curves for all filters are plotted in
Figure~\ref{fig:filters} .

\subsection{HST}
\label{ss:hst_ima}


\subsubsection{ACS Optical imaging}

The \hst/ACS F435W, F606W, F775W, and F850LP images used in our
catalog are the version v3.0 of the mosaicked images from the GOODS
\hst/ACS Treasury Program. They consist of data acquired prior to the
\hst\ Servicing Mission 4, including mainly data of the original GOODS
\hst/ACS program in \hst\ Cycle 11 \citep[GO 9425 and 9583;
see][]{goods} and additional data acquired in \hst/ACS F606W and F814W
as part of the CANDELS survey and during the search for high redshift
Type Ia supernovae carried out during Cycles 12 and 13 \citep[Program
ID 9727, P.I. Saul Perlmutter, and 9728, 10339, 10340, P.I. Adam
Riess; see, e.g.,][]{riess07}.


\subsubsection{WFC3 IR imaging}

\begin{figure}[ht]
\center{
\includegraphics[scale=0.47, angle=0]{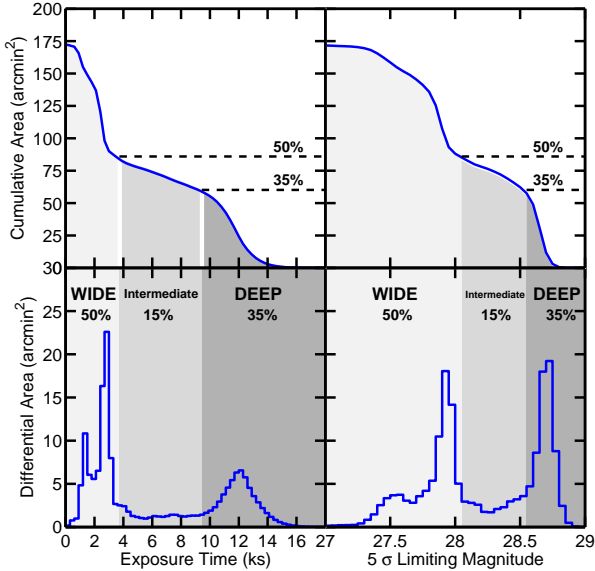}}
\caption[]{Distributions of exposure time and limiting magnitude of
  the F160W mosaic used as the detection image of our catalog. The
  {\it left} column shows the cumulative (upper panel) and
  differential (lower) distributions of the exposure time, while the
  {\it right} column shows the same distributions of 5$\sigma$
  limiting magnitude of the image.
\label{fig:limitmag}}
\vspace{-0.0cm}
\end{figure}

The CANDELS survey observed the GOODS-N field in 3 \hst/WFC3 IR
filters F105W, F125W, and F160W following a ``wedding-cake'' observing
strategy similar to that in the CANDELS/GOODS-S field but with only 2
layers, deep and wide (i.e., there is no ultra-deep region). The deep
region consists of a rectangular grid of 3$\times$5 pointings that
covers the central one-third of the mosaic (see
Figure~\ref{fig:footprint}) with an approximated area of $\sim$55
arcmin${\rm ^2}$ ($\sim35\%$ of the mosaic). The observations were
done over 10 epochs at 6 to 8 orbit depth in F125W and F160W. The wide
region covers the northern and southern two-thirds of the field
($\sim50\%$ of the mosaic) with 2$\times$4 pointings in both filters
and has approximately 2-orbit exposures. The distributions of exposure
time and limiting magnitude of the F160W mosaic are shown in
Figure~\ref{fig:limitmag}.  An intermediate-depth region (between
$\sim$4 and 9ks) is defined by the overlapping area between the wide
and deep regions. The CANDELS F105W observations consist only of the
deep and wide regions with a exposure gap in the intermediate
region. See \citet{candelsgro} and \citet{candelskoe} for more details
of CANDELS \hst/WFC3 observations and data reduction.  We also include
in our catalog the 2-orbit depth F140W images taken as a part of the
G141 AGHAST survey GO: 11600 (PI: B.\ Weiner; see next section) and
GO:12461 COLFAX (PI: Riess).

The WFC3 mosaics used in this paper have been reduced following the
same data reduction strategy described in the previous CANDELS data
release papers for the other fields.  The images in all bands are
drizzled to 0\farcs06/pixel to match the default CANDELS pixel scale
\citep[see][for details]{candelskoe}.

\subsubsection{WFC3 G102 and G141 grism spectroscopy}

The GOODS-N field was observed in the {\it HST}/G141 grism at a
2-orbit depth as a part of the AGHAST program (GO:11600; PI:
Weiner). The 28 pointings of the program were reduced, analyzed and
incorporated to the 3D-HST survey (\citealt{3dhst};
\citealt{3dhstgrism}), which uses a similar observing strategy over
the other 4 CANDELS fields. Each pointing was observed for two orbits,
with $\sim$800~s of direct imaging in the F140W filter and 4511-5111~s
with the G141 grism per orbit. The observations were arranged in a
4$\times$6 grid. There is no imaging or grism spectra in the
northwestern edge of the field (dark blue line in
Figure~\ref{fig:footprint}). In this paper we make use of the 3D-HST
spectra released in their v4.1.5 data products described in
\citet{3dhstgrism}.

Furthermore we present complementary {\it HST}/G102 observations
(GO:13179; PI: Barro) that were designed to follow the same tiling
strategy of AGHAST program in other to maximize the number of galaxies
with simultaneous grism coverage. The observations consist of 28 two
orbit depth pointings with 400~s of direct imaging in the F105W filter
and $\sim$5000~s with the G102 grism per orbit following the same
4-point dither pattern of 3D-HST survey. The observations were
processed using the 3D-HST data reduction pipeline described in
\citet{3dhstgrism}. The pipeline combines the individual G102
exposures into mosaics using AstroDrizzle \citep{gonzaga12}.  These
individual exposures are aligned using \texttt{tweakreg} and grism sky
backgrounds are subtracted using master sky images as described by
\citep{brammer15}. Each exposure is then interlaced to a final image
with a pixel size of $\sim0\farcs06$. Before sky-subtraction and
interlacing each individual exposure was checked and corrected for
elevated backgrounds due to the He Earth-glow using the
script\footnote{\url{https://github.com/gbrammer/
 wfc3/blob/master/reprocess\_wfc3.py}} described by \citep{brammer14}.

From the final G102 mosaics, the spectra of each individual object are
extracted by predicting the position and extent of each
two-dimensional spectrum based on the \texttt{SExtractor} \citep{sex}
segmentation of the CANDELS F160W image. As this is done for every
single object, the contamination, i.e., the dispersed light from
neighboring objects in the direct image field-of-view is estimated and
accounted for. We also carried out visual inspections of the
individual 2D and 1D extractions for a magnitude limited subset of the
data (F105W$<23$~mag) in order to flag catastrophic failures. The
automated redshift determination and the emission line measurements
based on these G102 and G141 datasets are presented in
\S~\ref{ss:emission_lines}.

\begin{table*}[ht]
\centering
\caption{Image sources}\label{table:ancil_data}
\begin{tabular}{lllll}
\hline \hline
\noalign{\smallskip}
Filters & Telescope/Instrument & Survey & Reference \\
\hline
\noalign{\smallskip}
 $U$  & KPNO 4m/Mosaic & Hawaii HDFN & \citet{capak04}\\
 $U'$  & LBT/ LBC & -  & \citet{grazian17}\\
 25 medium-band optical  &  GTC / OSIRIS &  SHARDS & \citet{shards}\\
 F435W, F606W, F775W, F850LP   & HST/ACS & GOODS  &  \citet{goods} \\
 F814W  & HST/ACS       & CANDELS &\citet{candelsgro, candelskoe}\\
 F105W, F125W, F160W    & HST/WFC3 & CANDELS &\citet{candelsgro, candelskoe}\\
 F140W   & HST/WFC3     & AGHAST   &   GO: 11600 (PI: B.\ Weiner)\\
 $K_s$  & Subaru/MOIRCS & MODS &  \citet{kajisawa11}\\
 $K$  & CFHT/Megacam    & - & \citet{Hsu19}\\
 3.6, 4.5 $\mu$m & Spitzer/IRAC & SEDS & \citet{ashby13} \\
 5.8, 8 $\mu$m & Spitzer/IRAC & GOODS &  Dickinson et al. (2003)\\
24, 70 $\mu$m & Spitzer/MIPS & GOODS/FIDEL &  Dickinson et al. (2003)\\
100, 160 $\mu$m & Herschel/PACS & PEP & \citet{berta11}, \citet{lutz11}\\
250, 350, 500 $\mu$m & Herschel/SPIRE & GOODS/Herschel, HerMES & \citet{oliver12},\citet{magnelli13}\\
\hline
\end{tabular}
\end{table*}

\subsection{Ground-Based Imaging}
\label{ss:ground_ima}
\subsubsection{Ultraviolet}

The $U$-band image was taken with the Mosaic camera on the Kitt Peak
4-m telescope by the Hawaii Hubble Deep Field North project
\citep{capak04}.\footnote{\url{http://www.astro.caltech.edu/~capak/hdf/index.html}}

In addition to the Kitt Peak imaging, an LBT Strategic Program (PI
A. Grazian) was approved on 2012B, with the aim of obtaining ultra
deep imaging in the U band of the CANDELS/GOODS-N field using the LBC
instrument at the prime focus of the LBT telescope
(\citealt{giallongo08}; \citealt{rothberg16}). The program consisted
on approximately 25 hours on a single pointing of the LBC camera. The
LBC field of view is larger than the whole CANDELS/GOODS-N field, and
it covers approximately 600 sq. arcmin. with homogeneous
coverage/depth. The same area has been observed also by other LBT
partners (AZ, OSURC, and LBTO), for a total exposure time of 33 hours
in U band (seeing 1.1 arcsec). The detailed description of these data
is provided in a dedicated paper \citep{grazian17} summarizing all the
LBC deep observations available in the CANDELS fields. The relatively
long exposure time and the good seeing allowed to reach a magnitude
limit in the U band of 30.2 AB at S/N=1, resulting in one of the
deepest UV images ever obtained.

\subsubsection{SHARDS Optical medium-band survey}
\label{sss:SHARDS}
The Survey for High-z Absorption Red and Dead Sources
\citep[][SHARDS]{shards}, an ESO/GTC Large Program, targeted the
GOODS-N field with GTC/OSIRIS in 2012-2015 and obtained 220 hours of
ultra-deep imaging data through 25 medium-band optical filters. The
wavelengths covered a range from 500 to 950~nm with a spectral
resolution of $R\sim50$. The depth is 26.5~mag at the $4\sigma$ level
(at least) and the seeing was always below 1\arcsec~for every single
filter. SHARDS used 2 OSIRIS (FOV 7.8\arcmin$\times$7.8\arcmin)
pointings to cover most of the CANDELS region
(110~$\rm{arcmin}^2$). The SHARDS optical imaging data has a
particular characteristic that has to be taken into account to obtain
accurate spectral energy distributions: the passband of the filter
seen by different parts of the detector changes, getting bluer as we
move away from the optical axis, which is located about 1\arcmin~ to
the left of the FOV. Therefore, every galaxy detected by SHARDS counts
with a unique set of SHARDS passbands, which are defined by their
transmission curves (whose shapes does not change and, therefore, are
the same for all galaxies) and their central wavelengths (which change
and must be provided for each galaxy). We remark that this is an
optical effect which affects all filters, so the final spectral energy
distribution for each galaxy counts with the same spectral resolution,
$R\sim50$, but all the filters are offsetted from the nominal central
wavelength. In order to properly account for this effect, the SHARDS
photometry of the F160W sources (see \S~\ref{ss:midres_tfit}) is
provided in a separated catalog (see Table~\ref{table:shardsphot})
which includes the central wavelength for each galaxy and filter.
Furthermore, the SHARDS science images in each of the 25 filters,
which are released with this paper, are provided jointly with a map of
the central wavelength for each pixel that can be used to account for
the wavelength shift (see \citealt{shards} for more details).

\subsubsection{Near Infrarred}
Deep $K_s$-band images of the field were taken using Multi-Object
Infrared Camera and Spectrograph (MOIRCS) on Subaru as part of the
MOIRCS Deep Survey (MODS,
\citealt{kajisawa11}).\footnote{\url{http://www.astr.tohoku.ac.jp/MODS}}.
The data reach 5$\sigma$ total limiting magnitude for point sources of
$K_s = 24.2$ over a 103 arcmin$^{2}$ mosaic consisting of 4 MOIRCS
pointings. The central $\sim$28 arcmin$^{2}$ of the mosaic contains a
deeper region were the data reach $K_s = 25.1$. In this work we make
use of the publicly available ``convolved" mosaic in which each of the
four pointings have been homogenized to match the field with the worst
seeing (FWHM$\sim$0\farcs6).

In addition to the MOIRCS data, we also make use of a deep broad-band
$K_s$ mosaic based on observations with the CFHT WIRCam instrument
(Hsu et al. 2019). The final mosaic used in this paper covers $\sim$0.4
square degrees around the GOODS-N field. It has a 50\% completeness
limit for point sources between $K_s=24.6-24.8$~mag. The astrometry
was calibrated using the Two Micron All Sky Survey (2MASS) catalog
\citep{2mass} with a final internal accuracy of $\sim$0\farcs1.

\subsection{Spitzer/Herschel mid-to-far IR}
\label{ss:fir_ima}
\subsubsection{IRAC S-CANDELS}

GOODS-N was observed by \spitzer/IRAC \citep{fazio04} during the
cryogenic mission in four bands (3.6, 4.5, 5.8, and 8.0 $\mu$m) for
two epochs with a separation of six months (February 2004 and August
2004) by the GOODS Spitzer Legacy project (PI: M. Dickinson). Each
epoch contained two pointings, each with total extent approximately 10
arcmin on a side. The exposure time per band per sky pointing was
approximately 25 hours per epoch and doubled in the overlap region. We
use the 5.8 and 8.0 $\mu$m imaging from this program in our catalog.

The 3.6 and 4.5\,$\mu$m photometry was measured on the mosaics from
the {\sl Spitzer} -CANDELS (S-CANDELS, PID 80216; \citealt{ashby15})
survey, which combines the original cryogenic data with that taken
from the warm mission phase. The resultant 3.6~$\mu$m and 4.5~$\mu$m
mosaics fully cover the WFC3 F160W area of the CANDELS survey to a
depth of at least 50 hours.  The IRAC data in all 4 bands were
reprocessed and mosaicked using the same CANDELS HST tangent-plane
projection and with a pixel scale of 0\farcs06/pixels, to prepare them
appropriately for further photometric analysis (see also
\citealt{ashby15}).

\subsubsection{MIPS GTO, PEP \& GOODS-Herschel}

The GOODS-N field has been observed in the mid-IR wavelengths with
{\it Spitzer}/MIPS at 24~$\mu$m and 70~$\mu$m as part of the GTO and
GOODS surveys \citep[][see also \citealt{frayer06}]{dickinson03}. Here
we use the photometric catalogs in both bands described in
\citet{pg05,pg08} which are based on the reduced and mosaicked
data. Furthermore, far-IR observations with the Photodetector Array
Camera and Spectrometer (PACS; \citealt{pacs}) and the Spectral and
Photometric Imaging REceiver (SPIRE; \citealt{spire}), on board the
Herschel Space Observatory were obtained as part of the PACS
Evolutionary Probe (PEP; \citealt{berta11}; \citealt{lutz11}),
GOODS-Herschel \citep{magnelli13} and HerMES \citep{oliver12}
surveys. The 5$\sigma$ detection limits of the far-IR data are
provided in Table~\ref{table:photometry_ir}. The mid-to-far IR
photometry probes the rest-frame wavelengths close to the peak of the
dust IR emission of galaxies up redshifts of $z\sim3$. Therefore it
provides a very useful SFR indicator, complementary to the UV
luminosity, for large number of galaxies. See \S~\ref{ss:lowres_ir}
and appendix~\ref{ap:SFRvalidation} for a more detailed description of
the IR data and the photometric measurements.

\begin{figure}[t]
\centering
\includegraphics[width=9cm,angle=0.]{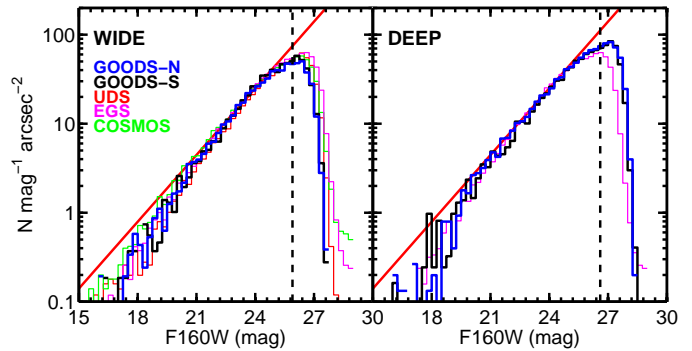}
\caption{\label{fig:nocounts} Differential number counts for objects
  detected in the GOODS-N F160W catalog on the wide (left) and deep
  (right) regions. The different histograms show the number counts in
  all 5 CANDELS fields for comparison. The UDS, COSMOS and EGS fields
  were only observed to ``wide'' depth. The solid red line in each
  panel shows the best power-law fit to the number counts in the
  magnitude range where the sample is complete.}
\end{figure}

\subsection{Value-added data}
\label{ss:added_value}
\subsubsection{Spectroscopic Redshifts}

A number of different spectroscopic observations were conducted in the
GOODS-N field over the course of the last 20 years. Here we include
redshift compilations based primarily on large spectroscopic surveys
using the {\it Keck}/DEIMOS optical spectrograph: the ACS-GOODS
redshift survey (\citealt{cowie04}; \citealt{barger08}), the Team Keck
Redshift Survey \citealt[TKRS][]{wirth04}) and the DEEP3 galaxy
redshift survey (\citealt{deep3}). We also included redshifts from a
number of other smaller surveys that often targeted specific types of
objects or small regions defined by observations with new instruments:
Lyman-Break galaxies \citep{reddy06}, bright-IR galaxies
\citep{pope08}, sub-mm galaxies \citep{chapman05} or the ACS-grism
PEARS survey \citep{ferreras09}. Furthermore, we complemented these
optical redshifts with results from recent NIR spectroscopic campaings
using the {\it Keck}/MOSFIRE spectrograph that are critical to
increase the number of secure spectroscopic redshifts beyond
$z\sim1.5$ : the 1st epoch of the MOSFIRE Deep Evolution Field
(MOSDEF) survey \citep{mosdef} and the TKRS2 \citep{wirth15}. The
extensive spectroscopic campaigns in GOODS-N yield a total of
$\sim$5000 unique redshifts within the CANDELS F160W mosaic coverage,
and $\sim$3000 of those with a highly reliable quality flag. The
counterparts to the spectroscopic sources were identified using a
crossmatch radius of 0\farcs8 (if more than one object falls within
the matching radius, the closest match with the highest confidence
flag was adopted). All spectroscopic identifications are listed in the
catalogs, but only those with reliable quality flags are used in the
analysis of galaxy properties.

\subsubsection{X-ray}

We used X-ray data from the Chandra 2 Ms source catalog by
\citet{chandra2m}, covering the entire surveyed region of the F160W
mosaic in GOODS-N, to select candidates to harbor an AGN within our
sample.  The most likely X-ray counterparts to the CANDELS sources
were identified using a cross-matching radius of 2\farcs5 between the
CANDELS F160W catalog and the X-ray catalog of \citet{chandra2m}.  We
identify a total of 316 X-ray sources with a reliable F160W
counterpart. This makes $\sim3\%$ of all the sources in the F160W
catalog down to $H<24.5$.
\begin{table*}[ht]
\centering
\caption{GOODS-N Optical-to-NIR Imaging}\label{table:gn_image_data}
\begin{tabular}{llllllll}
\hline \hline
\noalign{\smallskip}
Band & $\lambda_{\mathrm{central}}$ & $A_{\lambda}$ & Zero Point & FWHM & ZP-corr & $5\sigma$ Depth \footnote{Based on aperture photometry with radius equal to the FWHM of the PSF in each band}\\
 & ($\mu m$) & (mag)  &(AB) & (arcsec) & (flux) & (mag)\\
\hline
\noalign{\smallskip}
U & 0.35929& 0.052& 31.369 & 1.26 & 0.88 & 26.7\\
U'& 0.36332& 0.052& 26.321 & 1.10 & 1.07 & 28.2\\
F435W& 0.43179& 0.044& 25.689  & 0.10& 1.03 & 27.1\\
SHARDS\footnote{25 medium bands, see Table~A3 in the appendix and \citet{shards} for more details} & 0.50-0.94 &-& - & -& - & -\\
F606W& 0.59194& 0.030& 26.511 & 0.10& 0.97 & 27.7\\
F775W& 0.76933& 0.020& 25.671 & 0.11& 0.98 & 27.2\\
F814W& 0.76933& 0.020& 25.671 & 0.11& 0.97 & 28.1\\
F850LP& 0.90364& 0.015& 24.871 & 0.11& 1.02 & 26.9\\
F105W& 1.24710& 0.009& 26.230 & 0.18& 1.03 & 26.4\\
F125W& 1.24710& 0.009& 26.230 & 0.18& 1.01 & 27.5\\
F140W& 1.39240& 0.007& 26.452 & 0.18& 1.04 & 26.9\\
F160W& 1.53960& 0.006& 25.946 & 0.19& 1.03 & 27.3\\
K&  2.13470& 0.004& 26.000 & 0.60& 0.92 & 24.4 \\
Ks& 2.15770& 0.004& 26.000 & 0.60& 0.96 & 24.7\\
IRAC1& 3.55690& 0.000& 21.581 & 1.7 & 0.93 & 24.5\\  
IRAC2& 4.50200& 0.000& 21.581 & 1.7 & 0.90 & 24.6\\ 
IRAC3& 5.74500& 0.000& 20.603 & 1.9 & 0.87 & 22.8\\ 
IRAC4& 7.91580& 0.000& 21.781 & 2.0 & 0.80 & 22.7\\ 
\noalign{\smallskip}
\hline
\noalign{\smallskip}
\end{tabular}
\end{table*}

\section{Photometry}
\label{s:photometry}
This section discusses the methods used to assemble the UV-to-FIR
multi-wavelength photometric catalog. The following subsections
describe the procedures to identify and characterize all the sources
detected in the WFC3/F160W image and to obtain self-consistent
photometric measurements in the high, intermediate, and low resolution
photometric datasets.

\subsection{High resolution HST data}
\label{ss:highres_hst}
\subsubsection{WFC3 F160W detection and photometry}
We follow a similar approach as in the previous four CANDELS data
papers (\citealt{guo13}; \citealt{galametz13}; \citealt{stefanon17};
\citealt{nayyeri17}). We identify sources using the reddest NIR band,
WFC3/F160W, mosaic. Both source detection and photometry were
performed using a slightly modified version of \texttt{SExtractor}
v2.8.6 \citep{sex} that fixes some known issues that often cause the
inclusion of false detections in the final catalog merged with real
sources, and a sky over-subtraction that could affect faint extended
sources \citep[see][]{galametz13}.

\begin{figure*}[t]
\includegraphics[scale=0.4, angle=0]{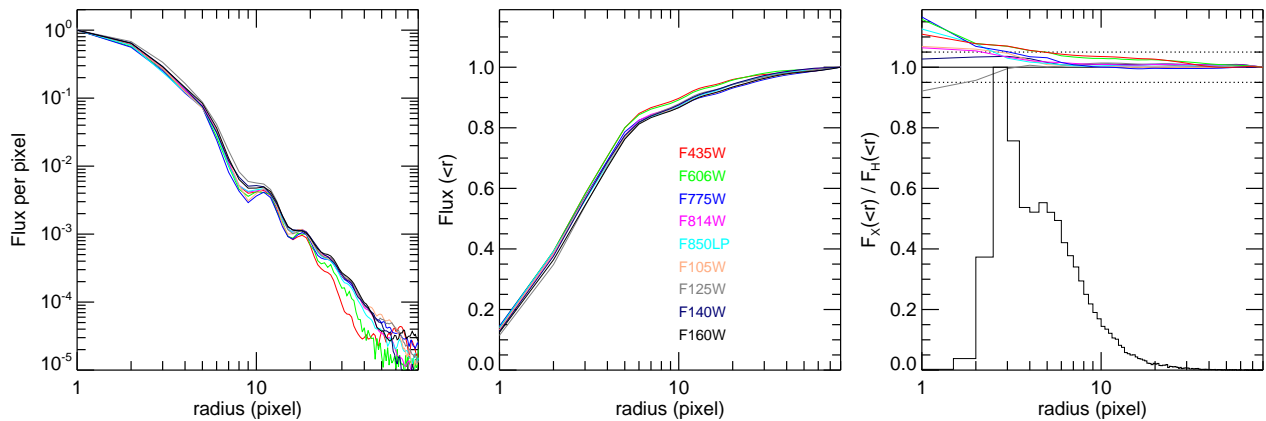}
\caption[]{Accuracy of PSF matching between other \hst\ bands and
  F160W. {\it Left}: the light profile of matched PSFs for each
  band. {\it Middle}: the curve of growth of each matched PSF. {\it
    Right}: the curve of growth of each matched PSF normalized by the
  curve of growth of the F160W PSF. In this panel, curves with values
  greater than unity are under-smoothed, and vice versa. All curves
  are color coded as labels in the middle panel show. Dotted lines in
  the right panel show the 5\% relative error. The solid histogram in
  the right panel shows the distribution of isophotal radii of all
  objects in our catalog.
\label{fig:psfcurve}}
\end{figure*}

The source detection is based on the 2-step ``cold'' plus ``hot''
strategy described in more detail in the CANDELS UDS
\citep{galametz13} and GOODS-S \citep{guo13} papers. Briefly, we ran
\texttt{SExtractor} twice using two different parameter configurations
(see Table~\ref{table:sex}) aimed at: detecting bright/large sources
without over-deblending them ({\it cold-mode}), or pushing the
detection limit to recover faint sources close to the limiting depth
of the mosaic ({\it hot-mode}). Then, we merge the cold and hot
catalogs following a similar approach as the \texttt{GALAPAGOS} code
(see \citealt{barden12} for more details). All cold-mode detected
sources are included in the merged catalog, but only those hot-mode
sources whose segmentation map does not overlap with the photometric
\citep{kron80} ellipse of a cold-mode source are included, i.e.,
hot-mode sources that are clearly overlapping with a cold-mode
detection or the result of excessive shredding are excluded from the
merged catalog. We detect 35445 sources in the F160W mosaic. Among
them, 27293 sources are detected by the cold mode and 8152 sources by
the hot mode.


Figure~\ref{fig:nocounts} shows the distribution of detected sources
in the F160W mosaic as a function of magnitude (i.e., the differential
number counts). The left panel depicts the number counts in the
GOODS-N wide region compared to those measured in regions of similar
depth in the other 4 CANDELS fields. All the measurements are in good
agreement up to $H\sim$~24~mag.  As pointed out in \citet{stefanon17},
the number counts in the wide region of the GOODSs and COSMOS fields
are slightly below those in the UDS and EGS measurements in the
$24<H<26.5$ range, most likely due to the slightly deeper data
(0.2~mag) in EGS compared to those fields. The right panel of
Figure~\ref{fig:nocounts} compares the number counts in the deep
regions of GOODS-N and GOODS-S, which are consistent up to
$H\sim29$~mag.

As expected, the bulk of the number counts are consistent with those
in the wide region (the counts in EGS are shown again as reference),
while the number of detections at fainter magnitudes $H\gtrsim26$
increases. The differential variation of the number counts in the
faint end can be used to assess the completeness of the catalog.
Following the approach in \citet{guo13}, we fit the number counts in
the region were the catalog is expected to be complete ($20<H<24$)
with a power-law of slope $\gamma = 0.20 \pm 0.06$. Then, we find the
$\sim50\%$ completeness limit by computing the magnitude where the
relative difference between the observed counts and the power-law
reaches a factor of two: $H\sim25.9$ and 26.6~mag in the wide and deep
regions, respectively (dashed lines in Figure~\ref{fig:nocounts}). These
values agree with the completeness limits of the CANDELS/GOODS-S
catalog at similar depths (see also Fig.~3 of \citealt{duncan14}). We
refer the reader to \citet{guo13} for a more detailed discussion on
the dependence of the completeness limit with the surface brightness
profiles of the galaxies in GOODS-S. Given the similar depths of the
GOODSs fields, those results are directly applicable to the GOODS-N
catalog.

\begin{table}[htbp]
\tabletypesize{\scriptsize}
\begin{center}
\caption{SExtractor Parameters in Cold and Hot Modes \label{table:sex}}
\begin{tabular}{ccc}
\hline\hline
        &   Cold Mode  & Hot Mode \\
\hline
DETECT\_MINAREA &  5.0   & 10.0 \\
DETECT\_THRESH  &  0.75  & 0.7 \\
ANALYSIS\_THRESH & 5.0   & 0.8 \\
FILTER\_NAME  &   tophat\_9.0 & gauss\_4.0 \\
DEBLEND\_NTHRESH & 16 & 64 \\
DEBLEND\_MINCONT & 0.0001 & 0.001 \\
BACK\_SIZE    &   256    &  128 \\
BACK\_FILTERSIZE  & 9 & 5 \\
BACKPHOTO\_THICK & 100  & 48 \\
MEMORY\_OBJSTACK & 4000  & 4000 \\       
MEMORY\_PIXSTACK & 400000 & 400000 \\
MEMORY\_BUFSIZE  & 5000 & 5000 \\
\hline
\end{tabular}
\end{center}
\end{table}

\subsubsection{Photometry Flags}

At this stage we also assigned a photometry flag to every object in
the catalog. The flagging system is the same adopted in previous
CANDELS papers and discussed in detail by \citet{galametz13}. Briefly,
the flagging scheme is based on the properties of the F160W mosaic. We
use a zero for sources with reliable photometry and assigned a value
of one either for bright stars or spikes associated with those
stars. The radius of the star's masks range between $3-5\arcsec$ for
$\sim$20 intermediate brightness stars and 10$\arcsec$ for the 2
brightest stars in the field. A photometric flag of two is associated
with the lower exposure edges of the mosaic or defects as measured
from the F160W RMS maps. This is a very conservative flag assigned
only to pixels with extreme ($>$1E5) values of the RMS map.

\subsubsection{Optical/NIR HST photometry}

The photometry in all other \hst\ bands $-$ ACS F435W, F606W, F775W,
F814W, F850LP and WFC3 F105W, and F125W was measured running
\texttt{SExtractor} in dual mode using the F160W mosaic as reference
to ensure that the colors are measured within apertures of the same
size. This means that the multi-band photometry is computed only for
the sources detected in the F160W mosaic. We follow the same cold+hot
routine described in the previous section by running
\texttt{SExtractor} twice per band. In order to take into account the
variations in spatial resolution as a function of wavelength (see the
typical FWHMs of the HST bands in Table~\ref{table:gn_image_data}) all
HST images were previously smoothed to the lower spatial resolution of
F160W (FWHM$\sim0\farcs18$) using the \texttt{IRAF/PSFMATCH} package
with kernels that matched the multi-band PSFs with that of F160W. We
computed semi-empirical PSFs in the WFC3 bands by combining a stack of
isolated, unsaturated stars from across the mosaic with synthetic PSFs
generated with TinyTim \citep{tinytim}. We used the central pixels
from the synthetic models and the wings of stacked stars (see
\citealt{vdw12} for more details). The ACS PSFs were based on purely
empirical models computed by stacking well-detected stars, without any
artifacts, in each ACS band.

The left panel of Figure~\ref{fig:psfcurve} compares the stacked light
profile of several high S/N stars, extracted from the deep region, in
all HST bands after running \texttt{PSFMATCH}. The central and right
panels shows also the curves of growth (fraction of light enclosed as
a function of aperture size) in each band and the fraction of enclosed
light relative to that of the F160W PSF. All profiles converge quickly
to unity after a few pixels, and the relative photometric error in all
\hst\ bands is less than $5\%$ for apertures larger than two pixels
(0\farcs12), which is larger than the typical isophotal radius for the
bulk of the sources ($\gtrsim97\%$).

We computed several different photometric measurements available in
the \texttt{SExtractor} configuration, namely, \texttt{FLUX\_AUTO},
measured on Kron elliptical apertures, \texttt{FLUX\_ISO} measured on
elliptical isophotes, and \texttt{FLUX\_APER} measured in a series of
11 circular apertures (see appendix~\ref{ap:photometry} for a
description of all the measurements included in the photometric
catalog). As discussed in the previous CANDELS data papers, we adopt
\texttt{FLUX\_AUTO} as the default ``total'' photometry for all the
sources in the F160W band, while for the other bands we determine the
total flux scaling \texttt{FLUX\_ISO} by the ratio of
\texttt{FLUX\_AUTO} / \texttt{FLUX\_ISO} in each band. This ratio is
used to convert their isophotal fluxes and uncertainties into the
total fluxes and uncertainties.  The isophotal correction ensures that
the flux is measured within the same isophotal area in all bands
(defined by the F160W segmentation map) and maximizes the S/N for
faint sources. This method provides an accurate estimate of colors and
fluxes subject to the prior assumption that the PSF-convolved profile
is the same in all bands. We verify the quality of the multi-band SEDs
in \S~\ref{s:quality} by performing both internal and external checks,
comparing to another catalog.

\subsection{Intermediate resolution ground- and space-based data: \texttt{TFIT}}
\label{ss:midres_tfit}

We computed multi-wavelength photometry in all the ancillary ground
based data and in the {\it Spitzer}/IRAC bands using the \texttt{TFIT}
code \citep{tfit} and following the same methodology described in the
previous CANDELS data papers. \texttt{TFIT} is a template-fitting
software conceived to overcome the issues related with obtaining
consistent photometry across large datasets that exhibit significant
differences in spatial resolution. The code uses accurate positional
information of the sources in the highest resolution band (in this
case HST/F160W) to create PSF-matched models (``templates'') of the
sources in the intermediate resolution bands (e.g., ground-based
K-band or IRAC). These ``templates'' are computed on an
object-by-object basis by smoothing the high resolution cutouts to
low-resolution using a convolution kernel (see e.g.,
\citealt{galametz13}). Then, the code fits iteratively for the
photometry by comparing the real to the modeled images in those
bands. With the simultaneous fitting approach, the code can take into
account the flux contamination for each source due to their
neighboring objects. The details of the software are described in
detail in \citet{tfit}, \citet{papovich01} and \citep{lee12} which
includes a set of simulations to validate the photometric measurements
and quantify its uncertainties. See also \citet{merlin15,merlin16} for
further tests and improvements on the code, branched as {\tt
  T-PHOT}. For this paper we chose to use the original {\tt TFIT} for
consistency with all the previous CANDELS catalogs and also with early
internal releases of the GOODS-N catalog.

In the following, we briefly summarize the main steps involved in the
\texttt{TFIT} photometric measurements. Before running the fitting
code, we perform an additional background subtraction step of the
intermediate resolution images to ensure that there are no
inhomogeneous regions that could potentially bias the photometry. The
iterative background fitting script is based on an IRAF script
``acall'' originally developed for GOODS (M. Dickinson 2013, private
communication; see \citealt{guo13} for more details).  Then, the
images are re-sampled to a pixel scale that is a multiple of the F160W
pixel scale (e.g., $\sim10\times$ for \spitzer/IRAC) using {\tt SWARP}
\citep{swarp}. Lastly, we identify and stack several bright, isolated
stars in each band to determine its average PSF and to compute the
transformation kernel required to match the PSF of the high resolution
F160W band.  The fitting ``templates'' for each galaxy are computed by
convolving their F160W segmentation maps with such kernels. As
discussed in the previous CANDELS data papers, we apply a small
``dilation'' correction to the F160W segmentation map to avoid an
artificial truncation of the light profiles of the sources. The
dilation factor was determined following the empirical relation in
Equation 3 of \citet{galametz13}.

\begin{figure}[t]
  \centering
\includegraphics[scale=0.60, bb=140 281 483 515]{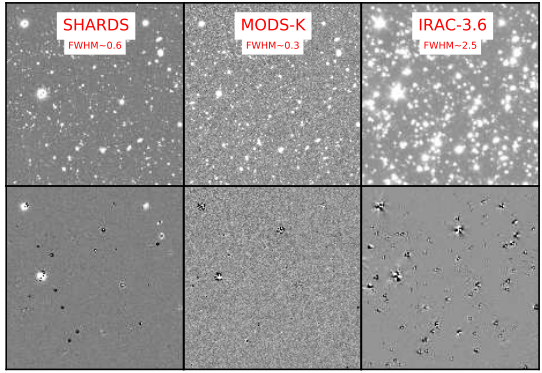}
\caption[]{\label{fig:resid} Example of the original image ({\it top})
  and the residual image after TFIT procedure ({\it bottom}) of
  several low-resolution bands as indicated in the upper panels in a
  representative sky region. Positive residuals in SHARDS and IRAC
  images are due to saturation around bright sources.}
\end{figure}

We run \texttt{TFIT} separately in all the ground based and {\it
  Spitzer}/IRAC images. As mentioned above, the flux for each object
is determined by fitting its template, and those of the neighboring
objects, to the intermediate resolution image, thus obtaining a direct
estimate of the possible flux contamination due to blending. The code
runs the fitting step twice, and the second iteration allows for small
shifts in the PSF-matched kernels to improve lower quality fits caused
by small image distortions in the intermediate resolution
images. Figure~\ref{fig:resid} shows examples of the \texttt{TFIT}
residual map i.e., the result of subtracting the best-fit object
templates from the original image, in three bands with different
spatial resolutions demonstrating that the fitting procedure was
successful.

With the ``dilation'' correction, included to avoid flux loss in the
outskirts regions of fainter objects, and assuming that the morphology
of the segmentation map is not strongly dependent on the wavelength,
we can consider the fluxes measured with \texttt{TFIT} analogous to
the ``total'' fluxes measured with \texttt{SExtractor}'s
\texttt{FLUX\_AUTO} (see also \citealt{lee12} and
\citealt{merlin15}). Therefore, we apply no further corrections to
these flux measurements. The merged photometric catalog combines the
HST fluxes measured with \texttt{SExtractor} and the \texttt{TFIT}
fluxes for the intermediate resolution bands. A quantitative analysis
of the quality of the photometric catalog is presented in
\S~\ref{s:quality}

\subsection{Low resolution mid-to-far IR data}
\label{ss:lowres_ir}

Here we describe the procedure to assign mid-to-far IR photometry to
the F160W sources. Given the significant differences in depth and
resolution between the optical/NIR and the IR imaging this procedure
consists of two steps. First, we build a self-consistent IR catalog
using only {\it Spitzer} and {\it Herschel} data. This merged IR
catalog combines prior-based extractions and direct detections
starting from the higher resolution {\it Spitzer} IRAC and MIPS bands
all the way up to the low resolution SPIRE bands. Second, we assign
those IR-fluxes to some of the CANDELS/F160W sources by crossmatching
the IR-only and F160W catalogs and identifying the most likely NIR
counterparts to the IR detections based on brightness and proximity
criteria.  In the following we briefly describe the main steps of the
method. A more detailed description is provided in
appendix~\ref{ap:SFRvalidation}.


We start by building merged, mid-to-far IR photometric catalogs using
the imaging datasets introduced in \S~\ref{ss:fir_ima}. The procedure
to carry out the source detection and to measure the photometry is
described in detail in appendix~\ref{ap:SFRvalidation}, as well as
several other previous works, \citet[][see also \citealt{rawle16} and
  \citealt{lrm19}]{pg10}. Briefly, the method consists of three steps:
(1) source identification in each of the IR bands starting from the
deeper and higher resolution bands at shorter wavelenghts and
progressing towards redder, lower resolution bands by using a
combination of priors and direct detections, (2) photometric
measurements based on PSF fitting and aperture photometry, and (3)
merging of the individual photometric catalogs to produce merged,
multi-band, MIPS, PACS, and SPIRE catalogs. Overall, the merged IR
catalog contains of the order of a few thousand detections at
24~$\mu$m and a few hundred in the PACS and SPIRE bands. This implies
that the multiplicity of F160W detections per IR source ranges between
5 to 10. Thus, in order to obtain a 1 to 1 match of the two catalogs,
it is necessary to identify the most likely counterparts on the basis
of their NIR brightness and their coordinates in the high resolution
images.

To do so, we run a crossmatching procedure sequentially from
high-to-low resolution bands, starting from F160W to MIPS, then MIPS
to PACS and lastly PACS to SPIRE. This method minimizes the
multiplicity of each crossmatch by chosing far IR pairs with
relatively small differences in resolution ($\sim1.5\times$). Then, we
choose the most likely counterpart from those within the matching
radius by prioritizing brightness and proximity to the low resolution
source. The crossmatch with the largest multiplicity is F160W to MIPS,
where the difference in resolution is almost $20\times$. However, in
this case the brightness in the reddest IRAC band at 8~$\mu$m is a
very effective discriminator, as it probes the rest-frame mid-IR
region (at $z\lesssim1.5$) which often exhibits a flux contribution
from the dust emission in addition to the stellar continuum. Based on
the sequential counterpart identification, each mid-to-far IR source
has a unique F160W counterpart in the final catalog. Nonetheless, we
provide supplementary IR catalogs (see appendix~\ref{ap:sfrcats})
which indicate all the secondary short-wavelength counterparts in each
of the IR bands involved in the crossmatching procedure. These
catalogs also indicate the crowdness, i.e., the total number of
counterparts to each long-wavelength, IR source, which can be used for
further diagnostics.

\begin{figure*}[t]
\includegraphics[scale=0.45, angle=0]{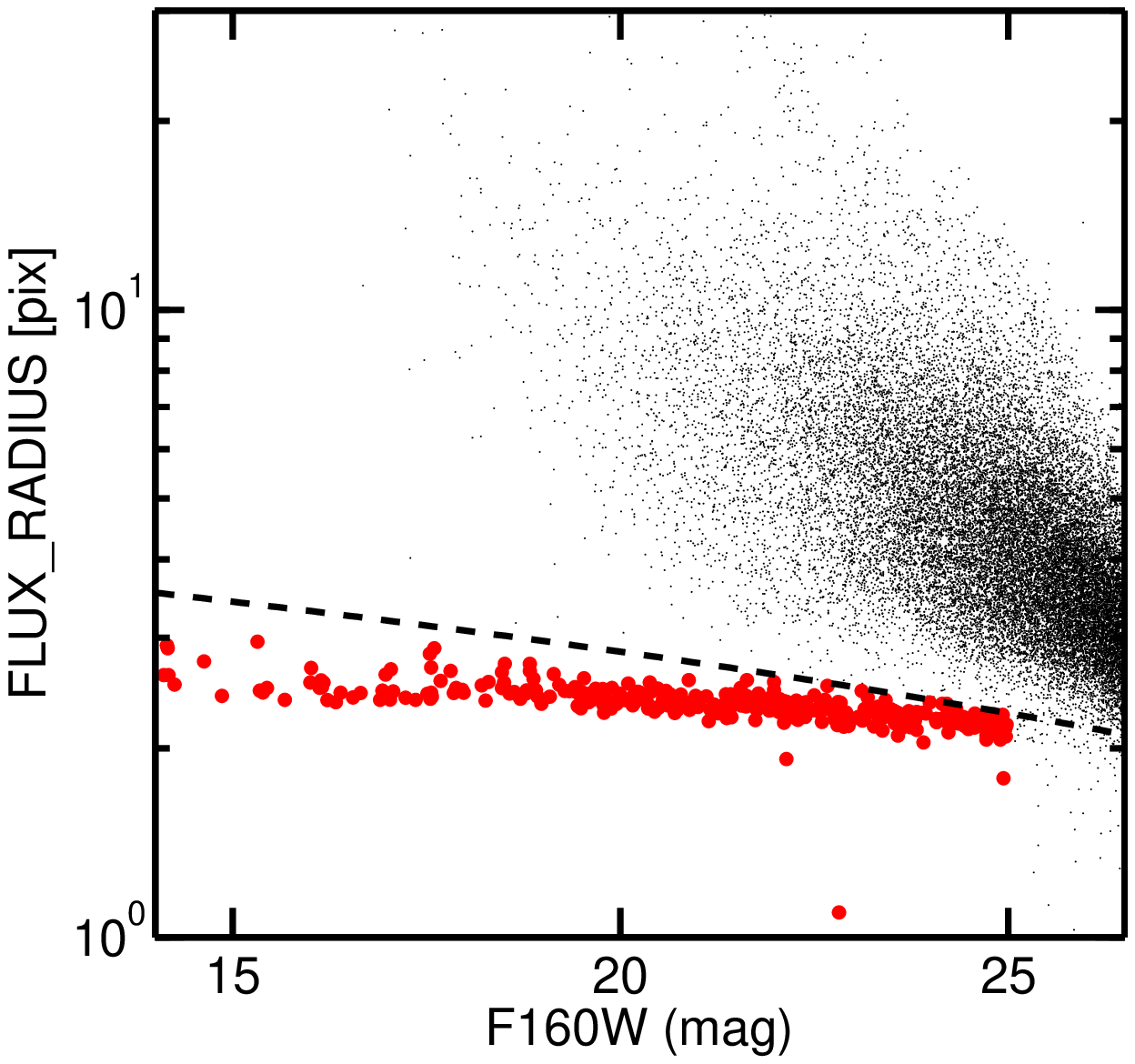}
\includegraphics[scale=0.45, angle=0]{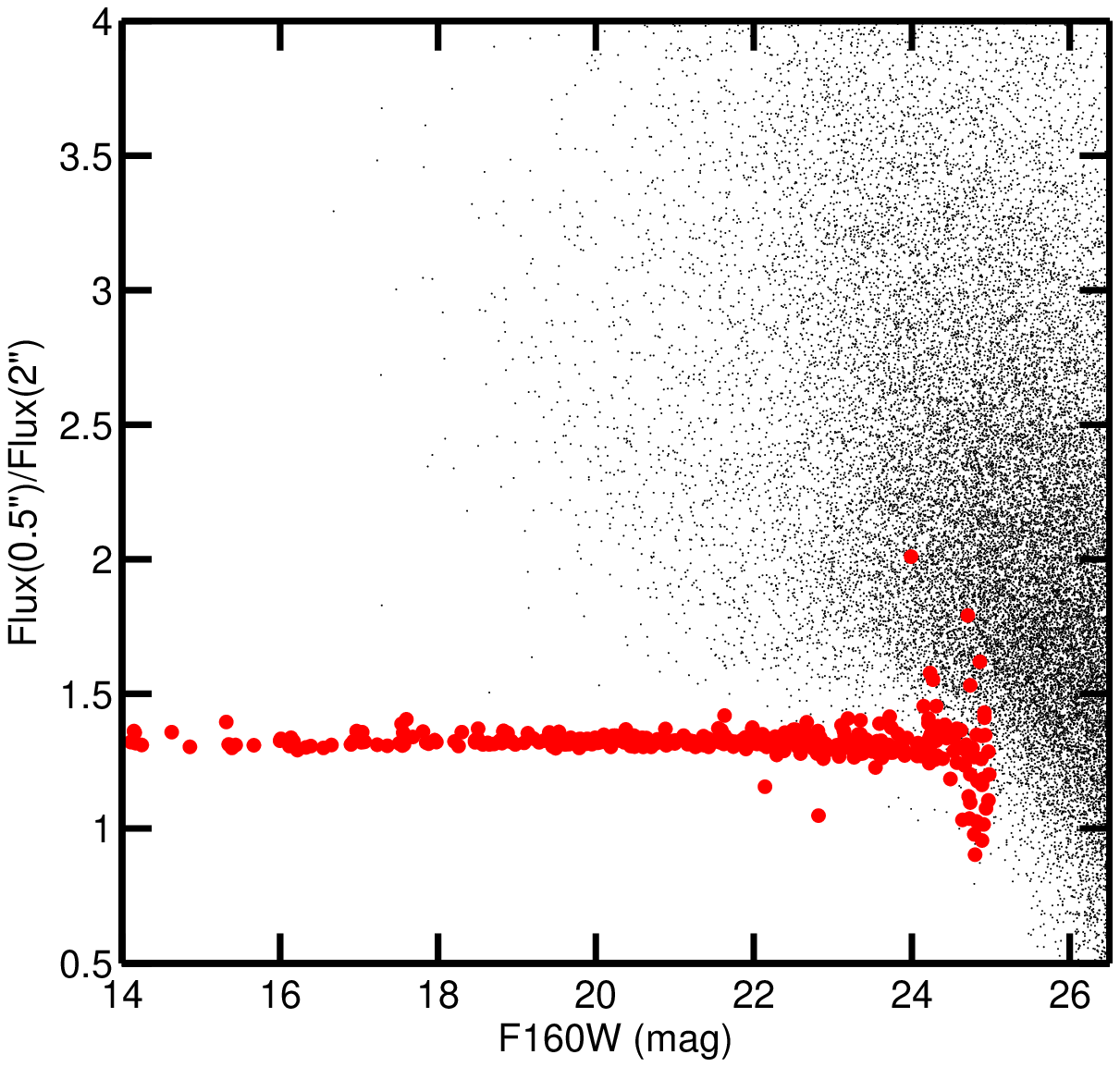}
\includegraphics[scale=0.45, angle=0]{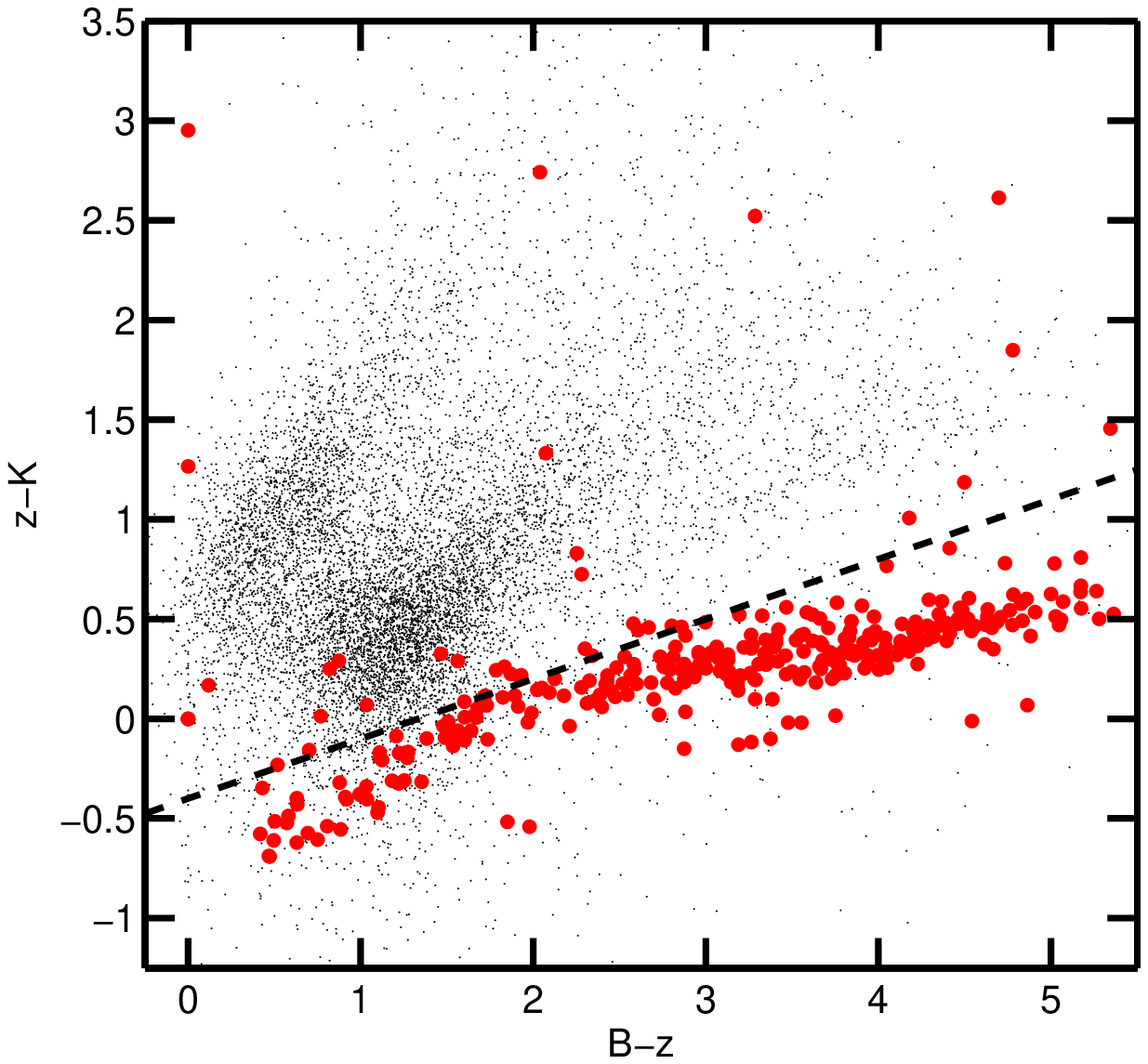}
\caption[]{{\it Left:} \texttt{SExtractor}'s \texttt{FLUX\_RADIUS}
  against total F160W magnitude. Objects classified as point sources
  in the catalog are shown with red circles. The black dashed line
  indicates the selection criterion. {\it Center:} alternative
  selection method based on the ratio of fluxes measured in large
  (2\farcs0) and small apertures (0\farcs5).  The stellar sequence
  (corresponding to the selection criterion of the left panel) at
  brighter magnitudes ($H<24$~mag) is tighter with this
  method. However, the separation from extended sources is less clear
  at fainter magnitudes. {\it Right:} $BzK$ color-color diagram
  showing another alternative method to identify stars (and galaxies;
  e.g. \citealt{daddi05}). The $BzK$ diagram shows an excellent
  agrement with the other two methods for bright stars (red). However,
  a color-based selection (dashed black line) would also include a
  significant fraction of faint and less realiable sources. Note also
  that the $BzK$ diagram relies on ground-based photometry which is
  shallower than the F160W mosaic.
  \label{fig:star_flag}}
\end{figure*}





\begin{figure}[t]
\includegraphics[scale=0.51, angle=0]{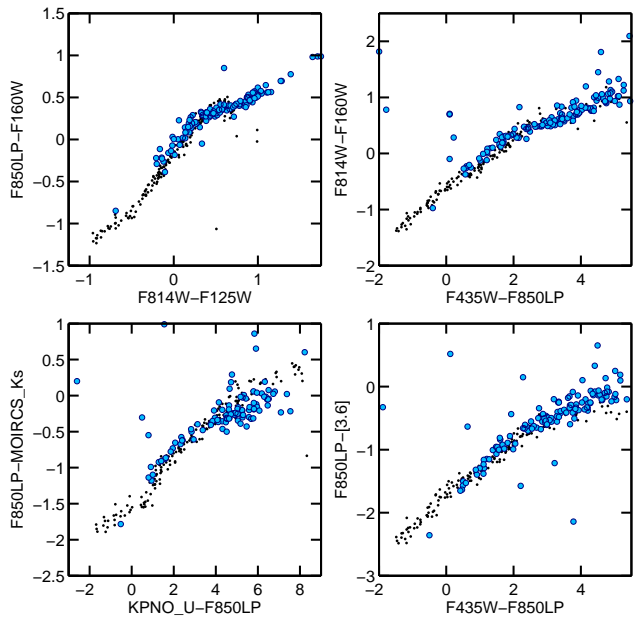}
\caption[]{ \label{fig:star-color} Color-color diagrams comparing the
  CANDELS GOODS-N photometry of unresolved sources classified as stars
  in blue (see \S~\ref{ss:stars}), and model stars in black from the
  \citet{gunn83} Atlas of stars. The model colors of stars are
  computed in each filter by integrating the model SED of stars from
  the library over the filter transmission curves.}
\end{figure}

\section{Quality Assessment}
\label{s:quality}

In this section we test the quality of the photometric catalog by (1)
comparing the observed colors of stars to those estimated from stellar
libraries, and (2) comparing the fluxes in our catalog to other
published catalogs in GOODS-N. Furthermore, in
Section~\ref{s:addedvalue} we also analyze several added value
properties computed from the fitting of the UV-to-FIR SEDs, which
depend on the quality of the photometric measurement described in the
previous section.

\subsection{Star identification and colors}
\label{ss:stars}

We compare the observed colors of the stars in our catalog to those
estimated from a stellar library. We use the synthetic models of stars
from the Bruzual-Persson-Gunn-Stryker Atlas of stars \citep{gunn83}
that we convolve with the response curves of the different
filters. Stars (unresolved sources) can be identified using a
size-magnitude diagram, as they form a tight sequence with fairly
constant small sizes as a function of magnitude. The left panel of
Figure~\ref{fig:star_flag} shows the \texttt{SExtractor}
\texttt{FLUX\_RADIUS} against the total F160W magnitude for all the
sources in the catalog. Point sources (red circles) can be cleanly
separated from extended sources down to $H \sim 25$~mag using the
following criterion: \texttt{FLUX\_RADIUS}~$< -0.115 H + 5.15$ (see
also \citealt{skelton14} for a similar approach).  We further verify
the accuracy of this selection by comparing it with two alternative
methods: 1) the ratio of the fluxes measured in large ($2\arcsec$) and
small ($0\farcs5$) apertures (central panel of
Figure~\ref{fig:star_flag}), which shows a similarly tight sequence at
bright magnitudes ($H\leq24$), and 2), the $BzK$ color-color diagram
\citep{daddi04}, which is often used to isolate distant galaxies at $z
> 1.4$ (right panel of Figure~\ref{fig:star_flag}), but it is also
very effective at isolating a clear stellar locus.

We note here that for some of the HST/ACS bands in our catalog,
particularly in F606W and F775W, the new mosaics created for this
paper include both pre-service mission data (from GOODS and other
smaller surveys) and new CANDELS data, which are separated in time by
more than 5 years. As a result, the photometry of stars, some of which
can have significant proper motions, is affected by systematics
effects such as: (a) they have moved enough that they are partially
falling out of the aperture defined by the F160W-band isophotes,
and/or (b) that they are getting partially rejected as cosmic rays due
to the motion. These effects are likely present as well in previous
version of the mosaics (e.g., v2), meaning that stellar photometry for
those stars in either set of mosaics is suspect.


Taking this into account, we compare the colors of stars, identified
with the method described above, to those of stellar models excluding
colors based on either the F606W and F775W bands (see next section for
a comparison of the fluxes of non-stellar sources in these bands to
the 3D-HST catalog). Figure~\ref{fig:star-color} shows four of these
diagrams. The observed colors of the point-like objects (blue circles)
are consistent with the general distribution predicted by the stellar
models showing no systematic biases. This further confirms the
accuracy of the photometry, specifically for the brighter sources.

\begin{figure*}[t]
  \centering
\includegraphics[scale=0.75, angle=0]{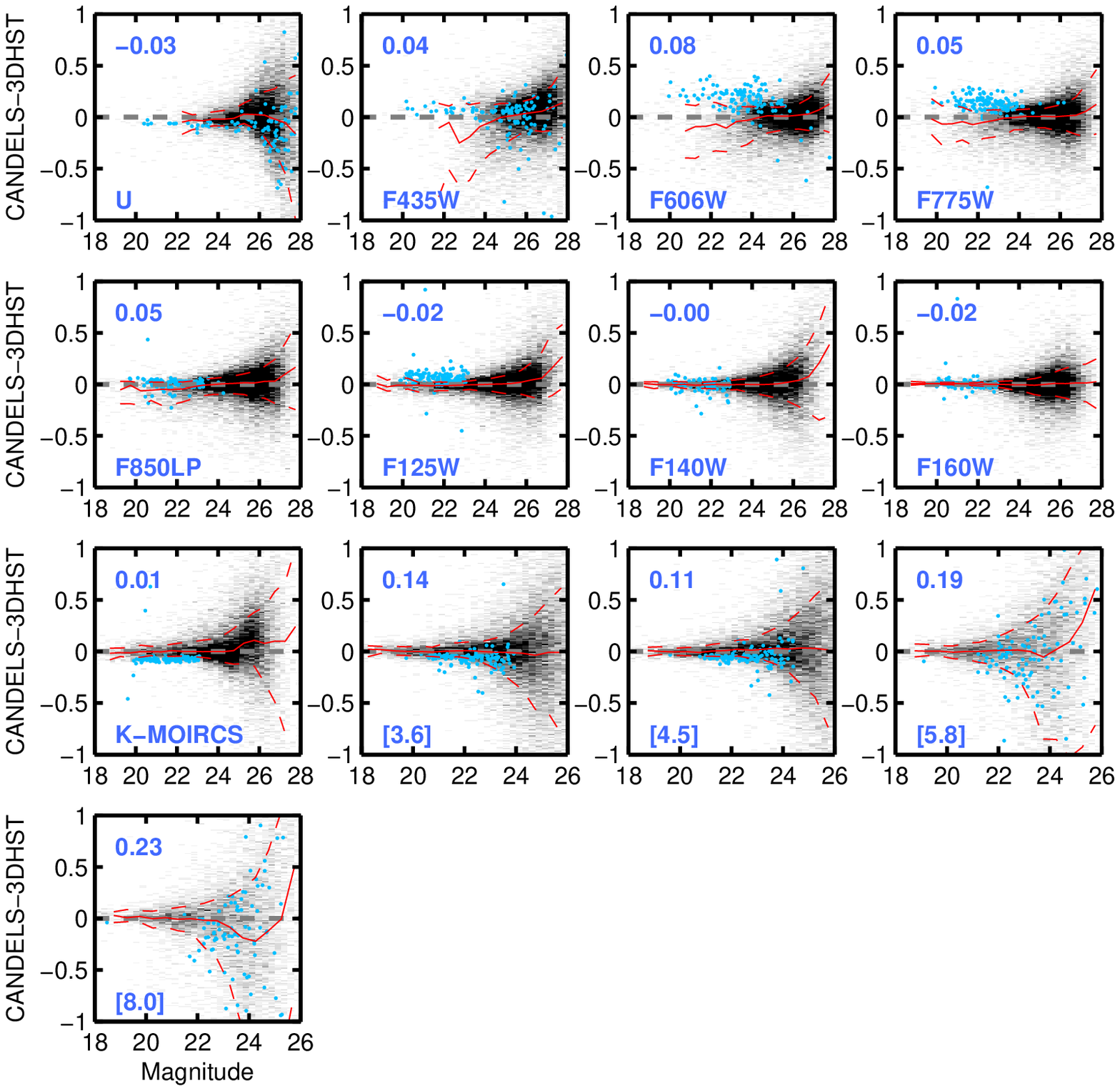}\\
\includegraphics[scale=0.75, angle=0]{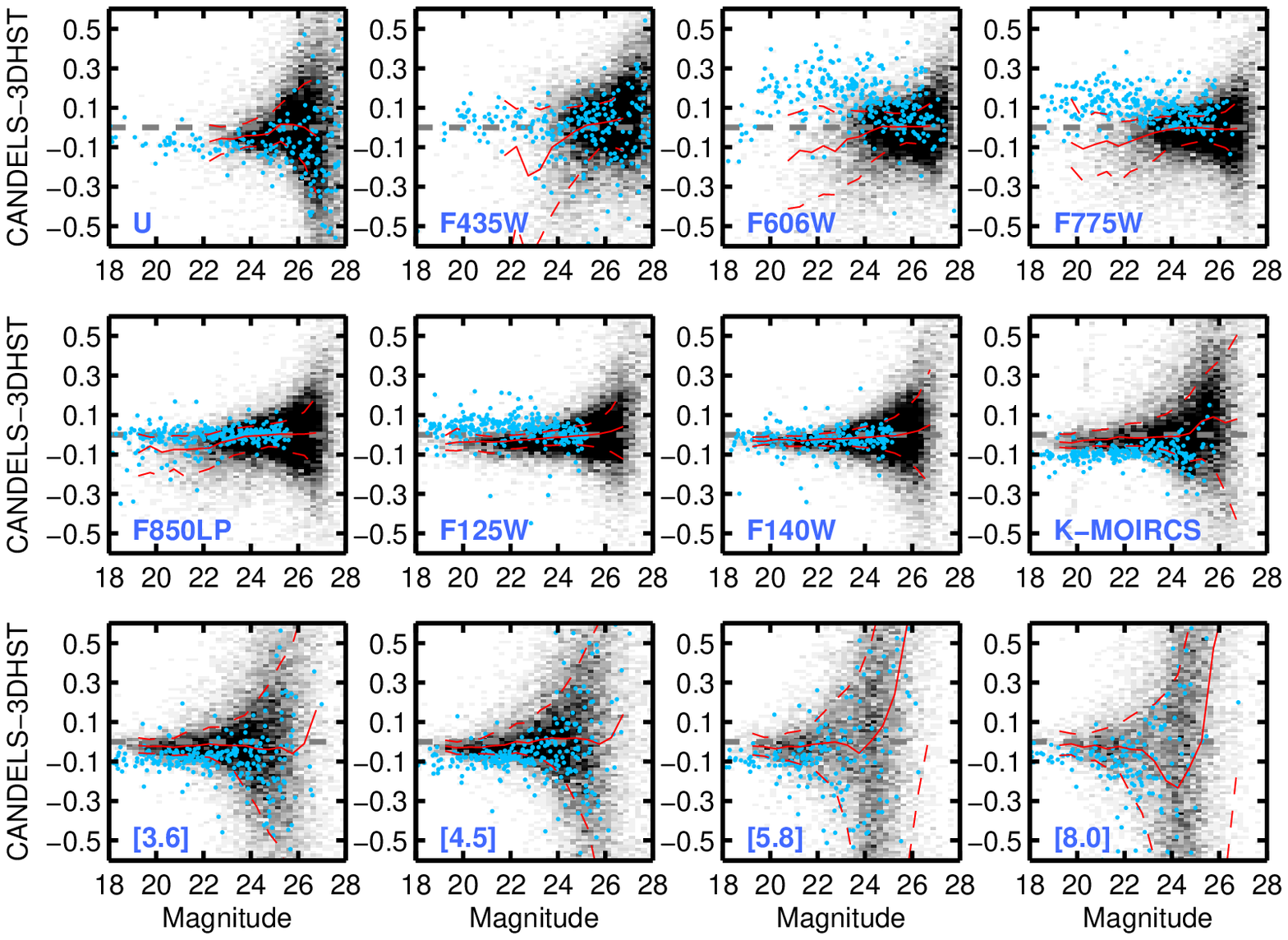}
\caption[]{\label{fig:comp_phot} Photometric comparison between the
  CANDELS and 3D-HST multi-wavelength photometry in the GOODS-N
  field. {\it Top: --} Magnitude difference (CANDELS-3DHST)
  vs. CANDELS magnitude in all the bands in common between the two
  catalogs. For each band, we only use sources with S/N$>$3 in both
  catalogs for comparison. The name of the bands is indicated in the
  bottom-left corner. The median of the magnitude difference computed
  in the bright magnitude range (m=20--24~mag) is shown in the
  top-left corner. The grey scale density map shows all sources and
  the cyan points show stars. Both sets are corrected for the median
  magnitude difference to center the distributions around zero.  The
  red solid line shows the running median (after a 3$\sigma$-clipping)
  of the magnitude difference as a function of magnitude. The upper
  and lower red dashed lines show 1$\sigma$ confidence level. {\it
    Bottom: --} Color difference in Band - F160W (CANDELS-3DHST)
  vs. CANDELS magnitude (e.g., top left panel is (U-F160W)$_{\rm
    CANDELS}$ -(U-F160W)$_{\rm 3D-HST}$ vs U$_{\rm CANDELS}$). The
  markers, lines and labels indicate the same as in the upper panel.}
\end{figure*}

\begin{figure*}[t]
\centering
\includegraphics[scale=0.375, angle=0]{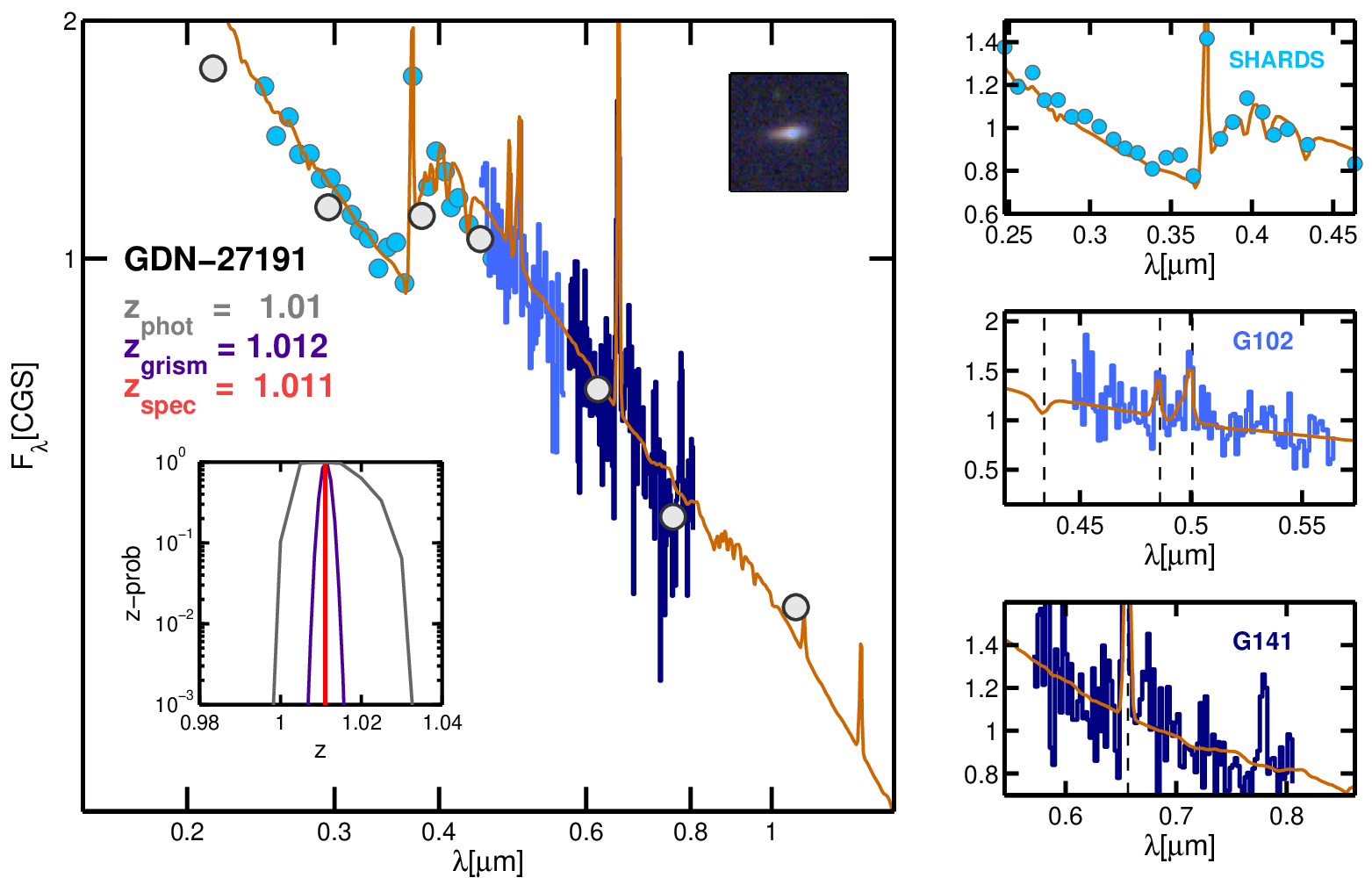}
\hspace{0.15cm}
\includegraphics[scale=0.375, angle=0]{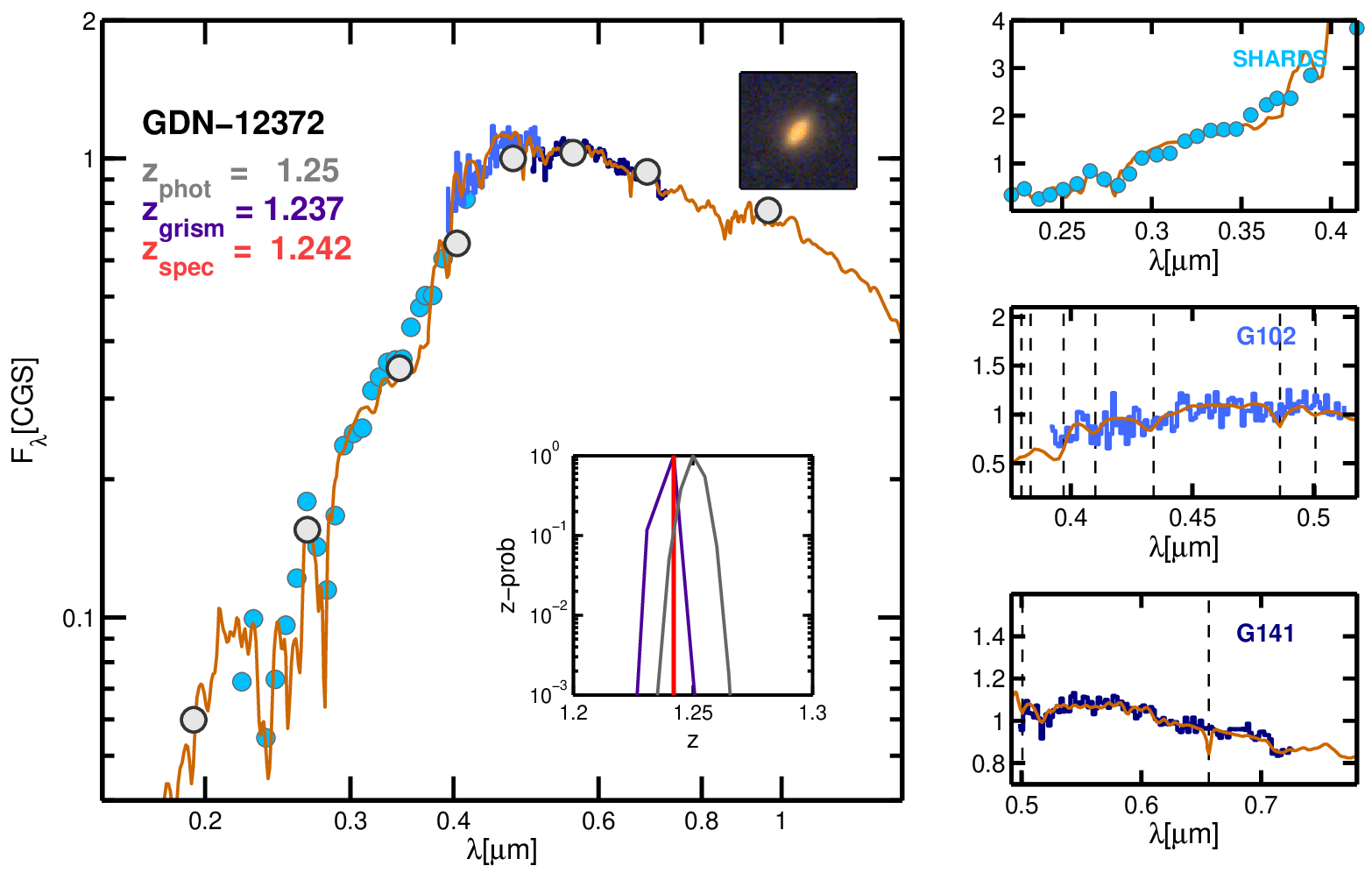}
\caption[]{\label{fig:sedfit} SEDs of a star-forming (left) and a
  quiescent (right) galaxy in the CANDELS/GOODS-N region. The grey
  circles indicate broad-band photometry.  The blue circles show the
  SHARDS medium-band data and the blue lines, from light to dark
  shading, the G102 and G141 grism spectroscopy. The rightmost panels
  in each figure show a zoom in into the key spectral regions for each
  of these datasets and highlight the most prominent emission and
  absorption features in the spectrum. The insets show the redshift
  probability distributions computed from the fitting to broad-band
  data (grey), and with the addition of SHARDS and grism data
  (blue). The red line indicates the spectroscopic redshift.}
\end{figure*}

\subsection{Comparison to other photometric catalogs}
\label{ss:othercats}
We compare our photometry with that of the 3D-HST multi-wavelength
catalogs in GOODS-N \citep{skelton14}. The 3D-HST catalog includes 22
different photometric bands. The main difference between the latter
and the CANDELS catalog in the optical-to-NIR bands is that CANDELS
includes photometry in 25 optical medium bands from the SHARDS survey
while the bulk of the optical ground-based data in the 3D-HST catalog
is based on the broad-band photometry from the Hawaii Subaru survey
\citep{capak04}. There is, however, direct overlap between the two
catalogs in the HST optical and NIR bands as well as in the
\spitzer/IRAC photometry and the $U$ and $K$ band data from Hawaii
Subaru survey and the NIR MODS \citep{kajisawa09}, respectively.

The photometry of the 3D-HST catalog was performed following a similar
methodology to ours (see \citealt{skelton14} for a full
description). Briefly, the photometry in the HST bands was computed
using \texttt{SExtractor} in dual-image mode. The fluxes were measured
in circular apertures and then corrected to total magnitudes based on
a factor derived from curve of growth of the F160W PSF. The photometry
in the lower-resolution bands was derived using a similar software to
\texttt{TFIT} (\texttt{MOPHONGO}; \citealt{labbe05,labbe06,labbe13}).
In addition to the aperture correction for the HST bands, two
additional corrections were applied to account for Galactic extinction
and small variations of the photometric zero-points. These two
corrections are removed from the 3D-HST photometry before the
comparisons described below.

We identify common sources between the CANDELS and 3D-HST catalogs by
crossmatching the source coordinates with a maximum matching radius of
0\farcs3. We only include in the comparison cleanly detected sources
(i.e., sources with good quality use-flag in both catalogs). The top
panels of Figure~\ref{fig:comp_phot} shows the magnitude difference
between the CANDELS and 3D-HST photometric catalogs for all the bands
in common between the two catalogs as a function of the magnitude in
each band. For each band, we only consider objects with S/N$>$3 in
both catalogs. Overall, the agreement is good, and the systematic
offsets (corrected and indicated in the upper-left corner) over the
high S/N magnitude range in most bands is of the order of a few
hundredth of a magnitude. The small differences likely stem from the
various systematic corrections that the 3D-HST catalog has applied.
The largest offsets of the order of $\Delta$m$\sim$0.1-0.2~mag are
found in the IRAC bands. These offsets are consistent with those found
in the similar comparisons between the CANDELS and 3D-HST catalogs
presented in previous CANDELS papers (\citealt{guo13},
\citealt{galametz13}, \citealt{nayyeri17}, \citealt{stefanon17}). Note
that, as indicated in the previous section, the stellar loci for the
HST/ACS bands F606W and F775W exhibits systematic deviations (cyan
circles) because the positions of some stars in the merged multi-epoch
mosaic have changed due to proper motions.

To further verify the accuracy of the photometry we also analyze the
difference in the colors as a function of magnitude between the two
catalogs, where the color is defined as the magnitude difference in a
given band minus F160W. In principle, a color comparison is more
straightforward as it should naturally factor out any dependence on
the aperture correction. This comparison is shown in the bottom panels
of Figure~\ref{fig:comp_phot}, and, again, we find an excellent
agreement. Overall, these tests indicate that the flux measurements in
both catalogs have been performed in a robust and self-consistent
manner.

\section{Added value properties}
\label{s:addedvalue}

In this section we present the added value properties for the galaxies
in the CANDELS GOODS-N catalog computed from the fitting of their
UV-to-FIR SEDs to stellar population synthesis models and dust
emission templates. We also present emission line measurements derived
from the WFC3 G102 and G141 grism spectroscopy.

\begin{figure*}[t]
\includegraphics[scale=0.45, angle=0]{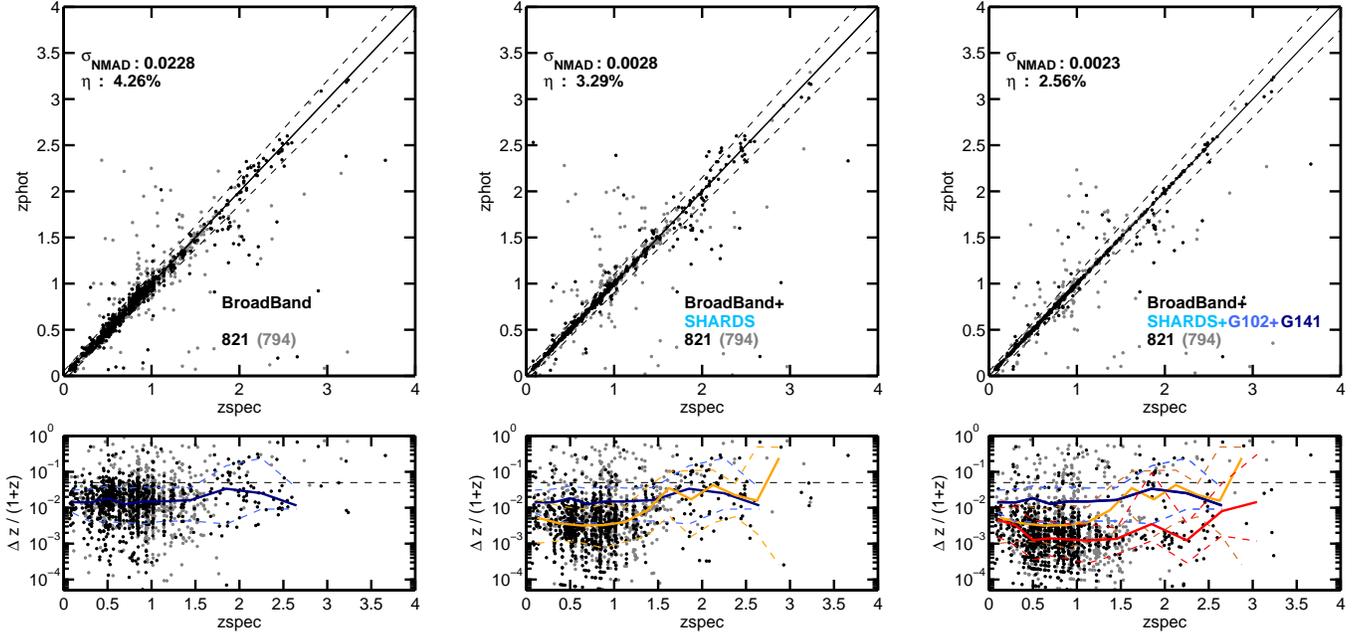}
\caption[]{\label{fig:redshift_tiers} Comparison of our multi-tiered
  photometric redshifts vs. spectroscopic redshifts for galaxies with
  both G102 and G141 spectra and good quality flag spectroscopic
  redshifts. The photometric redshifts are estimated using broad-band
  photometry only (tier 3, left), adding the SHARDS medium bands (tier
  2, center), and adding the G102 and G141 grism spectroscopy to SED
  fit (tier 1, right). The black and grey circles indicate sources
  with high- and medium- quality spectroscopic redshift flag,
  respectively. The bottom panels shows the overall accuracy ($\Delta
  z/(1+z)$) of the photometric redshift as a function of redshift. The
  colored solid and dashed lines show the running median and 68\%
  confidence regions. The black dashed line indicates the outlier
  threshold ($\Delta z/(1+z)=0.15$). Both the accuracy
  ($\sigma_{\mathrm{NMAD}}$) and the fraction of catastrophic outliers
  ($\eta$) improve with the addition of higher spectral resolution
  photometry.}
\end{figure*}

\subsection{Photometric redshifts}
\label{ss:photoz}
Here we discuss the photometric redshift estimates for the galaxies in
the CANDELS GOODS-N catalog computed from SED fitting. The main
difference between the galaxy SEDs in GOODS-N with respect to the
other 4 CANDELS fields is that this catalog includes photometry in 25
medium-bands of the SHARDS survey ($R\sim50$;
$\lambda=0.50-0.95~\mu$m) and HST/WFC3 grism observations in both G102
and G141, thus allowing for a continuous wavelength coverage from $0.9
\le \lambda \le 1.7\mu$m with a resolution of R$\sim210$ and 130,
respectively. Together, all these datasets provide remarkable spectral
resolution on a galaxy-by-galaxy basis that is uniquely suited to
provide high quality, SED-fitting based properties.

The use of higher spectral resolution photometric bands, such as
medium or narrow band filters has been shown to improve the accuracy
of the photometric redshifts up to the few percent level (e.g.;
\citealt{ilbert10}; \citealt{whitaker11}; \citealt{straatman16}).  The
inclusion of WFC3 grism spectroscopy provides even higher spectral
resolution capable of detecting emission lines, and thus provide
redshift estimates of similar quality as those from typical,
ground-based spectroscopic surveys (e.g.; \citealt{atek10};
\citet{3dhstgrism}; \citealt{treu15}; \citealt{cava15};
\citealt{bezanson16}).

Given that the number of available spectro-photometric datasets for
any given galaxy (i.e., whether they have SHARDS and/or grism data)
depends on its magnitude and its location within the WFC3 mosaic we
have implemented a three-tier classification for the photometric
redshift estimates with increasing spectral resolution data. Tier 3
consists of photometric redshifts determined from broad-band
photometry only. Although these redshifts are based on lower
resolution data, they can be computed for {\it all} the galaxies in
the catalog using the same set of input fluxes and therefore provide a
uniform, homogeneous set of baseline redshifts. The second tier
redshifts are based on the SED fitting to both broad-band and SHARDS
medium band data, and the first tier includes broad and medium band
data plus the WFC3 grism spectra. Roughly $\sim80\%$ of the galaxies
in the catalog lie in the region of GOODS-N covered by the SHARDS
medium-band survey, and a large fraction of those, $\sim60\%$ at
$H<24$~mag, have also grism detections in either G141 or G102 (more
details in \S~\ref{sss:tiers}). All these galaxies have a more
detailed SED coverage and, therefore, their photometric redshifts are
likely to be more precise. In the following we describe the methods
used to compute photometric redshifts for the galaxies in each of the
three quality tiers.





\begin{figure*}[t]
  \centering
  \includegraphics[scale=0.60, bb=105 69 857 599]{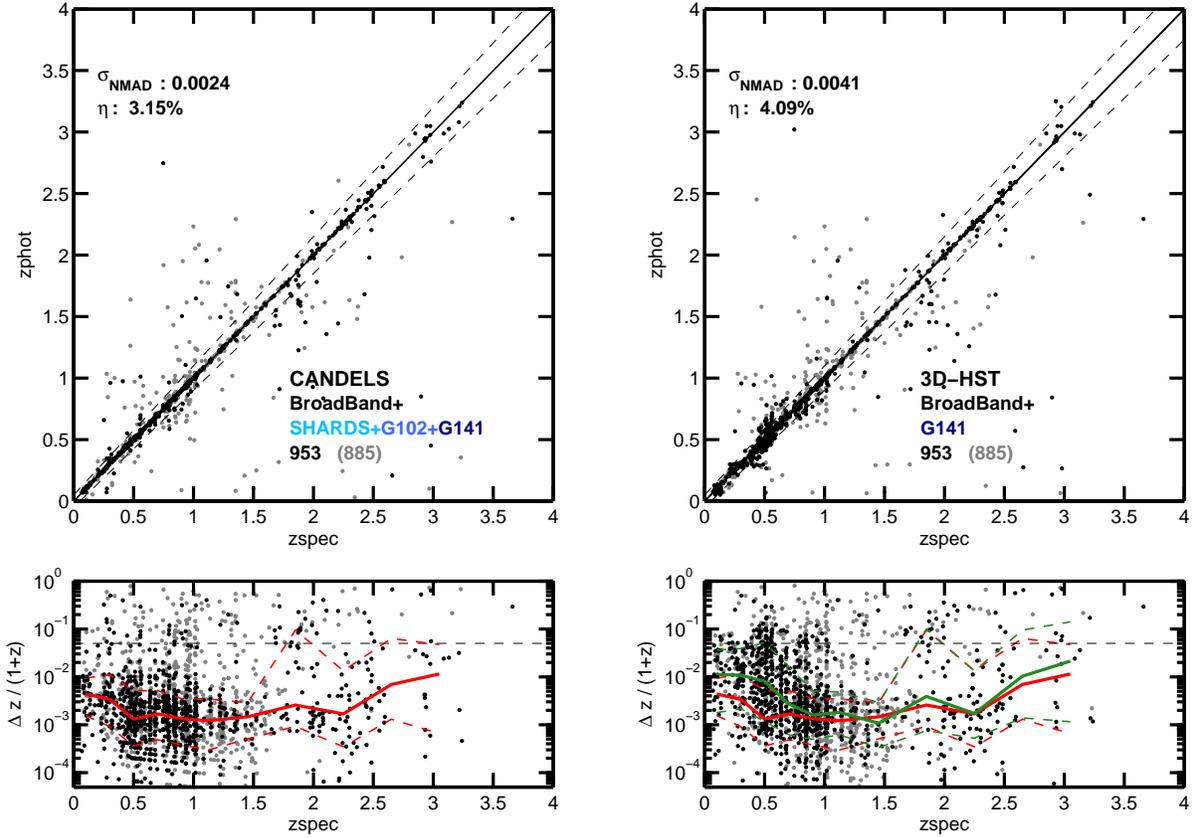}
  \caption[]{\label{fig:redshift_check2} Comparison of our photometric
    redshifts vs. spectroscopic redshifts (left) and the 3D-HST
    photometric redshifts (right) from \citet{3dhstgrism} which are
    based on broad-band photometry and G141 spectra. This comparison
    is restricted to galaxies with G141 spectra in both CANDELS and
    3D-HST and having good quality spectroscopic redshifts. The lines
    in the bottom panels show the redshift evolution of the median
    $\Delta z/(1+z)$ for the CANDELS (red) and 3D-HST (green) samples.
    The addition of medium bands and the blue grism improves the
    quality of the photometric redshifts predominantly at low-z
    ($z\lesssim0.7$), while both photometric redshift estimates are
    fully consistent at medium- and high-z. The colors of the points
    and lines have the same meaning as in
    Figure~\ref{fig:redshift_tiers}.}
\end{figure*}

\subsubsection{Tier 3: Broad-band based photometric redshifts}
\label{sss:bbphotoz}

Following the same approach as in previous CANDELS papers we computed
several estimates of the photometric redshifts using a number of
different codes, e.g. \texttt{EAZY} \citep{eazy}, \texttt{HyperZ}
\citep{hyperz}, \texttt{SpeedyMC} \citep{speedymc}, etc, based either
on $\chi^{2}$ and MCMC fitting methods and using different templates
and SED modeling assumptions (see Appendix~\ref{ap:redshift_masses}
for more details on all the different codes). As a common practice,
each of the methods fine-tuned the performance of the photometric
redshifts by computing small zeropoint corrections to the photometric
fluxes by minimizing the difference between the observed fluxes and
those expected from the best-fit templates.  Since these corrections
are dependent on the fitting codes and template libraries we followed
the same approach as in the previous CANDELS data papers and we did
not include such adjustments in the photometric catalog.  However, we
report the average photometric zeropoint offsets adopted by each group
in Table~\ref{table:gn_image_data}.

As shown in \citet{dahlen13}, using the median of multiple photo-z
estimates provides a more accurate prediction of the true redshift and
it helps mitigating some of the most common problems, such as
systematic offsets and catastrophic outliers. Here we compute the
median photometric redshift based on five different codes.  All these
codes used the same set of broad-band photometric data for all the
galaxies in the sample. We adopt these median values as the tier 3
redshift estimates. Note that, while the tier 2 and 1 photometric
redshift estimates presented in the following sections are
significantly more accurate than the tier 3 redshifts for many
galaxies, the tier 3 estimate is the only value available for those
galaxies without SHARDS and/or grism coverage. Furthermore, since the
improvement in the quality of the photo-z owing to the addition of
high spectral resolution data is magnitude dependent, the tier 3
photo-z's will also be very similar to tier 2 and 1 values for many
faint, typically high-z, galaxies.

\subsubsection{Tier 2: Broad and medium band photometry photometric redshifts}
\label{sss:shardsz}

The tier 2 photometric redshift estimates are based on the fitting of
the galaxy SEDs that include both broad band photometry and the 25
medium bands of the SHARDS survey. These redshifts are available to
the nearly 80\% of the galaxies which lie in the overlapping region
between the CANDELS and SHARDS mosaics (see
Figure~\ref{fig:footprint}).  The photometric redshifts are computed
using a slightly modified version of \texttt{EAZY} \citep{eazy}
adapted to take into account the spatial variation in the effective
wavelength of the SHARDS filters depending on the galaxy position in
the SHARDS mosaics (see \S~\ref{sss:SHARDS}).

\subsubsection{Tier 1: Broad and medium band photometry plus grism spectroscopy photometric redshifts}
\label{sss:grismz}

The tier 1 photometric redshift estimates are based on the fitting of
galaxy SEDs that include the broad and medium band photometry from
tier 2 and the WFC3 grism spectroscopy. The accuracy of the
grism-based photo-z's depends critically on whether any prominent
emission line falls within the observed spectral range (see e.g.,
Figure~\ref{fig:filters}), and if such line is detected with high
SNR. Given the limited spectral range of the grism, the majority of
the emission line detections in either G141 or G102 consist of only
one prominent line. However, if the SNR of that line is high enough
(SNR$\gtrsim5$), the use photo-z priors, such as the ones computed in
tier 2 or 3, can help break the redshift degeneracies and provide a
very precise redshift determination ($\Delta z\lesssim$1E-3;
e.g. \citealt{3dhstgrism}).

In order to take full advantage of both the broad and medium band
photometry and the grism spectroscopy, we computed the tier 1
photometric redshifts using the SED-fitting code developed by the
3D-HST survey and discussed in detail in \citet{3dhst} and
\citet{3dhstgrism}. This code was designed to estimate redshift
probability distribution functions (PDFz) based on the constraints
from the broad- and medium-band photometry as well as their G141 grism
spectroscopy. Here we use a slightly modified version of the code
which makes use of both the G102 and G141 spectroscopy in this
calculation. Briefly, the redshift determination is done iteratively
in three steps, first using only the photometric SED to obtain coarse
constraint on the PDFz, then fitting the grism data alone over a finer
redshift grid, and lastly fitting both together multiplying the
likelihood of all the redshift distributions.  The first iteration of
the fitting uses the preliminary photo-z estimate from the previous
section as a prior on the fit to broad and medium-band photometry. The
fit to the grism spectrum is done in 2D to take into account the
impact of the spatial extent of the source in the spectral
direction. This is done by using the SExtractor segmentation map of
each source in the direct F105W and F140W images for G102 and G141,
respectively.

An advantage of the iterative fitting method is that the resulting
PDFz defaults to the tier 3 or 2, photometry-only solution in all the
cases where there is no significant contribution to the probability
distribution from the fit to the grism data. Thus, the improvement in
the accuracy of the PDFz over the photometry-only case depends on the
significance of detected features on the grism spectra, e.g., strong
emission lines, continuum breaks or absorption lines (see
Figure~\ref{fig:sedfit} for two different examples of these possible
cases).

\begin{figure}[h]
\centering
\includegraphics[scale=0.60, angle=0]{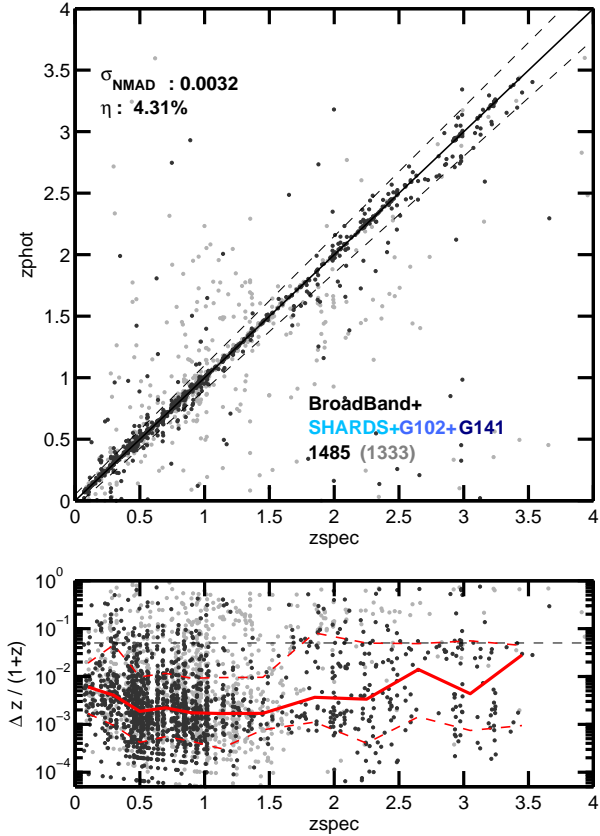}
\caption[]{\label{fig:allredshifts} Comparison between our three tier
  photometric redshifts vs. spectroscopic redshifts for galaxies with
  good quality spectroscopic redshift flag in the GOODS-N field. The
  redshift tier breakdown is approximately 20\%, 21\% and 59\% in tier
  3, 2 and 1, respectively. The colors of points and the lines in the
  bottom pabel have the same meaning as in
  Figures~\ref{fig:redshift_tiers} and \ref{fig:redshift_check2}.}
\end{figure}


\subsubsection{Quality assesment of the photometric redshifts}
\label{sss:photz_quality}
Figure~\ref{fig:redshift_tiers} compares our three tier photometric
redshift estimates versus spectroscopic redshifts for galaxies with
both G102 and G141 grism spectra and good quality spectroscopic
redshifts. Each panel illustrates the gradual improvement in the
overall accuracy of the photo-z with the addition of higher spectral
resolution data to the SED fit, starting with the broad-band only fits
(tier 3, left panel), and progressively including the SHARDS medium
bands (tier 2, middle panel), and the grism spectroscopy (tier 1,
right panel). The normalized median absolute deviation
($\sigma_{\mathrm{NMAD}}$) of $\Delta
z=z_{\mathrm{phot}}-z_{\mathrm{spec}}$, defined as
$\sigma_{\mathrm{NMAD}}=1.48\times\mathrm{median}\left(\left|\frac{\Delta
  z-\textrm{median}(\Delta z)}{1+z_{\mathrm{spec}}}\right|\right)$,
improves significantly by a factor of $\sim 10$ and 12 with the use of
medium bands and grism spectra, respectively. Similarly, the fraction
of outliers, defined as $\eta=\Delta~z/(1+z)>0.15$, decreases from
4.2\% to 3.3\% and 2.7\% in those cases. The bottom panels of
Figure~\ref{fig:redshift_tiers} show the dependence of the
$\Delta$z/(1+z) scatter with redshift for the 3 cases. The median
value of such scatter corrected by the median offset in $|\Delta z|$
is, by definition, $\sigma_{\mathrm{NMAD}}$. For the tier 3 redshifts,
the scatter is relatively constant up to $z\sim1.5$ and increases by a
few percent at higher redshifts. The addition of SHARDS photometry
significantly improves the accuracy of the tier 2 redshifts at $z<1.5$
by almost a factor of 7. However, the impact of the medium band data
at higher redshifts ($z>1.5$) is almost negligible. This is because
the most relevant spectral features (e.g., Balmer or 4000~\AA~break)
shift out of the SHARDS spectral range around that redshift, and thus
diminish the constraining effect of the medium-bands.  The addition of
HST grism spectroscopy does not significantly change the {\it overall}
$\sigma_{\mathrm{NMAD}}$ accuracy of the redshift with respect to the
tier 2 case. Nonetheless, it consistently reduces $\Delta$z/(1+z) to
$\sim0.01\%$ for galaxies with clear emission lines in the redshift
range $z=0.4-3$. As a result, the relative improvement of the tier 3
redshifts at low-z is smaller than a factor of 3, but it can increase
to almost a factor of 10 for high-z galaxies.

Figure~\ref{fig:redshift_check2} shows the comparison of our
photometric redshifts vs. spectroscopic redshifts (left) and vs. the
photometric redshifts from the 3D-HST survey (right) which also make
use of G141 spectra \citet{3dhstgrism}. This comparison is limited to
galaxies with G141 spectra in both catalogs and good quality
spectroscopic redshifts. The purpose of this comparison is twofold,
first showing the relative impact of adding SHARDS medium-band
photometry and G102 spectra vs. the G141-only case of 3D-HST, and
second, verify that our redshift estimates are consistent with theirs
for the galaxies in which the G141 data is the key contributor to the
quality PDFz. The redshift dependence on $\Delta$z/(1+z) shown in the
bottom panels of Figure~\ref{fig:redshift_check2} indicates that our
photometric redshift accuracy is slightly higher at $z\lesssim0.7$ due
to the additional constraints from SHARDS and G102, which are both
more effective at picking up emission lines at low-z (see
Figure~\ref{fig:filters}). At higher redshifts our estimates are in
excellent agreement with those from 3D-HST.



\begin{figure}[t]
\centering
\includegraphics[scale=0.385, angle=0]{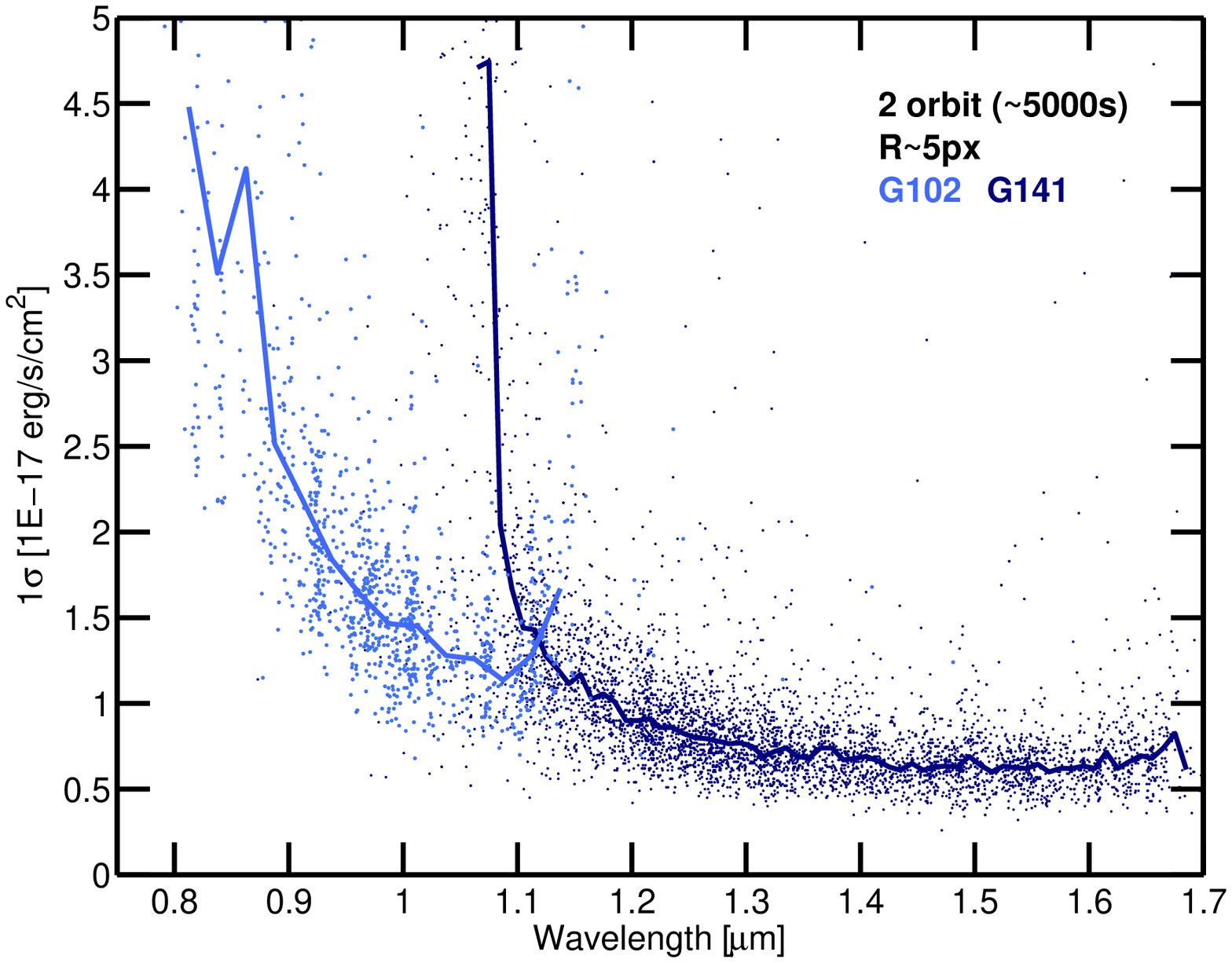}\\
\includegraphics[scale=0.385, angle=0]{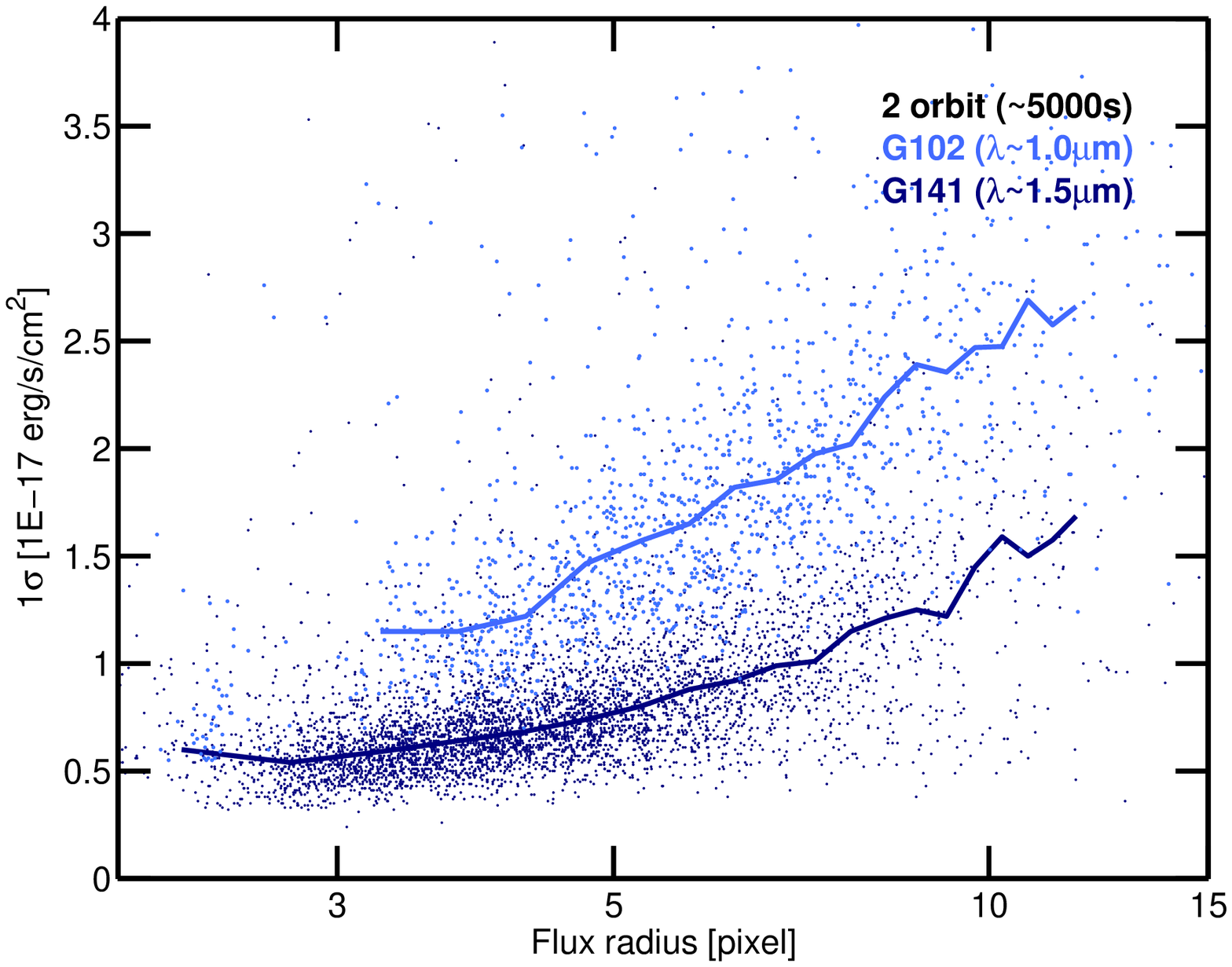}
\caption[]{\label{fig:linesensitivity} 1$\sigma$ emission line
  sensitivity in the G102 and G141 spectra as a function of the
  observed wavelength (upper panel) and the \texttt{SExtractor
    flux\_radius} size of the galaxies (lower panel). In order to
  illustrate the effect of the two main drivers of emission line
  sensitivity separately, the upper panel shows galaxies with similar
  sizes of $R\sim5$ pixels, typical for resolved galaxies, while the
  bottom panel shows galaxies with emission lines measured around the
  same wavelengths, $\lambda\sim$$1.0~\mu$m and $1.5~\mu$m for G102
  and G141. At those wavelengths, the average line uncertainty of the
  2-orbit depth grism spectra for a resolved galaxy are 0.75 and 1.5
  $\times10^{-17}$ erg s$^{-1}$ cm$^{-2}$ in G141 and G102,
  respectively.}
\end{figure}

Note that Figures~\ref{fig:redshift_tiers} and
\ref{fig:redshift_check2} include only spectroscopically confirmed
sources with clear emission lines. Therefore, the comparisons are
biased towards the best possible targets for redshift determination
using the HST grisms. This bias consequently boosts the accuracy
redshift estimates, i.e., if a galaxy has a confirmed optical emission
line it is easier for the NIR grism to pick up another one line in
different spectral range thus providing a high precision (0.01\%
level) redshift estimate. Figure~\ref{fig:allredshifts} compares
photometric and spectroscopic for {\it all} galaxies with reliable
spectroscopic flag regardless of their HST grism detectability. The
number of galaxies in the figure is more than 1.5$\times$ larger than
in the previous comparisons and it includes a significant amount of
galaxies from tier 2, i.e., galaxies for which the grism spectra do
not contribute decisively to the PDFz. Although still biased towards
galaxies with emission lines, this comparison provides a more
representative estimate of the overall quality of the photometric
redshifts for the whole, magnitude limited sample. The accuracy is
slightly lower than in the previous comparisons but it is still
significantly better than the $\sim$1\% precision typical of
broad-band only surveys ($\sigma_{\mathrm{NMAD}}$$\sim0.3\%$ with
$\eta\sim4\%$).

\begin{figure}[t]
\centering
\includegraphics[scale=0.6, angle=0]{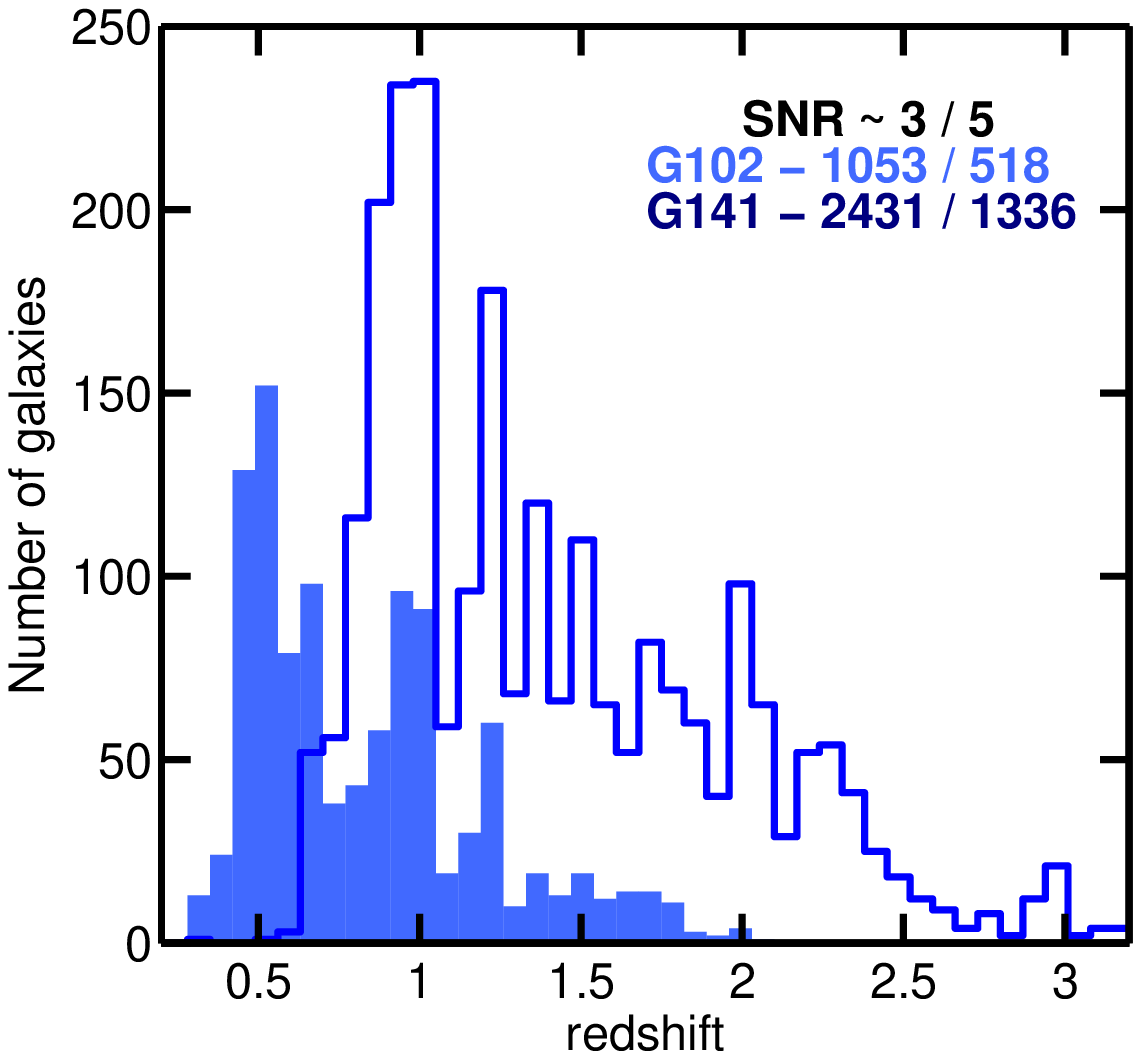}\\
\includegraphics[scale=0.6, angle=0]{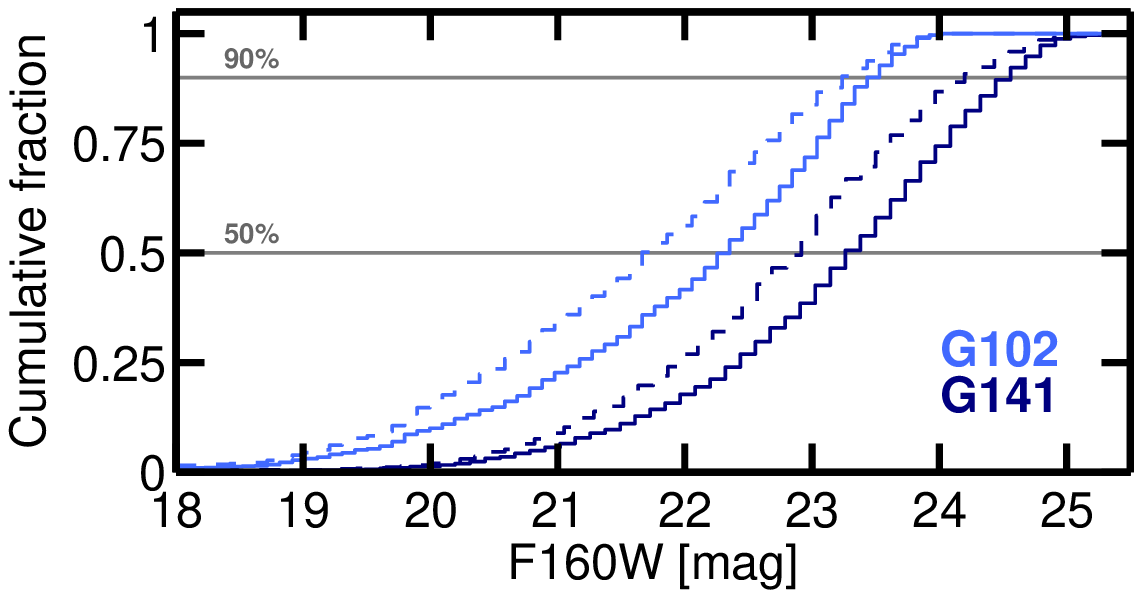}
\caption[]{{\it Top:} \label{fig:grismhisto} Redshift histogram for
  galaxies with clear emission lines (SNR$\geq$3) detected in the G102
  (empty light blue) and G141 spectra (filled dark blue). {\it
    Bottom:} Cumulative F160W magnitude distribution for the emission
  line galaxies in upper panel. The dashed and solid lines indicate
  the distributions down to SNR$=3$ and 5, respectively.}
\end{figure}

\begin{figure*}[t]
\centering
\includegraphics[scale=0.6, angle=0]{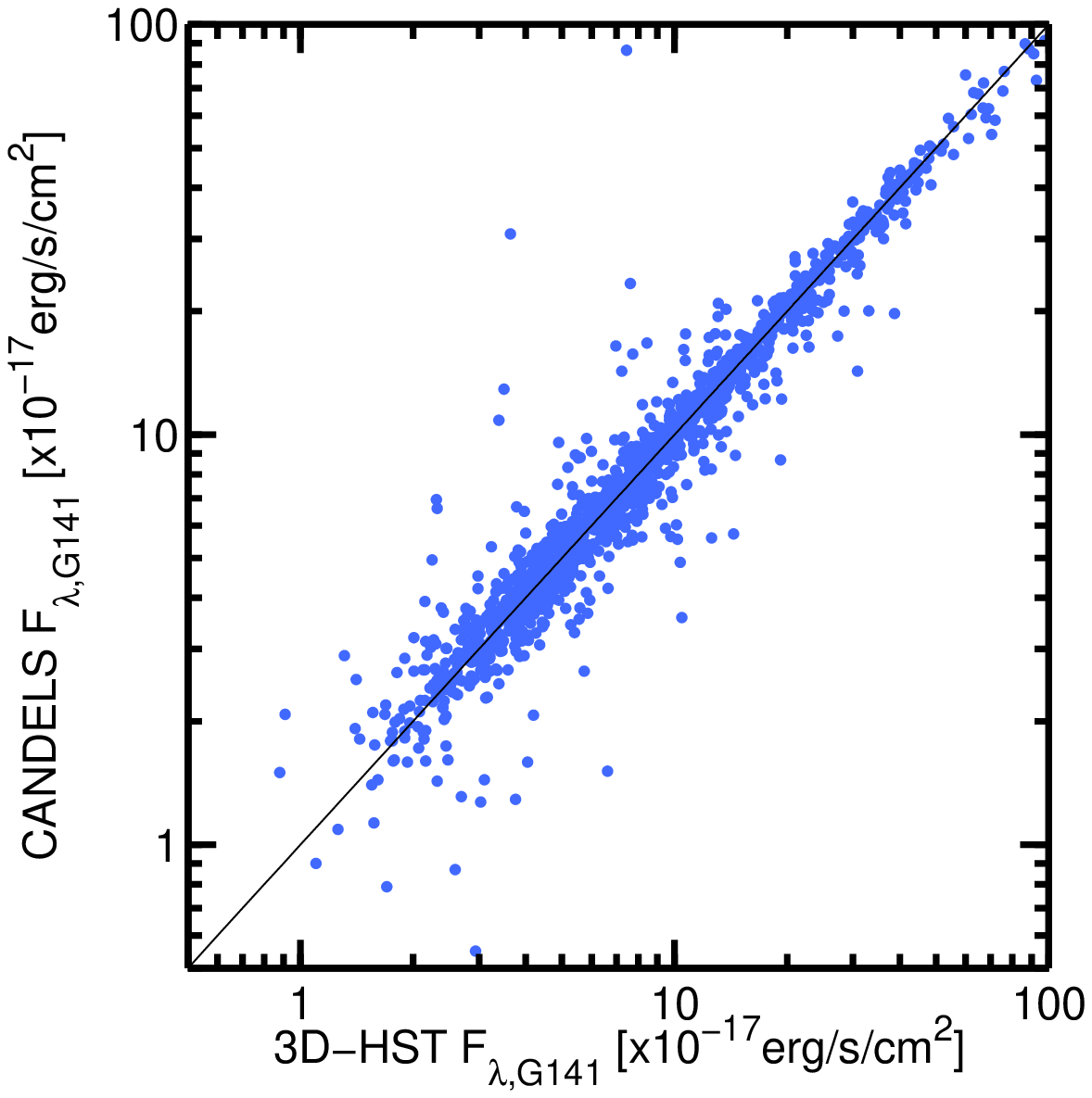}
\hspace{1cm}
\includegraphics[scale=0.6, angle=0]{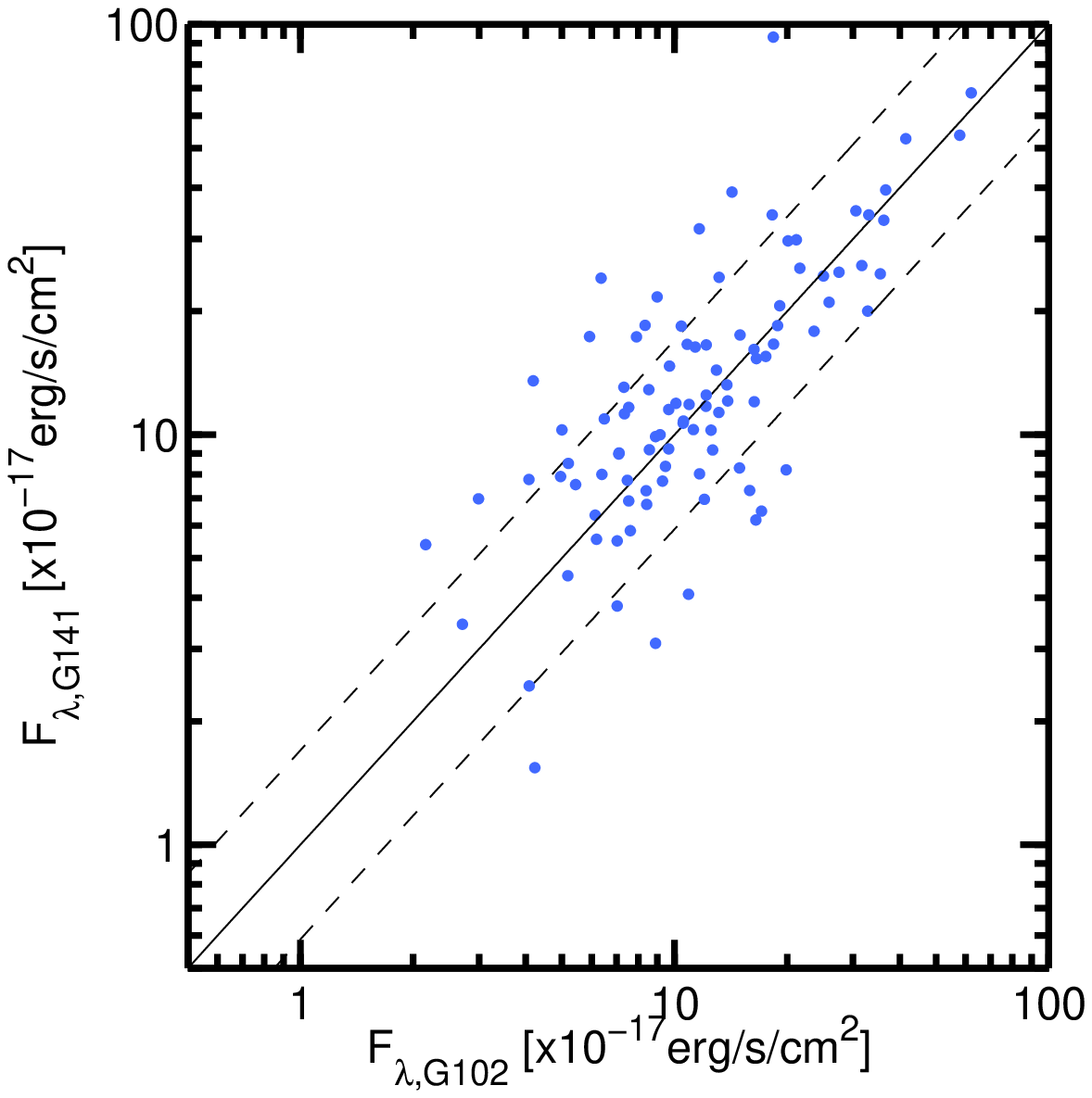}
\caption[]{\label{fig:grismfluxes} {\it Left:} Comparison of emission
  line fluxes measured in the G141 spectra by the 3D-HST survey
  \citet{3dhstgrism} and our measurements in this paper. Both
  measurements come from the same G141 images but using different
  source extractions and SED fits to estimate the line redshift and
  fluxes. The results are in excellent agreement and demonstrate the
  consistency of the measurements.  {\it Right:} Comparison of
  emission line fluxes for galaxies in which the same line (either
  H$\alpha$ or [OIII]/H$\beta$ can be detected simultaneously in both
  of our G102 and G141 measurements.  This is only possible for a
  small fraction of the emission line galaxies at narrow redshift
  ranges around z$\sim0.7$ (for H$\alpha$) and z$\sim1.3$ (for
  [OIII]/H$\beta$).}
\end{figure*}

\subsubsection{Breakdown of the photometric redshift tiers}
\label{sss:tiers}
Since the quality of the photometric redshifts depends on the data
used for the SED fit it is useful to report the relative fractions of
galaxies in the sample that have observations in each of the relevant
datasets or photo-z tiers discussed in the previous sections. In terms
of area coverage, approximately $\sim$80\% of the CANDELS F160W mosaic
is covered by the SHARDS medium band imaging, and an additional
$\sim$3\% of the non-SHARDS area is covered by the WFC3 G102 and/or
G141 mosaics. Thus less than $\sim$20\% of all galaxies have tier 3
redshifts, i.e., based on broad-band photometry only. Among the 80\%
of the sample with SHARDS observations, the relative fraction of
galaxies with tier 2 and tier 1 redshifts, i.e., the fraction of
galaxies with both SHARDS photometry and HST grism spectra is
magnitude dependent. For a magnitude limit of H$<24$~mag, the
breakdown is 26\% and 74\% in tiers 2 and 1, and 35\% and 65\% for
$H<25$~mag. Relative to the whole catalog, these numbers imply that
60\% and 52\% of all galaxies have tier 1 redshifts at H$<24$ and
25~mag, respectively.

For the galaxies with observations in both grisms, the SED fitting
procedure combines the G102 and G141 spectra for the redshift
determinations. However, given the lower sensitivity of the G102 grism
(see next section for more details) some galaxies might only have G141
detections. For galaxies detected in at least one of the grisms and a
magnitude limit of H$<24$~mag, the breakdown between galaxies with
both G141 and G102 spectra vs. only G141 spectra is approximately 55\%
to 33\%. The remaining 12\% of the galaxies have only G102
observations. The latter are typically located in a region with G141
coverage, however, differences in the orientation of the G102 and G141
observations can make the G141 spectra unavailable or severely
contaminated.


\begin{figure*}[t]
\centering
\includegraphics[scale=0.6, angle=0]{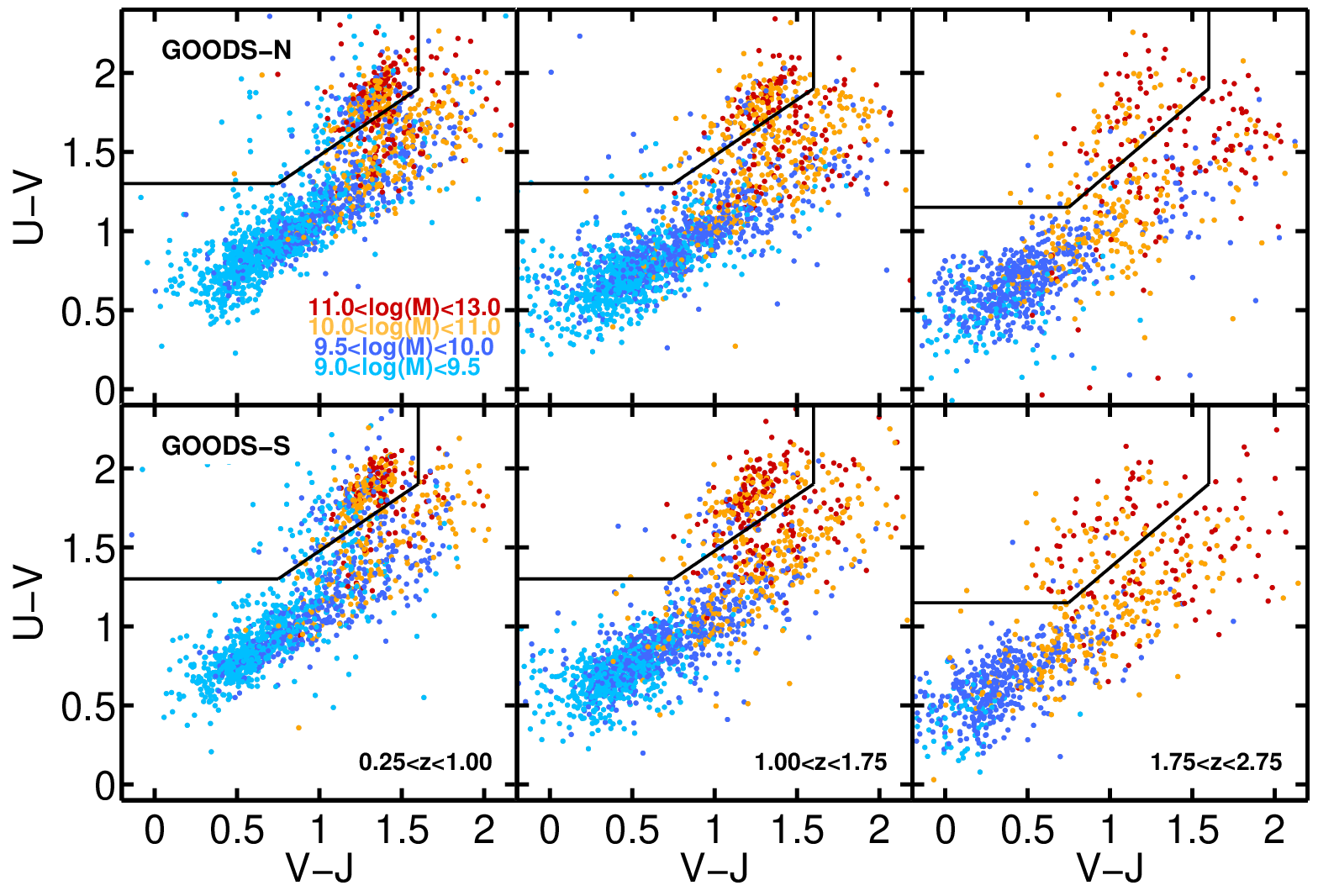}
\caption[]{\label{fig:uvj} Rest-frame UVJ diagram for galaxies in the
  deepest CANDELS fields, GOODS-N (this paper) and GOODS-S (from
  \citealt{guo13}). Each column presents the rest-frame colors for
  galaxies in the two fields at the same redshifts, with redshift
  increasing from left to right. The diagram is color-coded by mass,
  with the lowest mass galaxies in blue and the highest mass galaxies
  in red, as shown by the legend in the upper-left panel. The black
  lines indicate the selection box used to separate quiescent from
  star-forming galaxies, based on \citet{williams09}.}
\end{figure*}

\subsection{G102/G141 Emission line fits}
\label{ss:emission_lines}
We compute emission line fluxes and observed-frame equivalent widths
from the G102 and G141 spectra using the same software described in
the 3D-HST survey paper \citet{3dhstgrism}. Briefly, this code adopts
the 2D continuum template determined from photometric redshift fit to
build a 2D model and then adds Gaussian lines with unresolved line
widths of $\sigma=100$~km/s. Each potential line is treated separately
by means of independent line-template normalizations relative to the
continuum. The code fits the observed grism data to the model using
the \texttt{emcee} sampler \citep{mcmc} to determine the marginalized
posterior distribution functions of the parameters for the individual
line-template normalizations. These are converted directly to line
fluxes and observed-frame equivalent widths in physical units (i.e.,
$\mathrm{erg}~\mathrm{s}^{-1}~\mathrm{cm}^{-2}$ and \AA,
respectively). Only the lines that fall within the rest-frame spectral
range of the grism data, as determined from the grism redshifts, are
included in the model (see Table~4 of \citealt{3dhstgrism} for a
complete list of all the species included in the fit). The line fluxes
are implicitly normalized to the total broad band photometry, as the
spectra are scaled to match the photometric data.  Note that the
line-template normalization is not required to be positive, and
therefore it is not restricted to measure emission lines, i.e., it can
also provide equivalent width estimates for absorption lines.

The MCMC chains provide a robust estimate of the uncertainties in the
fit, which are primarily determined by two effects: 1) the wavelength
dependence of the grism throughput, and 2) the galaxy size (i.e., the
area of the effective aperture of the 2D spectrum
fit). Figure~\ref{fig:linesensitivity} illustrates these two effects
separately for the G102 and G141 spectra. The top panel depicts the
wavelength dependence of the sensitivity for sources with
\texttt{SExtractor} \texttt{FLUX\_RADIUS}$=3-5$~pixels while the
bottom panel depicts the dependence on the galaxy size at the peak
sensitivity wavelength of each grism. Overall, a typical resolved
galaxy exhibits a 1$\sigma$ flux uncertainty of
$\sim1.5\times10^{-17}$erg~s$^{-1}$~cm$^{-2}$ in G102 and
$\sim0.8\times10^{-17}$erg~s$^{-1}$~cm$^{-2}$ in G141 for 2-orbit
depth exposures. The lower sensitivity threshold of the G102 grism
compared to G141 is largely due to its higher spectral resolution
(i.e., the line flux spreads over more pixels and thus reaches a lower
the S/N per resolution element for similar exposure times). The noise
levels are in good agreement with previously published sensitivities
of the HST NIR grisms (e.g., \citealt{atek10}; \citealt{3dhst};
\citealt{trump13}; \citealt{treu15}).

Figure~\ref{fig:grismhisto} shows the redshift distribution and the
cumulative fraction as a function of magnitude for emission line
galaxies detected in the G102 and G141 spectra.  The lower sensitivity
of the G102 grism results in approximately half the number of emission
line detections as in G141 at SNR~$\gtrsim3$ or 5. Furthermore, its
bluer central wavelength implies that the majority of those detections
in G102 have lower median redshifts than those in G141. As indicated
in Figure~\ref{fig:filters}, the bluest of the most prominent emission
lines, \OII, shifts out of the G102 spectral coverage at $z\sim2$,
while it can be detected in G141 up to $z\sim3.5$. This is consistent
with the distributions shown in the histograms of
Figure~\ref{fig:grismhisto}. The overall brighter magnitudes of the
emission line galaxies detected in G102 implies that the majority of
those galaxies are also detected in G141 for redshifts $z\gtrsim0.6$
(i.e., when the H$\alpha$ line shifts into the G141 passband). This
naturally provides simultaneous detections of two relevant lines in
the combined dataset, for example H$\alpha$ and H$\beta$ at
$z=0.6-1.3$, or [OIII] and [OII] at $z=1.3-2.0$. The majority
($>90\%$) of the G102 emission line detections with SNR~$\geq$3 (5)
have magnitudes $H\leq23.5$ (23), while the G141 detections are about
1 magnitude fainter with SNR~$\geq3$ (5) at H~$\leq$~24.5
(24). Relative to the full galaxy catalog, these numbers imply that
$\sim$25\% of the galaxies with $H<24.5$ have at least one emission
line detected in the G141 grism, and $\sim$35\% of the galaxies with
$H<23.5$ have two emission lines detected, one on each of the G102 and
G141 grisms.

In order to validate the quality of the emission line extractions, we
compare the line fluxes measured in the G141 grism to those released
by the 3D-HST survey in \citet{3dhstgrism}. Note that while we make
use of the reduced G141 images released by the 3D-HST survey, the 2D
extraction of the spectra, the redshift determination and the line
measurements depend on our object detection procedure and on our
SEDs. Therefore, this is a useful quality check to verify that our
extraction and SED fitting procedures are accurate. This comparison is
shown in the left panel of Figure~\ref{fig:grismfluxes}, which
illustrates that the fluxes from both catalogs are in excellent
agreement even for the faintest emission lines with
f$_{\lambda}\sim1\times10^{-17}$erg s$^{-1}$ cm$^{-2}$ which have low
SNR~$\lesssim3$.

The right panel of Figure~\ref{fig:grismfluxes} extends this
validation test to the emission lines detected in G102 by comparing
the flux measurements for emission lines that are simultaneously
detected in both G102 and G141. This is only possible for a small
sub-sample of galaxies in narrow redshift ranges where the most
prominent lines lie in the reddest and the bluest sides of the G102
and G141 wavelength ranges, respectively ($z\sim0.7$ for H$\alpha$,
and $z\sim1.3$ for [OIII]/H$\beta$). In this case the comparison is
between fully independent measurements performed in different
datasets, and we also find a good agreement for all the emission lines
with SNR$\gtrsim3$. The scatter is consistent with the $\sim1.5\times$
dispersion reported in \citet{3dhstgrism} for the comparison between
grism and ground-based spectroscopic measurements. Note also that, in
order to measure the same line in both grisms, the fluxes are
typically measured near the edges of the spectra, around $\lambda_{\rm
  obs}\sim1.1\mu$m, where the sensitivities are lower (see
Figure~\ref{fig:linesensitivity}).

\subsection{Rest-frame colors and stellar population properties}
\label{ss:colors_masses}
We compute stellar population properties and rest-frame luminosities
by fitting the observed SEDs to galaxy templates and adopting the best
photometric redshift. We used the best available SED for every galaxy
including broad- and medium- band photometry, but not the grism
spectroscopy. First, we estimate stellar masses and other physical
properties of the galaxies, such as stellar ages, dust extinctions or
SFRs by fitting the SEDs with the codes {\tt FAST} \citep{kriek09} and
{\tt Synthesizer} \citep{pg05,pg08}. The redshift is fixed to best
redshift estimate, i.e., spectroscopic where available and photometric
otherwise. The modeling assumptions for both codes are as follows: we
use the \citet{bc03} stellar population synthesis models with a
\citet{chabrier} IMF and solar metallicity. We assume exponentially
declining star formation histories with a minimum e-folding time of
$\log_{10}(\tau/yr) = 8.5$, a minimum age of 40 Myr, $0 < A_V < 4$~mag
and the \citet{calzetti} dust attenuation law. The only difference
between the {\tt FAST} and {\tt Synthesizer} fits is that the latter
uses SED templates that include emission lines. In addition to the
stellar population properties, we also estimate rest-frame
luminosities and colors for all galaxies using {\tt EAZY}
\citep{eazy}. This code computes the rest-frame luminosity in a set of
typical photometric filters (see Table~\ref{table:sps_fast}), and then
derives rest-frame colors as the ratio of the luminosities in two of
those filters.

\begin{figure}[t]
\centering \includegraphics[scale=0.5, angle=0]{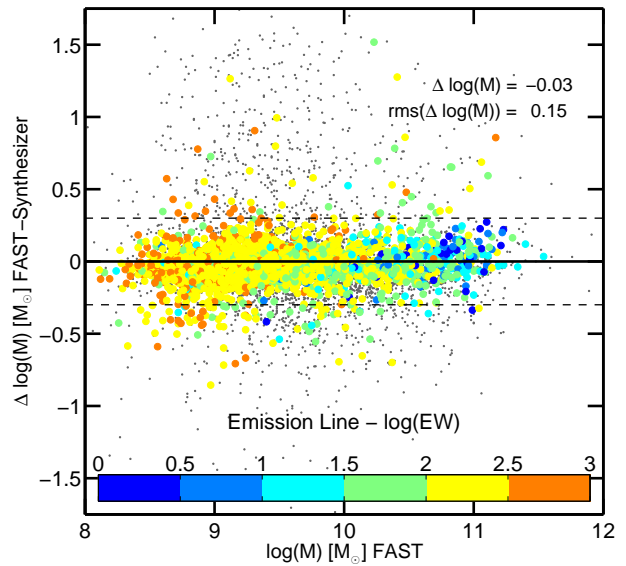}
\caption[]{\label{fig:massfastvsynth} Comparison of the stellar masses
  computed using {\tt FAST} and {\tt Synthesizer} with the same
  modeling assumptions (see text). The primary difference between the
  galaxy models is that {\tt Synthesizer} includes emission lines in
  the SEDs. The stellar mass estimates are fully consistent. The
  overall distribution is centered around zero with a scatter of
  $0.15$~dex. We find no systematic affects the stellar masses of
  galaxies with strong emission lines, identified using the WFC3 grism
  spectra. The color code indicates the EW of the \Ha~and \OIII
  emission lines in those galaxies.}
\end{figure}

\begin{figure*}[t]
\centering
\includegraphics[scale=0.55, angle=0]{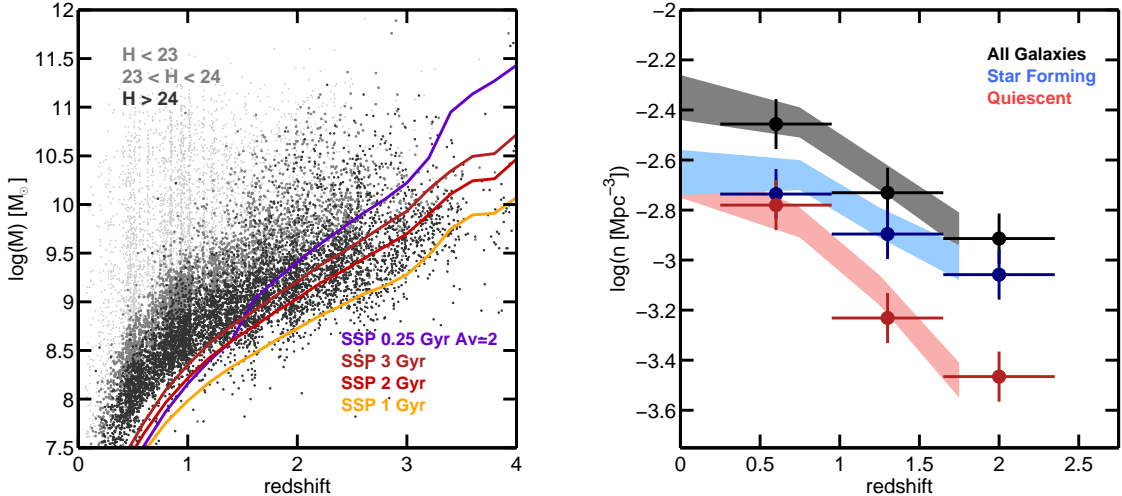}
\caption[]{\label{fig:masscompleteness} {\it Left:} Galaxy stellar
  masses as a function of redshift for the CANDELS GOODS-N sample with
  a grey-color intensity scale for the H-band magnitude.  We estimate
  the stellar mass completeness of the sample by showing the stellar
  mass threshold for a series of intrinsically red, quiescent or dusty
  (orange, red and purple lines), galaxy templates with an observed
  H-band magnitude equal to the SNR$\sim$5 detection limit of the
  catalog, H$\sim$26. {\it Right:} Comoving number density for
  galaxies more massive than \lmass$>10$ divided in three groups, star
  forming (blue), quiescent (red) and both together (black), and
  compared to the results from the ULTRAVISTA COSMOS sample for the
  same range in stellar mass (shaded regions). The catalog provides
  results which are consistent with the literature \citep{muzzin13smf}
  indicating that both our stellar masses and rest-frame colors are
  robust.}
\end{figure*}

Figure~\ref{fig:uvj} illustrates the consistency of these rest-frame
colors by comparing the distribution of galaxies in the UVJ
color-color diagrams \citep{williams09} based on the CANDELS/GOODS-N
catalog and the CANDELS/GOODS-S catalog of \citet{guo13} (i.e., the
deepest CANDELS fields) in 3 redshift bins. The color distributions
are qualitatively very similar and they are also consistent with the
UVJ diagram for the 3D-HST sample (Figure~26 of \citealt{3dhst}). The
mass distribution in the color-color diagram is also consistent with
previous results which showed that the majority of massive galaxies at
$z\gtrsim1$ tend to be intrinsically red ($U-V$), either because of
dust obscuration of because they host older stellar populations
\citep{brammer11}. The UVJ diagram is indeed particularly useful to
make this distinction because it breaks the degeneracy between the
dusty star-forming galaxies and quiescent galaxies with low levels of
star formation. Both the old and the dusty populations have red $U-V$
colors (upper left region), but dusty star-forming galaxies typically
have redder $V-J$ colors (e.g., \citealt{whitaker12}). The black lines
indicate the selection threshold that is typically used to distinguish
these two populations. In the following we adopt the UVJ criterion to
divide the GOODS-N sample in star-forming and quiescent galaxies at
all redshifts. We validate the accuracy of this selection criterion by
comparing the evolution in the number densities of these two
populations to the results from previous works (see below).

\begin{figure}[t]
\centering
\includegraphics[scale=0.5, angle=0]{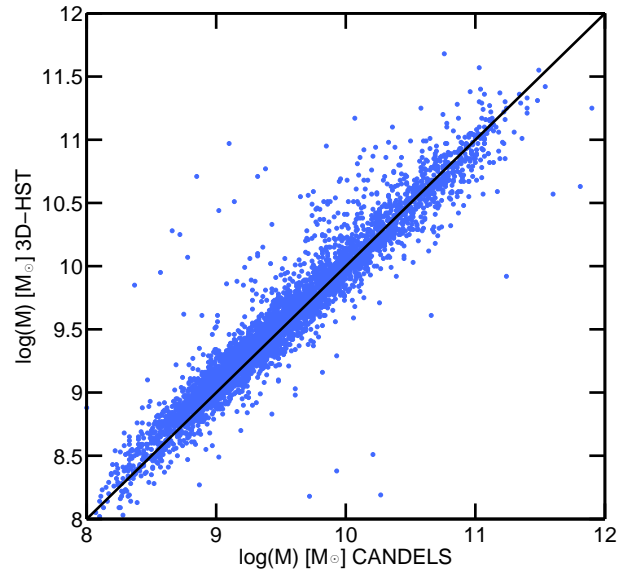}
\caption[]{\label{fig:3dhstmass} Stellar mass comparison for the
  galaxies in the GOODS-N CANDELS and 3D-HST catalogs. Both sets of
  stellar masses are consistent with each other showing almost
  negligible offsets, $\Delta$\lmass~$< 0.03$~dex, and a scatter of
  $\sim$0.3~dex, typical of the SED fit based stellar mass estimates.}
\end{figure}

Figure~\ref{fig:massfastvsynth} shows the comparison of the stellar
masses computed with {\tt FAST} and {\tt Synthesizer}. The overall
difference between these estimates for the whole galaxy sample is
consistent with zero within the usual scatter of $\sim0.3$~dex typical
of the comparison of stellar masses derived with different codes
(e.g., \citealt{mobasher15}, \citealt{nayyeri17}). Furthermore, we
find no obvious systematic differences in the stellar masses of
galaxies with strong, high EW emission lines, identified in the G102
and G141 spectra (\Ha~or \OIII; colored circles), as a result of using
galaxy templates with or without emission lines in the SED fitting
procedure.  Nonetheless, we release with this paper the best-fit SEDs
computed with both {\tt FAST} and {\tt Synthesizer} to enable further
investigations in specific subsets of emission line galaxies. Since
there are no obvious advantages to the use of either set of stellar
mass estimates we choose the values computed with {\tt FAST} as our
fiducial stellar masses for the remainder of this work. This choice
allows a more direct comparison to the stellar masses computed by the
3D-HST survey using the same fitting code and modeling assumptions
(see below).

The left panel of Figure~\ref{fig:masscompleteness} shows the stellar
mass distribution for all galaxies in our catalog as a function of
redshift using a grey-color intensity scale to indicate the H-band
magnitude of the galaxies. We use this diagram to study the mass
completeness of the sample. The completeness of magnitude limited
samples, such as this one, decreases with redshift and is typically
lower for red galaxies, either because they have intrinsically old
stellar populations with larger mass-to-light ratios, or because the
dust attenuation makes the galaxies fainter than unobscured SFGs of
the same mass. Therefore, we characterize the mass completeness of the
sample by estimating the mass threshold for the reddest galaxies with
H-band magnitudes equal to the SNR$\sim$5 detection limit of the
survey ($H\sim26$).  Red galaxies fainter than this threshold will be
undetected at the depth of the survey. The orange and red lines show
the mass completeness limit for 3 galaxy templates of quiescent
galaxies with ages ranging between $1-3$~Gyr, i.e., the age of a
recently quenched galaxy at any redshift, and the age of maximally old
galaxies at $z=2-3$. The purple line shows the detection limit for a
young (250~Myr), dusty (Av$=2$) star-forming galaxy. In agreement with
previous estimates of the mass completeness for the CANDELS catalogs
in other fields (e.g., \citealt{tal14}; \citealt{nayyeri17};
\citealt{stefanon17}), we find that our catalog is complete to
\lmass$\gtrsim10$ up to redshift $z\sim3$ except perhaps for the most
extreme dusty galaxies (Av$\gg2$; see e.g., \citealt{wang16} for a
study of dusty H-band drop outs). Interestingly, massive, recently
quenched galaxies (also called post-starburst) can be reliably
detected up to $z\sim4$, as shown for example in
\citealt{straatman14}.

We further verify the quality of the stellar mass estimates by
studying the comoving number density of massive (\lmass$>$~10)
star-forming and quiescent galaxies as a function of redshift. The
right panel of Figure~\ref{fig:masscompleteness} shows that these
number densities agree with the results from the ULTRAVISTA sample
\citep{muzzin13smf}, which covers a larger area but to a shallower
depth, at the same redshifts and follow the predicted trend at higher
z. The CANDELS/GOODS-N sample is likely more susceptible to cosmic
variance effects at the lowest redshift bin, but owing to its deeper
limiting magnitude, it is possible to follow the evolution of both
blue and red galaxies up to higher
redshifts. Figure~\ref{fig:3dhstmass} shows one last quality check
which compares our stellar mass estimates vs. those from 3D-HST
catalog for the same galaxies. We find that the average offset, and
the scatter are in excellent agreement as reported in similar
comparisons presented in previous CANDELS data papers for the other 4
fields.

\begin{figure}[t]
\centering
\includegraphics[scale=0.45, angle=-90]{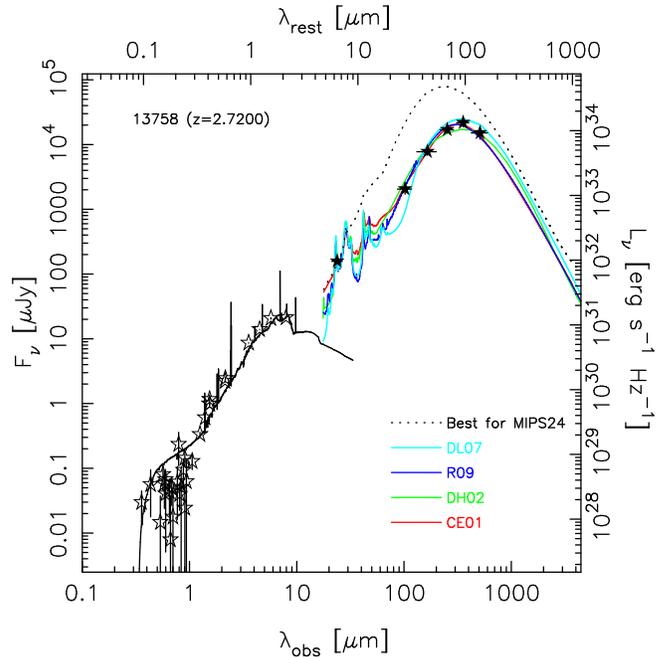}
\caption[]{\label{fig:IRSED} Example of a UV-to-FIR SEDs.  The black
  line shows best-fit BC03 stellar population model for the photometry
  up to 8$\mu$m rest-frame (open black stars), which provide an
  estimate of the stellar population properties and the dust
  attenuation.  The filled balck stars show the mid-to-far IR data
  from {\it Spitzer} MIPS and Herschel PACS and SPIRE. The red, green,
  blue and cyan lines show the best-fit dust emission models from the
  libraries of \cite{ce01}, \cite{dh02}, \cite{rieke09} and
  \cite{dl07}, which provide similar results. The dashed line shows
  the fit to only the MIPS 24~$\mu$m data. The significant difference
  between this and the other templates illustrates why using an
  average luminosity template, as in the \citep{wuyts11b} method, is a
  better approach for MIPS only SEDs.}
\end{figure}

\begin{figure*}[t]
\centering
\includegraphics[scale=0.6, angle=0]{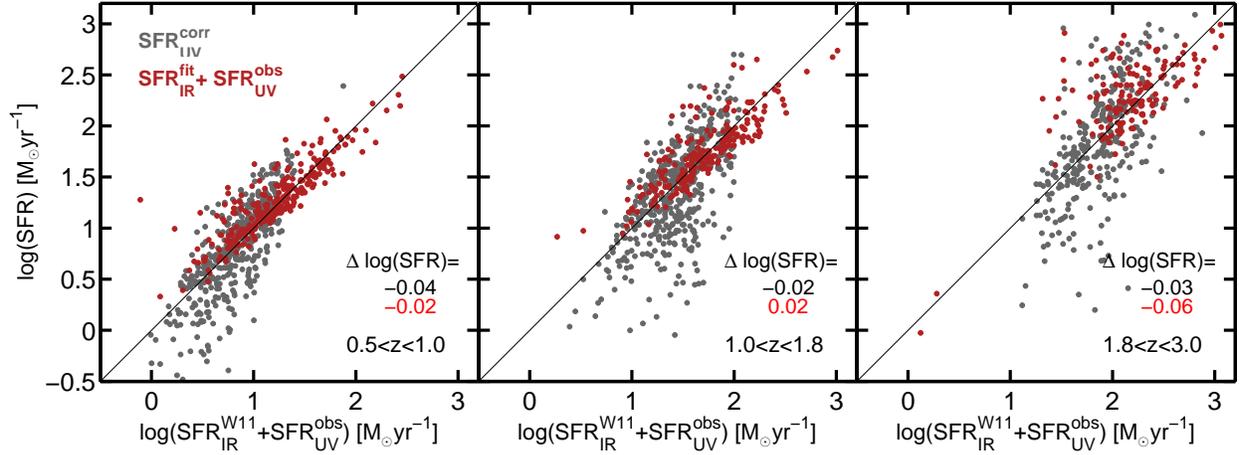}
\caption[]{\label{fig:sfrladder}Comparison between different total SFR
  indicators at different redshifts. The x-axis show the SFRs derived
  from the combination of the observed (unobscured) UV SFR$_{\rm
    UV}^{\rm obs}$ and the IR-based SFR determined from the MIPS
  24~$\mu$m flux using the prescription of \citet{wuyts11b}, SFR$_{\rm
    IR}^{\rm W11}$. They y-axis shows the total SFR derived either
  from the UV luminosity, corrected for extinction (SFR$_{\rm UV}^{\rm
    corr}$; grey circles), or the combination of SFR$_{\rm UV}^{\rm
    obs}$ and the IR-based SFRs computed from the fit of {\it all} the
  available FIR data to dust emission templates (red circles). The
  median difference between SFR estimates, $\Delta\log$SFR, is
  indicated for the UV corrected (black) and IR-based (red) SFRs.
  Overall, the two different estimates exhibit a good agreement
  showing that is it possible to combine them in a SFR ladder, that
  spans a range of more than 2~dex in SFRs.}
\end{figure*}


\subsection{UV+IR SFRs}
\label{ss:sfrs}
\subsubsection{The ``ladder'' of SFR indicators}
\label{sss:sfrladder}

In this section we use the UV-to-FIR SEDs to provide SFR estimates for
the galaxies in the sample. The depth of the optical/NIR photometry
guarantees accurate measurements of the rest-frame UV emission
($\lambda\sim1500-2500$~\AA), which is an excellent tracer of the
ongoing star formation, for all galaxies up to the highest
redshifts. However, the ubiquitous presence of dust in star forming
galaxies, typically embedded in the same gas from which stars are
formed, implies that the {\it intrinsic} UV emission of the galaxies
if frequently attenuated by dust absorption. In galaxies with
low-to-mid attenuations, it is possible to correct for the effect of
dust absorption using different methods such as the slope of the UV
emission, which is correlated with the Av (e.g.; \citealt{meurer97}),
or by estimating such Av from the optical-to-NIR SED fitting to
stellar population models, which provides also a direct estimate of
the intrinsic SFR. Nonetheless, several works have shown that the
strong dust attenuations found in massive galaxies at $z\gtrsim1.5$
can bias these corrections downwards, thus underestimating the SFRs
(e.g., \citealt{daddi07}; \citealt{reddy08,reddy10};
\citealt{wuyts11a}). The availability of FIR photometry provides a
direct probe into the emission of the dust particles, responsible for
the optical attenuation, which are being heated by the UV photons and
re-radiate this energy at longer wavelengths. Hence the FIR photometry
provides a useful SFR indicator for severely obscured galaxies, which
make a significant fraction of the massive galaxy population at high
z. The main drawback of this method is that the shallower depth of the
IR observations, compared to the optical and NIR data, limits the SFR
detection threshold, which becomes increasingly higher with redshift
and eventually leads to incompleteness even at the high mass end (see
e.g., Figure~2 of \citealt{scoville16}).

\begin{figure*}[t]
\centering
\includegraphics[scale=0.55, angle=0]{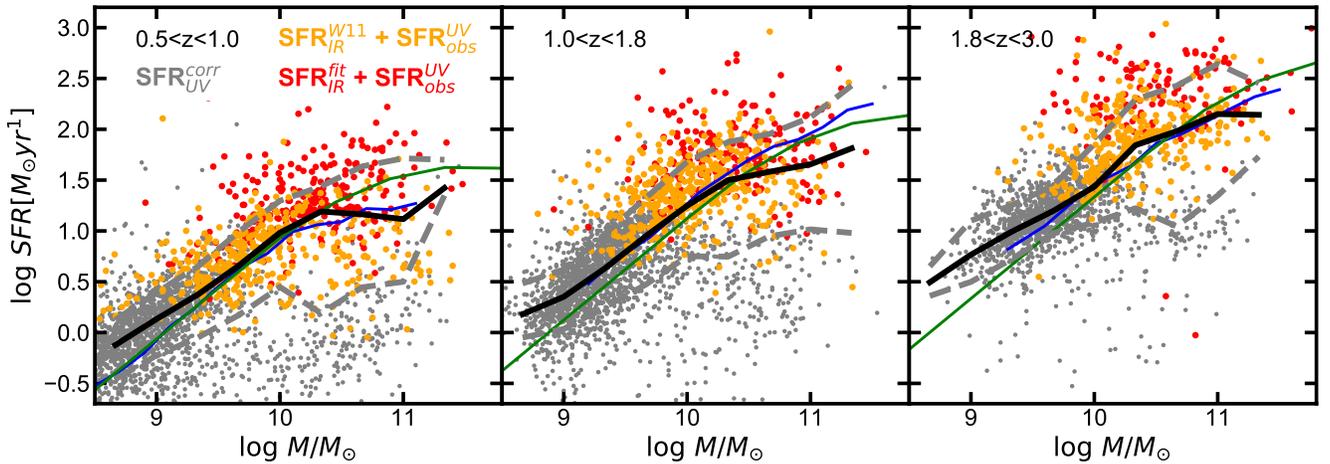}
\caption[]{\label{fig:sfrseq} SFR vs. stellar mass diagram (i.e., the
  SFR sequence; \citealt{mainseq}) as a function of redshift compared
  to results from the literature. The SFRs in the y-axis are computed
  by combining the different SFR indicators in
  Figure~\ref{fig:sfrladder} (grey, orange and red circles) to
  assemble the SFR ladder. The black and grey lines show the running
  median and 1$\sigma$ scatter of the star forming galaxies identified
  with the UVJ criterion. The blue and green lines show the SFR
  sequence from \citet{whitaker14} and \citet{coren15} which are
  roughly consistent with the overall distributions at all redshifts.}
\end{figure*}

A way forward to overcome the issues in both the UV and IR based SFR
estimates is to combine the constraints coming from both methods by
using a {\it ladder} of SFR indicators that can be cross-calibrated on
relatively massive galaxies with intermediate dust attenuations and
low IR fluxes. Here we follow such approach by using a method similar
to the one described in \citet{wuyts11a}. Briefly, the SFR ladder
usually consists of three steps which differ on the amount of SFR
indicators that are available for each galaxy, namely, UV, which is
availabe for all galaxies, mid-IR, available only for a subset of
massive (\lmass$\gtrsim$10) up to $z\sim3$, and far IR, available only
for a subset of those mid-IR detected galaxies.

The first step of the ladder is a UV-based SFR derived from the
rest-frame UV luminosities of the galaxies. Owing to the breadth of
optical and NIR photometric data the rest-frame UV luminosity (i.e.,
the monochromatic luminosity at 280~nm) can be measured for galaxies
at $z\gtrsim0.3$. Thus, it provides a general, homogeneous SFR
estimate for the whole sample. To compute the SFR from the UV
luminosity, we use the relation of \citet{ken98}, applying a
correction for dust attenuation, SFR$_{\rm TOT}(M_{\odot}~\rm yr^{-1})
= SFR_{\rm UV}^{\rm corr} =
1.09\times10^{-10}\times10^{0.4*A_{UV}}[3.3~L_{\rm
    UV}(280)/L_{\odot}]$, where $L(280)$ and A$_{\rm UV}$ are the UV
luminosity and dust attenuation at $\lambda=280$~nm, respectively.
The UV attenuation can be inferred directly from the the best-fit
model to the overall SED, which assumes a \citet{calzetti} attenuation
law (i.e., A$_{\rm UV}$=A(280)=1.8Av).  However, rather than assuming
these values, here we estimate $A_{\rm UV}$ on a galaxy-by-galaxy
basis from the comparison of the slope of the UV continuum emission
($\beta$; $f_{\lambda}\propto\lambda^{\beta}$) to the ratio of UV to
IR luminosities ($L(8-1000)/L_{\rm UV}\equiv$IRX), or upper limits of
the latter. The IRX-$\beta$ diagram provides a useful diagnostic to
characterize the dust attenution in each galaxy (\citealt{meurer97};
\citealt{kong04}; \citealt{2007ApJS..173..404B}). This is particularly
relevant to avoid overestimating the dust attenuation in blue, low
mass galaxies, undetected in the IR, for which the {\it starburst}
like \citet{calzetti} attenuation law is not a good match. A more
detailed description of this method is presented in
appendix~\ref{ap:SFRvalidation}.

The second and third steps of the SFR ladder include only galaxies
with mid-to-far IR detections. For those galaxies, we use a hybrid
estimate that combines UV and IR tracers. Firstly, we estimate the
contribution of the dust obscured SFR by fitting the FIR SEDs with
dust emission templates (see next section) and computing the
integrated IR luminosity, from 8 to 1000\,$\mu$m, L(8-1000). Then, we
obtain the {\it total} SFR by adding the contributions from the
obscured (IR) and the observed, unobscured (UV) SFR, SFR$_{\rm
  TOT}=~$SFR$_{\rm IR}+$SFR$_{\rm UV}^{\mathrm{obs}}$ (e.g.,
\citealt{ken98}; \citealt{bell05}),
\begin{equation}\label{sfrtot}
{\rm SFR}_{\rm TOT}=1.09\times10^{-10}[L(8-1000)+3.3~L_{\rm UV}(280)]
\end{equation}
where SFR is given in $\mathrm{M_\odot\,yr^{-1}}$, the luminosities
are in L$_{\odot}$ and the conversion factor from luminosity to SFR
assumes a \citet{chabrier} IMF.

\subsubsection{Mid and far IR based SFR}
\label{sss:irbased_sfr}

For galaxies with IR detections we distinguish between those with only
mid-IR detections and those with both mid- and far- IR detections. As
discussed in Section~2, our catalog includes photometry both from {\it
  Spitzer} and Herschel. The {\it Spitzer}/MIPS 24~$\mu$m photometry
is deeper than Herschel PACS and it has higher spatial resolution than
that of SPIRE. The downside of MIPS, is that it probes shorter, mid-IR
wavelengths, whereas the Herschel data spans a broader, FIR wavelength
range from $\lambda=100-500$~$\mu$m. Nonetheless, owing to its depth,
a larger fraction of the sample is detected at 24~$\mu$m than in any
Herschel band and thus provides SFR estimates for a large fraction of
the massive galaxies up to $z\sim3$. For galaxies detected only in
MIPS 24~$\mu$m, the modeling of the FIR emission with dust templates
is uncertain, as it is only constrained by one point, and it is prone
to overestimate the SFR at high-z due to a bias towards ULIRG
templates for relatively normal galaxies (\citealt{daddi07};
\citealt{symeonidis08}; \citealt{kartaltepe10}; \citealt{barro11b};
\citealt{elbaz11}). In order to avoid this problem, we use the
analytic conversion from MIPS 24~$\mu$m fluxes to L(TIR) from
\citet{wuyts08,wuyts11b}, SFR$_{\rm IR}^{\rm W11}$. This calibration
uses only one template generated by averaging multiple dust emission
templates and comparing to stacked {\it Spitzer} and Herschel
photometry for multiple galaxies (see also \citealt{rieke09} or
\citealt{rujopakarn13} for similar calibrations).

For galaxies with detections in multiple FIR photometric bands (e.g.,
galaxy SED in Figure~\ref{fig:IRSED}), we fit the thermal IR emission
(those bands redder than 5~$\mu$m rest-frame, which are supposed to be
dominated by dust emission) to the templates from the libraries of
\citet{ce01}, \citet{dh02}, \citet{rieke09} and \citet{dl07}. Then, we
compute the integrated luminosity L(TIR) as the average value of the
four sets. The typical differences in the predictions from the
different models are roughly a factor 2, consistent with the typical
uncertainties in similar kind of studies
(\citealt{2006ApJ...640...92P}, \citealt{daddi07},
\citealt{magnelli09}), SFR$_{\rm IR}^{\rm fit}$. We also
compute L(TIR) using the \citet{wuyts08} formula for these galaxies to
have an additional estimate of the SFR that can be used to validate
the accuracy of the IR SED fitting.

\subsubsection{Comparison of the SFRs}
\label{sss:sfr_checks}

Following \citealt{wuyts11b}, we cross compare the different SFR
estimates to validate their accuracy and
consistency. Figure~\ref{fig:sfrladder} shows the comparison between
the total SFR based on the UV luminosity, corrected for extinction
(SFR$_{\rm UV}^{\rm corr}$; grey), and the total SFR determined from
the fitting of all the far IR data to dust emission templates
(SFR$_{\rm IR}^{\rm fit}$+SFR$_{\rm UV}^{\rm obs}$; red) vs. the total
SFR based on the MIPS~24~$\mu$m fluxes and the \citet{wuyts11b}
relation (SFR$_{\rm IR}^{\rm W11}$+SFR$_{\rm UV}^{\rm obs}$). The
comparison to SFR$_{\rm UV}^{\rm corr}$ is based on galaxies with
relatively weak MIPS detections, f(24$\mu$m)$>20-150$~$\mu$Jy, while
the comparison to the IR SED-fitting SFRs includes all galaxies with
at least one detection in a Herschel band.  Overall, both comparisons
are in qualitatively good agreement showing that the SFR ladder method
can be used to derive self-consistent SFRs across a wide range of
galaxy masses and redshifts. The comparison to SFR$_{\rm UV}^{\rm
  corr}$ shows a small offset and scatter,
$\Delta$SFR$\sim0.03\pm0.32$~dex, that are consistent with the typical
dispersion expected in this kind of comparisons ($\sim0.3$~dex; e.g.,
\citealt{wuyts11b}). The comparison between the two IR-based SFRs
shows a similar offset but a much smaller scatter
($\lesssim$0.02$\pm0.17$~dex). The excellent agreement for the latter
is not surprising, considering that the MIPS data is used in both SFR
estimates, whereas the comparison to the UV-corrected SFRs is based on
completely different tracers (UV vs. IR).

We further analyze the consistency of the SFR ladder by comparing the
locus of our SFR main sequence, i.e., the correlation between SFR and
stellar mass (e.g., \citealt{mainseq}; \citealt{elbaz11};
\citealt{whitaker14}), at different redshifts, with previous results
from the literature. Figure~\ref{fig:sfrseq} shows how the combination
of the different tiers of the SFR ladder (grey, orange and red
circles) reproduce the characteristic distribution of the SFR
sequence, which exhibits a slope close to unity at lower masses, and
flattens towards the high-mass end (e.g., \citealt{speagle14}). The
black and grey lines show the running median and 1$\sigma$ scatter of
the distribution for star-forming galaxies selected using the UVJ
diagram (see figure~\ref{fig:uvj}). The overall distribution exhibits
a good agreement with the results of \citet[][blue line]{whitaker14}
and \citet[][green line]{coren15}, also based on far-IR photometry,
and shows a similar curvature as a function of mass. We further
discuss the properties of the SFR main sequence derived from our SFR
estimates in appendix~\ref{ap:SFRvalidation}, which also expands the
analysis to include data from the 5 CANDELS fields.

\section{Summary}
\label{s:summary}

In this paper we present an HST/WFC3 F160W (H-band) selected sample in
the CANDELS/GOODS-N field characterized with UV-to-FIR SEDs. The
sources are selected from an F160W mosaic which combines data from the
deep and wide observations of the CANDELS program over a total area of
171~arcmin$^{2}$. The photometric catalog includes the newly observed
HST data in F105W, F125W, and F160W, taken as part of the CANDELS
survey, with all the previous HST optical observations in ACS F435W,
F606W, F775W, F814W, and F850LP from GOODS and other smaller
programs. In addition to the HST data we include deep photometric data
in 25 optical medium-band filters from the SHARDS survey
\citep{shards}, which constitutes one of the deepest
($i$-band~$\sim28$~mag) and highest resolution (R$\sim$50) medium-band
surveys, surpassing the depth, spectral coverage and resolution of
similar medium band observations in other deep fields such as COSMOS
or GOODS-S. Furthermore, we include ancillary data in the ultraviolet:
LBC/U and KPNO/U', near IR: MOIRCS/K, CFHT/K' and {\it Spitzer}/IRAC
3.6, 4.5, 5.8, 8.0 $\mu$m and a wealth of far IR photometry at 24\mic,
and 70\mic\, from {\it Spitzer} MIPS and at 100\mic, 160\mic, 250\mic,
350\mic, and 500\mic, from Herschel PACS and SPIRE. In addition, we
complement and expand the broad and medium band photometry with
WFC3/IR grism spectroscopy in G102, from new observations (GO:13179),
and in G141 from AGHAST (GO:11600), as reduced by 3D-HST
\citep{3dhstgrism}. The combined grism observations make GOODS-N the
first among the deep cosmological fields to have similar areal
coverage in the contiguous grisms, which enables a simultaneous
analysis of several emission lines at different redshifts.

As in previous CANDELS papers, the source detection is based on a
slightly modified version of SExtractor which runs and combines the
output of two configurations (“cold” and “hot”) to optimally detect
and extract all sources ranging from the largest, most extended, and
brightest ones to the faintest and smallest. Similarly, we use {\tt
  TFIT} to measure uniform multi-band photometry in the lower
resolution optical-to-NIR bands. {\tt TFIT} uses priors from the
positions of the F160W detections to construct object templates that
are PSF-matched and fitted to the low-resolution images to measure the
photometry. The FIR photometry is measured independently and
self-consistently on the \spitzer and Herschel images using MIPS
24$\mu$m priors. Then, each IR source is assigned a single F160W
counterpart based on coordinates and IRAC fluxes. The F160W-selected
catalog contains $>$35,000 sources down to a 5$\sigma$ limiting depth
of $H=$27.4 and 28.2~mag in the wide and deep regions of the mosaic,
respectively.


We show that the overall SEDs present the level of consistency
required to characterize the intrinsic stellar populations of the
galaxy, and we perform a galaxy-by-galaxy fitting of the UV-to-FIR
SEDs to stellar population and dust emission models and a fit of their
grism spectra for emission and absorption lines. From the best fitting
optical and IR templates, we estimate: (1) photometric redshifts, (2)
stellar population properties (i.e., stellar masses, Av, etc.) and (3)
SFRs. From the line-fitting of the G102 and G141 HST grism data, we
estimate line fluxes and equivalent widths.  Then, we analyze the
accuracy and reliability of these estimates with respect to different
parameters. A summary of the most important results of this analysis
follows.

\begin{itemize}
\item The use of medium-band photometry and grism spectroscopy
  significantly improves the quality of the photometric redshifts
  relative to the case of SED fitting to broad-band photometry
  only. The comparison to 1,485 spectroscopic redshifts up to $z\sim3$
  yields a $\sigma_{\rm NCMAD}=0.0032$ and an outlier fraction of
  $\eta=$4.3\%, nearly a factor of 7 more precise than the typical,
  broad-band based redshifts.

\item Owing to the deep limiting magnitude of the F160W mosaic, our
  catalog is nearly complete for stellar masses above \lmass$=10$ up
  to redshift $z\sim3$, except perhaps for the most extreme dusty
  galaxies (Av$\gg$2).

\item The line catalog contains nearly 2,000 emission lines detected
  at SNR$>3$, with 30\% and 70\% of them identified in the G102 and
  G141 spectra, respectively. The G102 detections are typically in
  brighter and lower redshift galaxies, $H<23.5$ and $\bar{z}=0.7$,
  while the G141 detections are in fainter and higher redshfit
  galaxies, $H<24.5$ and $\bar{z}=1.5$. For magnitudes $H<23.5$,
  roughly 35\% of all galaxies have two emission lines detected,
  typically \Ha(G141) and \Hb(G102) at $z=0.6-1.3$.
  
\item We obtain robust SFRs for all the galaxies in the catalog by
  using in each case the best available SFR tracer from either the
  rest-frame UV or the far IR emission. The overall SFR {\it ladder}
  is computed self-consistently verifying the consistency in the SFRs
  for galaxies with multiple indicators, and showing that the
  resulting SFR main sequence is in good agreement with previous works
  at all redshifts over a range of 2~dex in stellar mass.
\end{itemize}




The multi-band photometry and the added-value catalogs (see
appendices) presented in this paper are further complemented with
additional data products from the CANDELS survey which allow even
richer studies the galaxy populations by adding, for example,
structural properties \citep{vdw14} or visual morphologies
\citep{huertas15}, which enables a wide array of additional science in
one of the premier cosmological fields.

\section*{Acknowledgments}

Support for Program number HST-GO-12060 was provided by NASA through a
grant from the Space Telescope Science Institute, which is operated by
the Association of Universities for Research in Astronomy,
Incorporated, under NASA contract NAS5-26555. G.B., S.M.F, and
D.K. acknowledge support from HST-GO-13420 and NSF grants AST-0808133
and AST-1615730. P.G.P.-G. acknowledges support from grant
AYA2015-63650-P. This work has made use of the Rainbow Cosmological
Surveys Database, which is operated by the Centro de Astrobiología
(CAB/INTA), partnered with the University of California Observatories
at Santa Cruz (UCO/Lick, UCSC). This work is based in part on
observations made with the Spitzer Space Telescope, which is operated
by the Jet Propulsion Laboratory, California Institute of Technology
under a contract with NASA. Support for this work was provided by
NASA.


\begin{appendix}
\setcounter{figure}{0}
\setcounter{table}{0}
\renewcommand{\thetable}{A\arabic{table}}

\section{Photometric catalogs}
\label{ap:photometry}
This appendix features CANDELS GOODS-N multi-band photometric catalog
entries with optical observations from KPNO, LBC, HST/ACS, near
infrared data from HST/ WFC3, CFHT and Subaru, and infrared
observations from \spitzer/IRAC (Table~\ref{table:phot}). This catalog
also contains useful SExtractor outputs for the HST bands as in the
catalogs released with the previous CANDELS papers.

\begin{table*}[htbp]
\setlength{\tabcolsep}{0.025in} \tabletypesize{\scriptsize}
\caption{Description of the CANDELS GOODS-N Photometric Catalog \label{table:phot}}
\begin{adjustbox}{max width=\textwidth}
\begin{threeparttable}
\begin{tabular}{lll}
\hline\hline
Column No. & Column Title & Description \\
\hline
1 & ID & Object identifier, beginning from 1 \\
2 & IAU Name & \\
3,4 & RA, DEC & Right ascension and declination (J2000.0; decimal degrees) \\
5 & FLAGS & SExtractor F160W flag used to designate suspicious sources that fall in contaminated regions\\
6 & CLASS\_STAR & SExtractor CLASS\_STAR parameter in the F160W band \\
7 & X-ray & X-ray ID from \citet{chandra2m}\\
8--41 & Flux, Flux\_Err        & Flux and flux in each filter. Sources that are not observed have (-99.00, -99.00, 0). \\
      &                        & Filters are included in order: KPNO\_U, LBC\_U, F435W, F606W, F775W, F814W, F850LP, F105W, \\
      &                        & F125W, F160W, MOIRCS\_Ks, CFHT\_Ks, 3.6 $\mu$m, 4.5 $\mu$m, 5.8 $\mu$m, and 8.0 $\mu$m \\
\hline
1 & ID & Object identifier, beginning from 1 \\
2--10  & FLUX\_MAX                & in HST bands\\
11--28 & FLUX\_ISO, FLUXERR\_ISO  & Isophotal flux and flux error in 9 HST bands\\
29--46 & FLUX\_ISOCOR, FLUXERR\_ISOCOR  & Isophotal flux and flux error in 9 HST bands\\
47--64 & FLUX\_AUTO, FLUXERR\_AUTO & AUTO flux and flux error in HST 9 bands\\
65--82 & FLUX\_PETRO, FLUXERR\_PETRO & PETRO flux and flux error in HST 9 bands\\
83--100  & FLUX\_BEST, FLUXERR\_BEST & BEST flux and flux error in HST 9 bands\\
101--280 & FLUX\_APER, FLUXERR\_APER & APER flux and flux error in HST 9 bands and 10 circular apertures of radius 1.47, 2.08, 2.94, 4.17, 5.88, 8.34, 11.79, 16.66, 23.57, 33.34, 47.13\\
\hline
1 & ID & Object identifier, beginning from 1 \\
2 & X\_IMAGE    & Object position along x [pixel] \\
 3 &  Y\_IMAGE  & Object position along y [pixel]\\
 4 & XPEAK\_IMAGE &x-coordinate of the brightest pixel [pixel]\\
 5 & YPEAK\_IMAGE &y-coordinate of the brightest pixel [pixel]\\
 6 & XMIN\_IMAGE &Minimum x-coordinate among detected pixels [pixel]\\
 7 & YMIN\_IMAGE &Minimum y-coordinate among detected pixels [pixel]\\
 8 & XMAX\_IMAGE &Maximum x-coordinate among detected pixels [pixel]\\
 9 & YMAX\_IMAGE &Maximum y-coordinate among detected pixels [pixel]\\
 10&  X2\_IMAGE &Variance along x [pixel**2]\\
 11&  Y2\_IMAGE &Variance along y [pixel**2]\\
 12&  XY\_IMAGE &Covariance between x and y [pixel**2]\\
 13&  CXX\_IMAGE &Cxx object ellipse parameter [pixel**(-2)]\\
 14&  CYY\_IMAGE &Cyy object ellipse parameter [pixel**(-2)]\\
 15&  CXY\_IMAGE &Cxy object ellipse parameter [pixel**(-2)]\\
16,17 & A\_IMAGE, B\_IMAGE & F160W profile RMS along major and minor axis (pixel) \\
18,19 & ERRA\_IMAGE, ERRB\_IMAGE & F160W profile RMS along major and minor axis (pixel) \\
20,21 & THETA\_IMAGE, ERRTHETA\_IMAGE & F160W position angle (degree) \\
22    & ISOAREAF\_IMAGE & SExtractor F160W Isophotal area (filtered) above Detection threshold (${\rm pixel^2}$) \\
23--31& ISOAREA\_IMAGE     & SExtractor Isophotal area (filtered) above Detection threshold (${\rm pixel^2}$) in HST bands \\
32--40& BACKGROUND         & Background at centroid position in HST bands \\
41--49& FLUX\_RADIUS\_1            & RADIUS\_1 with the 0.2 fraction of light in HST bands\\
50--58& FLUX\_RADIUS\_2            & RADIUS\_2 with the 0.5 fraction of light in HST bands\\
59--67& FLUX\_RADIUS\_3            & RADIUS\_3 with the 0.8 fraction of light in HST bands\\
68--76& FWHM\_IMAGE & FWHM of the image of an object, in unit of pixel (1 pixel = 0\farcs06) in HST bands\\
77    & KRON\_RADIUS & F160W band Kron radius from SExtractor (in unit of A\_IMAGE or B\_IMAGE) \\
78    & PETRO\_RADIUS & F160W band Petrosian radius from SExtractor (in unit of A\_IMAGE or B\_IMAGE) \\
\hline
\end{tabular}
\end{threeparttable}
\end{adjustbox}
\end{table*}

A separated catalog containing the photometry in the 25 medium bands
of the SHARDS survey is also provided (see
Table~\ref{table:shardsphot}). The medium-band photometry is shown
separately to explicitly indicate the variation in the central
wavelength of each passband (see Table~\ref{table:shardswav}) due to
the location of the sources in the SHARDS mosaic (see
\S~\ref{sss:SHARDS} and \citet{shards} for more details).

\begin{table*}
\begin{center}
  \caption{Description of the SHARDS photometry for the CANDELS
    catalog \label{table:shardsphot}}
\label{tab:shards}
\begin{threeparttable}
\begin{tabular}{l l l}
\hline
\hline
Column No. & Column Title & Description \\
\hline
1 & ID & Object identifier, beginning from 1 \\
2--76& Flux, Flux\_Err, Eff\_wav        & Flux, flux error and central effective wavelength (CWL) in each of the 25 SHARDS filters.\\
&  & Values of Flux $>$ 0 with Flux\_Err $=$ 0 indicate upper limits for non-detected sources.\\
&  & Values of Flux $=$ 0 and Flux\_Err $=$ 0 indicate that the source is not observed.\\
&  & The CWL is a function of the position of the object in the SHARDS mosaic (see \S~2.2.2)\\
\hline
\end{tabular}
\end{threeparttable}
\end{center}
\end{table*}

\begin{table}
\begin{center}
  \caption{Nominal Central Wavelenghts of the SHARDS
    filter set \label{table:shardswav}}
\label{tab:shards}
\begin{threeparttable}
\begin{tabular}{l l l}
\hline
Filter & CWL (nm) & Width (nm) \\
\hline
F500W17&500.0& 15.0\\
F517W17&517.0& 16.5\\
F534W17&534.0& 17.7\\
F551W17&551.0& 13.8\\
F568W17&568.0& 14.4\\
F585W17&585.0& 15.1\\
F602W17&602.0& 15.5\\
F619W17&619.0& 15.8\\
F636W17&638.4& 15.4\\
F653W17&653.1& 14.8\\
F670W17&668.4& 15.3\\
F687W17&688.2& 15.3\\
F704W17&704.5& 17.1\\
F721W17&720.2& 18.2\\
F738W17&737.8& 15.0\\
F755W17&754.5& 14.8\\
F772W17&770.9& 15.4\\
F789W17&789.0& 15.5\\
F806W17&805.6& 15.6\\
F823W17&825.4& 14.7\\
F840W17&840.0& 15.4\\
F857W17&856.4& 15.8\\
F883W35&880.3& 31.7\\
F913W25&913.0& 27.8\\
F941W33&941.0& 33.3\\
\hline
\end{tabular}
\begin{tablenotes}
\item[\url{https://guaix.fis.ucm.es/~pgperez/SHARDS/Filters}]
\end{tablenotes}
\end{threeparttable}
\end{center}
\end{table}

\section{Optical/NIR SED fitting based catalogs: photometric redshifts and stellar properties}
\label{ap:redshift_masses}

This appendix describes the content of the tables with the photometric
redshifts and stellar properties for all the objects in the catalog
derived from the fitting of their SEDs.  The redshift
Table~\ref{table:photoz_tier} contains the best-effort redshift
estimate based either on good quality spectroscopic redshift or on the
3-tier photometric redshift method, i.e., computed either with
broad-band data, broad-band and SHARDS medium-band data or broad and
medium-band plus WFC3 grism.

The redshift Table~\ref{table:photoz_team} contains supplementary
photometric redshift estimates computed by 5 different investigators
using only broad-band data. For this particular estimates homogeneity
in the input photometric data was preferred over quality of the
photo-z.  The latter can be improved by adding higher spectral
resolution data (as in the 3-tier method), but such data is only
available for smaller subsets of the whole sample. The codes an
assumptions used by each different investigator are the following:
Finkelstein, using {\tt eazy} based on the standard templates with
emission lines plus an additional high-z galaxy template (BX14 from
\citealt{erb10}); Salvato, using {\tt Lephare} \citep{lephare} based
on BC03+Polleta AGN templates without emission lines; Fontana, using
{\tt zphot} \citep{fontana00} based on BC03 templates with emission
lines; Wuyts, using {\tt eazy} based on the standard templates with
emission lines; Wiklind, using {\tt WikZ} from \citet{wiklind08} based
on BC03 templates without emission lines.

Table~\ref{table:sps_fast} describes the content of the catalog with
the stellar population properties computed with {\tt FAST} and the
rest-frame absolute magnitudes derived with {\tt
  EAZY}. Table~\ref{table:sps_synth} describes the content of the
catalog with the stellar population properties computed with {\tt
  Synthesizer}.

\setcounter{table}{0}
\renewcommand{\thetable}{B\arabic{table}}
\begin{table*}
\tabletypesize{\scriptsize}
\begin{center}
\caption{Description of the photometric redshift catalog \label{table:photoz_tier}}
\label{tab:photoz}
\begin{threeparttable}
\begin{tabular}{l l l}
\hline
\hline
Column No. & Column Name & Description\\
\hline
1     & id       &	Object identifier\\
2     & zspec    &      Spectroscopic redshift\\
3     & zpecflag &      Spectroscopic redshift quality flag (1-- most reliable, 2 -- reliable, 3 -- unreliable)\\
4     & zref     &      Original reference for the spectroscopic redshift\\
5     & ztier    &      Best photometric redshift from the 3-tiered estimate\\
6     & ztier\_err   &  Uncertainty in ztier computed from the 68\% confidence region of the PDFz\\
7     & ztier\_class &  Classification of ztier:\\
      &              &  1- Broad-band only, 2-Broad-band and SHARDS, 3-Broad-band, SHARDS and WFC3 grism\\
8     & zbest        &  Best redshit estimate from zspec, if available and with flag $<3$, or ztier\\
\hline
\end{tabular}
\end{threeparttable}
\end{center}
\end{table*}

\begin{table}
\begin{center}
\caption{Description of the supplementary photo-z from different investigators \label{table:photoz_team}}
\label{tab:fastheader}
\begin{threeparttable}
\begin{tabular}{l l}
\hline
\hline
Column No.   & Column ID. \\
\hline
1 &id      \\      
2 &zphot\_finkelstein      \\	  
3 &zphot\_finkelstein\_inf68\\    
4 &zphot\_finkelstein\_sup68\\          
5 &zphot\_salvato      \\	  
6 &zphot\_salvato\_inf68\\        
7 &zphot\_salvato\_sup68\\              
8 &zphot\_fontana      \\	  
9 &zphot\_fontana\_inf68\\        
10&zphot\_fontana\_sup68\\              
11 &zphot\_wuyts      \\	  
12 &zphot\_wuyts\_inf68\\         
13 &zphot\_wuyts\_sup68\\               
14 &zphot\_wiklind      \\	  
15 &zphot\_wiklind\_inf68\\       
16 &zphot\_wiklind\_sup68\\       
\hline
\end{tabular}
\end{threeparttable}
\end{center}
\end{table}

\begin{table}
\begin{center}
\caption{Stellar population properties catalog from FAST \label{table:sps_fast}}
\label{tab:fastheader}
\begin{threeparttable}
\begin{tabular}{l l l}
\hline
\hline
Column No. & Column ID. & Description\\
\hline
1 &id&			Object identifier\\
2 &z&      		=zbest from Table B1\\
3 &ltau&     		log[tau/yr] \\
4 &metal&      		metallicity (fixed to 0.020)\\
5 &lage&        	log[age/yr]\\
6 &Av&     		dust reddening\\
7 &lmass&      		log[$M/M_{\Sun}$]\\
8 &lsfr&     		log[SFR/($M_{\Sun}$/yr)]\\
9 &lssfr&      		log[sSFR(/yr)]\\
10&la2t&      		log[age/$\tau$]\\
11&chi2&		minimum $\chi^2$\\
12&M$_{U}$&              abs. magnitude in Johnson U\\
13&M$_{V}$&              abs. magnitude in Johnson V\\
14&M$_{J}$&              abs. magnitude in 2MASS J\\
\hline
\end{tabular}
\end{threeparttable}
\end{center}
\end{table}

\begin{table}
\begin{center}
\caption{Stellar population properties from Synthesizer \label{table:sps_synth}}
\label{tab:synthheader}
\begin{threeparttable}
\begin{tabular}{l l l}
\hline
\hline
Column No. & Column ID. & Description\\
\hline
1 &id&			Object identifier \\
2 &z&      		=zbest from Table B1\\
3 &ltau&     		log[tau/yr] \\
4 &metal&      		metallicity (fixed to 0.020)\\
5 &lage&        	log[age/yr]\\
6 &Av&     		dust reddening\\
7 &lmass&      		log[$M/M_{\Sun}$]\\
\hline
\end{tabular}
\end{threeparttable}
\end{center}
\end{table}

\section{WFC3 grism spectroscopy: line catalogs}
\label{ap:linefluxes}
This appendix describes the content of the WFC3 grism line catalogs
derived from the G102 and G141 spectra. As discussed in
\S~\ref{ss:emission_lines}, the line fluxes and equivalent widths
presented in Table~\ref{table:line_cat_cols} are measured using the
same code introduced by the 3D-HST survey \citep{3dhstgrism}. This
code fits the spectra to gaussian line templates, normalized to the
total broad band photometry, using an MCMC method. The names and
nominal wavelengths of the lines are given in
Table~\ref{table:em_lines} (also Table~4 from \citealt{3dhstgrism}).

\setcounter{table}{0}
\renewcommand{\thetable}{C\arabic{table}}
\begin{table*}[!th]
\centering
\caption{Emission Line Catalog Based On The G102 And G141 Spectra}\label{table:line_cat_cols}
\begin{tabular}{p{2cm}l}
\hline \hline
\noalign{\smallskip}
Column name & Description \\
\hline
\noalign{\smallskip}
id & Object identifier \\
z  & Grism redshift used in the emission line fit, identical to {\tt ztier} in the redshift catalog\\
X\_FLUX   & Emission line flux in units of $10^{-17}$ $ergs~s^{-1}~cm^{-2}$\\
X\_ERR    & Error in the emission line flux  in units of $10^{-17}$ $ergs~s^{-1}~cm^{-2}$\\
X\_SCALE  & Multiplicative scaling factor to correct the flux of the emission line to the photometry\\ 
X\_EQW    & Emission line equivalent width in \AA\\
\noalign{\smallskip}
\hline
\noalign{\smallskip}
\multicolumn{2}{l}{NOTE: X = emission line name, as given in accompanying table.}
\end{tabular}
\vspace{+0.5cm}
\end{table*}

\begin{table}[ht]
\centering
\caption{Emission Lines}
\label{table:em_lines}
\begin{tabular}{llcc}
\hline \hline
\noalign{\smallskip}
~Line & Catalog ID & Rest wavelength [\AA] & Ratio    \\
\noalign{\smallskip}
\hline
\noalign{\smallskip}
~Ly$\alpha$ & Lya & 1215.400 &\nodata \\
~\ion{C}{4} & CIV & 1549.480 & \nodata \\
~\ion{Mg}{2} & MgII & 2799.117 & \nodata \\
~\ion{Ne}{5}  & NeV & 3346.800 & \nodata \\
~\ion{Ne}{6}  & NeVI & 3426.850 & \nodata \\
~[\ion{O}{2}]  & OII & 3729.875 & \nodata\\
~[\ion{Ne}{3}]  & NeIII & 3869.000 & \nodata \\
~\ion{He}{1} & HeIb & 3889.500 & \nodata \\
~H$\delta$ & Hd & 4102.892 & \nodata\\
~H$\gamma$ & Hg & 4341.680 & \nodata\\
~[\ion{O}{3}] & OIIIx & 4364.436 & \nodata\\
~\ion{He}{2} & HeII & 4687.500 & \nodata\\
~H$\beta$ & Hb & 4862.680 & \nodata\\
~[\ion{O}{3}] & OIII & 5008.240, 4960.295 & 2.98:1 \\
~\ion{He}{1} & HeI & 5877.200 & \nodata\\
~[\ion{O}{1}] & OI & 6302.046 & \nodata \\
~H$\alpha$ & Ha & 6564.610 & \nodata \\
~[\ion{S}{2}] & SII & 6718.290, 6732.670 & 1:1\\
~\ion{S}{3} & SIII & 9068.600, 9530.600 & 1:2.44\\
\noalign{\smallskip}
\hline
\noalign{\smallskip}
\end{tabular}
\end{table}

\section{Self-consistent far-IR photometry and SFRs for galaxies in the 5 CANDELS catalogs}
\label{ap:SFRvalidation}

The first section of this appendix describes in detail the method to
measure mid-to-far IR photometry and how to assign those fluxes to the
most likely counterparts in the F160W selected CANDELS catalogs in 5
fields, namely: UDS \citep{galametz13}, GOODS-S \citep{guo13}, EGS
\citep{stefanon17}, COSMOS \citep{nayyeri17}, and GOODS-N (this
paper). Then, we review the procedure to compute self-consistent SFRs
for all galaxies in the catalog using the SFR-{\it ladder} method,
which is based on a combination of UV- and IR- SFR tracers. This
method is also described in \S~\ref{ss:sfrs}. Here we provide
additional details on the procedure to estimate the UV dust
attenuation from the slope of the UV SED ($\beta$) and the ratio
between the observed UV and IR luminosities (IRX). Next, we assess the
quality of the IR photometry and SFRs by comparing our estimates to
those from other works in the literature. Lastly, we describe the
content of the tables presenting the IR fluxes, dust attenuations and
SFRs for all galaxies, and the quality flag tables for the IR
photometry which are aimed to help diagnose potentially problematic
sources.

\subsection{mid-to-far IR catalogs and matching to F160W sources}
\label{ap:IRphotometry}
\setcounter{table}{0}
\renewcommand{\thetable}{D\arabic{table}}
\setcounter{figure}{0}
\renewcommand{\thefigure}{D\arabic{figure}}

We build merged, mid-to-far IR photometric catalogs using {\it
  Spitzer}/IRAC and MIPS, and Herschel/PACS and SPIRE datasets
presented in \citet{pg05,pg08,pg10}, PEP + GOODS-Herschel
\citep{lutz11,magnelli13} and HerMES
\citep{oliver12}. Table~\ref{table:photometry_ir} sumamrizes the
5$\sigma$ limiting fluxes in each band and field. The method to
extract IR sources from the MIPS, PACS, and SPIRE mosaics and to
measure their photometry is described in detail in \citet[][see also
  \citealt{rawle16} and \citealt{lrm19}]{pg10}. Briefly, the method
consists of three steps: (1) source identification in each of the
individual IR bands by using a combination of priors and direct
detections, (2) photometric measurements based on PSF fitting and (3)
merging of the individual photometric catalogs to produce multi-band
MIPS, PACS, and SPIRE catalogs.

\subsubsection{mid-to-far IR photometric measurements}

\begin{table*}
\centering
\caption{Limiting fluxes (5$\sigma$) of the \textit{Spitzer} and \textit{Herschel} photometry used in this work.}
\label{table:photometry_ir}
\begin{tabular}{l c cc ccccc} 
\hline
& \multicolumn{2}{c}{$\mathcal{F}_{\mathrm{lim}}$ [$\mu$Jy]} & & \multicolumn{5}{c}{$\mathcal{F}_{\mathrm{lim}}$ [mJy]} \\
\cline{2-3} \cline{5-9}
& \multicolumn{2}{c}{$Spitzer$/MIPS} & & \multicolumn{2}{c}{$Herschel$/PACS} & \multicolumn{3}{c}{$Herschel$/SPIRE}\\
\rotatebox{0}{Field} & \rotatebox{0}{24$\mu$m}& \rotatebox{0}{70$\mu$m} & & \rotatebox{0}{100$\mu$m} & \rotatebox{0}{160$\mu$m} & \rotatebox{0}{250$\mu$m} & \rotatebox{0}{350$\mu$m} & \rotatebox{0}{500$\mu$m}\\
(1) & (2) &  (3) & & (4) & (5) & (6) & (7) & (8)\\
\hline
GOODS-N    &  30 &2500 &  & 1.6  & \,\,\,3.6  & 9.0  & 12.9 & 12.6\\
GOODS-S    &  30 &2500 &  & 1.1  & \,\,\,3.4  & 8.3  & 11.5 & 11.3\\
EGS        &  45 &3500 &  & 8.7  & \,\,\,13.1 & 14.7 & 17.3 & 17.9\\
COSMOS     &  70 & - &  & 2.9  & \,\,\,6.6  & 11.0 & 9.6  & 11.2\\
UDS        &  70 & - &  & 14.4 & \,\,\,26.7 & 19.4 & 19.2 & 20.0\\
\hline
\end{tabular}
\end{table*}

Given the overall higher sensitivity and spatial resolution of MIPS24,
the source identification in this band is based on direct detections
(i.e., without IRAC priors). For the lower-resolution and sensitivity
PACS and SPIRE bands we use a combination of IRAC and MIPS priors to
reduce the effects of confusion. For these priors, we use
IRAC-selected catalogs drawn from {\it Spitzer}/IRAC mosaics (see
\citealt{pg08}), i.e., not the TFIT-deblended sources based on F160W
described in \S~\ref{ss:midres_tfit}. Before extracting sources in
PACS and SPIRE, we remove from the prior catalog those sources that
are too close to be spatially resolved in each of those bands. Then,
the prior-based idenfications are complemented with direct detections
identified by running the source detection iteratively at different
detection levels, starting from bright to faint sources. This approach
improves the detecion of faint sources located close to very bright
sources \citep{pg05}.

The source detection catalog in each of the IR bands is used to
measure the photometry by fitting PSFs at the given positions using
the daophot package in {\tt IRAF}, and allowing for one pixel
centering offsets. The PSF models for each band are created using
bright, well-detected sources. The tailored PSFs allow to account for
data reduction effects such as drizzling and a repixelization of the
original images. The total fluxes are estimated from the flux
densities measured in optimum circular apertures and applying aperture
corrections (see section 2 of \citealt{pg10} for more details). To
ensure the robustness of the photometric measurement, all sources
detected below the $3\sigma$ detection limit of each band (see
Table~\ref{table:photometry_ir}) are removed from the single-band
catalogs.

Lastly, all the single-band catalogs are unified into 3 merged
catalogs for MIPS (24 and 70), PACS (100-to-160) and SPIRE
(250-to-500), respectively, which will then be used to assign IR
fluxes to the sources in the CANDELS/F160W catalog. We note that there
is little ambiguity in assigning MIPS24 counterparts to the PACS100
detections. In fact, 95\% of the PACS100 sources have a single
possible MIPS24 counterpart, and the remaining 5\% have only
two. Similarly, PACS160 has a lower spatial resolution, but also a
lower sensitivity (i.e., fewer detections). This means we can assign
reliable PACS100 counterparts for all the PACS160 sources and from
there tie them to MIPS. For the SPIRE bands, we find that each
SPIRE350 and SPIRE500 source can be identified with a single
counterpart in the deeper and higher resolution SPIRE250
band. Nonetheless, the multiplicity of the SPIRE250 sources in bluer
bands (down to IRAC) is equal or larger than 2 for 85\% of the
sources.

\subsubsection{Matching IR fluxes to F160W sources}
Based on the IR-only catalogs described above, we assign mid-to-far IR
fluxes to the CANDELS sources following the method described in
\citet{lrm19} to identify the most likely F160W counterpart for each
mid- and far-IR source. Briefly, we use a crossmatching procedure
based on the celestial coordinates which runs in 3 steps: first
identifying F160W counterparts to MIPS sources within 2\farcs5, then
MIPS counterparts to PACS sources within 3\farcs0 and finally PACS
counterparts to SPIRE sources within 9\farcs0.  Before the crossmatch,
we set SNR~$>3$ lower limits for the IR fluxes below which the sources
are excluded from the rest of the analysis.

If the multiplicity of the F160W to MIPS match is larger than 1, the
primary counterpart is determined by assigning the highest priority to
the F160W source with the highest flux in the reddest IRAC band,
typically 8.0$\mu$m, but sometimes 3.6/4.5$\mu$m for faint IRAC
sources.  If multiple counterparts have similar IRAC fluxes within
1$\sigma$ the primary is the one closest in distance to the MIPS
source. For the MIPS to PACS, and PACS to SPIRE identifications the
primary counterpart is determined again by prioritizing the brightness
in the reddest available band, e.g., the primary counterparts to a
PACS160 (SPIRE250) source is the brightest MIPS24 (PACS160)
neighbor. In the few cases where a PACS source has no MIPS
counterparts, we crossmatch directly to F160W and we assign the
primary counterpart based on the IRAC fluxes determined with {\tt
  TFIT}.

After this sequential process, only the primary counterpart in each
crossmatch is used in the following step. As a result, each mid-to-far
IR detection has a unique F160W counterpart in the final catalog.
Only those F160W sources with IR detections are used in the next
section to compute IR-based SFR estimates. Nonetheless, this paper
provides supplementary IR catalogs (see appendix~\ref{ap:sfrcats})
which indicate the all the secondary short-wavelength counterparts in
all the bands involved in the crossmatching procedure. These catalogs
also indicate the multiplicity, i.e., the total number of counterparts
to each long-wavelength source, which can be used for further
diagnostics.

\subsection{SFRs and dust attenuations from the SFR-ladder method}
\label{ap:sfr}

In the following we describe our method to obtain Star Formation Rates
(SFRs) for all the galaxies in the CANDELS catalogs. Our empirical
approach is based on an auto-consistent combination of three tracers:
the ultraviolet, the mid-, and the far-infrared emission. With them,
we analyze the extinction properties of the CANDELS $H$-band-selected
galaxies in order to provide a robust estimation of the SFR in a
galaxy-by-galaxy basis, taking into account both the evolving
intrinsic properties of galaxies as well as the limitations of the
data currently available in the CANDELS fields (in the relevant
wavelengths). Several tests have been carried out to check the
consistency of the different estimations. We describe the method and
tests in detail in the following paragraphs.

For each galaxy in our sample we aim at estimating its SFR taking into
account both the direct emission from young stars as well as from
stars obscured by interstellar dust. With this in mind, the total SFR
of a galaxy, $\mathrm{SFR}_\mathrm{TOT}$, can be calculated in two
different ways:

\begin{itemize}
\item[1.-]adding $\mathrm{SFR}_\mathrm{UV}^\mathrm{obs}$, the SFR
  linked to unobscured stars (provided by the observed total UV
  emission), to $\mathrm{SFR}_\mathrm{IR}$, the SFR associated with
  the emission from young stars which is absorbed by dust and
  reradiated in the IR:
\begin{equation}
  \mathrm{SFR}_\mathrm{TOT}=\mathrm{SFR}_\mathrm{UV}^\mathrm{obs}+\mathrm{SFR}_\mathrm{IR}
\end{equation}
\item[2.-]using the observed total UV emission,
  $\mathrm{SFR}_\mathrm{UV}^\mathrm{obs}$, and applying an attenuation
  correction to obtain the total SFR:
\begin{equation}
  \mathrm{SFR}_\mathrm{TOT}=\mathrm{SFR}_\mathrm{UV}^\mathrm{corr}=10^{0.4\mathrm{A_{UV}}}\mathrm{SFR}_\mathrm{UV}^\mathrm{obs}
\end{equation}
\end{itemize}

We note that joining the previous two equations we can obtain a
expression of the attenuation in terms of the SFRs:

\begin{equation}
  \mathrm{A_{UV}}=2.5\log(\mathrm{SFR}_\mathrm{IR}/\mathrm{SFR}_\mathrm{UV}^\mathrm{obs}+1)
\end{equation}

Concerning the first method, the observed UV-based SFR is typically
estimated from monochromatic luminosities, after applying a bolometric
correction to obtain the emission at all UV wavelengths. In this work
we consider $\mathrm{SFR}_\mathrm{UV}^\mathrm{obs}$ estimations based
on the luminosity at 160 and 280~nm rest-frame, transformed into a
bolometric UV emission and a SFR following the calibrations found in
\citet{ken98} and \citet{bell05}, respectively. For the first
transformation, we correct the original factor calculated for a
Salpeter (1955) IMF to a \citet{chabrier} IMF dividing by a 1.6
factor.  The UV-to-SFR calibrations are:

\begin{equation}
  \mathrm{SFR}_\mathrm{UV}^\mathrm{obs}=8.8\times10^{-29}\,L_\nu(160)
\end{equation}

\begin{equation}
  \mathrm{SFR}_\mathrm{UV}^\mathrm{obs}=1.0\times10^{-28}\,L_\nu(280)
\end{equation}

In both equations, the SFRs are given in $\mathrm{M_\odot\,yr^{-1}}$,
and the luminosity densities in $\mathrm{erg\,s^{-1}\,Hz^{-1}}$.

Concerning the IR-based SFR, $\mathrm{SFR}_\mathrm{IR}$, we estimate
it in two different ways. For galaxies which are detected (at least at
a $5\sigma$ level) by the {\it Spitzer} MIPS instrument at 24\,$\mu$m
and by one or several {\it Herschel} bands with the PACS and/or SPIRE
instruments, we fit the flux points to dust emission models. We use
the libraries published by \citet{ce01}, \citet{dh02}, \citet{rieke09}
and \citet{dl07}, and calculate the total IR luminosity integrating
the best fitting model for each library from 8 to 1000\,$\mu$m to
obtain L(8-1000). The typical scatter in L(8-1000) for the different
template sets is 0.05~dex. This bolometric IR luminosity is
transformed into a SFR using the calibration in \citet{ken98},
assuming a \citet{chabrier} IMF:

\begin{equation}
  \mathrm{SFR}_\mathrm{IR}^\mathrm{fit}=2.8\times10^{-44}\,L(8-1000)
\end{equation}

\noindent where L(8-1000) is given in $\mathrm{erg\,s^{-1}}$ and the
SFR in $\mathrm{M_\odot\,yr^{-1}}$,

For galaxies which are only detected by MIPS at 24~$\mu$m, the
calculation of L(8-1000) involves an extrapolation which is larger,
more uncertain, and affected by significant systematic errors as we
move to higher redshifts (see, e.g., \citealt{papovich07}).
Consequently, for these galaxies we used the transformations between
24~$\mu$m emission and SFR presented in \citet[][W11]{wuyts11a},
$\mathrm{SFR}_\mathrm{IR}^\mathrm{W11}$, who tested them against {\it
  Herschel}-based SFRs (see also the discussion in the next
paragraph). We also considered the transformations from MIPS 24~$\mu$m
to SFR proposed in \citet[][R13]{rujopakarn13}, but decided to use
Wuyts et al.'s recipe based on the better performance (see below).

\begin{figure*}
\begin{center}
\includegraphics[angle=-90,width=16cm]{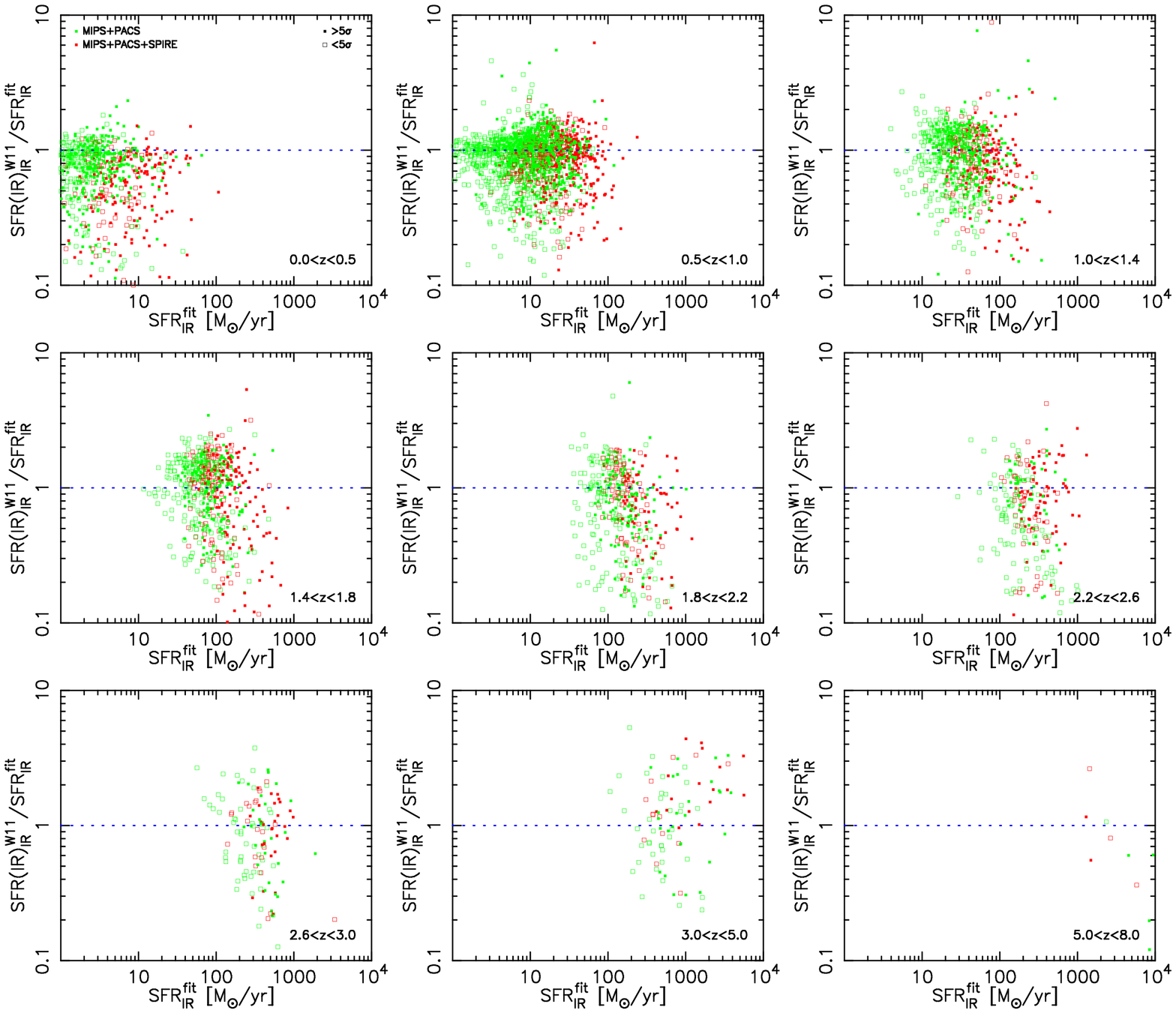}
\figcaption{\label{fig:sfr_herschel_W11}Comparison of the SFR
  estimations based on the MIPS 24~$\mu$m fluxes alone (using the
  \citealt{wuyts11a} recipe; $\mathrm{SFR}_\mathrm{IR}^\mathrm{W11}$)
  with the SFRs calculated by fitting the MIPS and Herschel flux data
  points to dust emission models (see text for details;
  $\mathrm{SFR}_\mathrm{IR}^\mathrm{fit}$) for all the IR detections
  in the five CANDELS fields. We divide the sample in galaxies
  detected, apart from MIPS, by both PACS and SPIRE (red points) or
  only by PACS (green points). Filled points refer to galaxies with at
  least $5\sigma$ detections, open points represent less significant
  detections.}
\end{center}
\end{figure*}

In order to test the reliability of the SFRs obtained from the
24~$\mu$m flux point alone, we compared them with the SFRs obtained
from the fits to dust emission models for the galaxies with {\it
  Herschel} detections. The results for GOODS-N are presented in
Figure~\ref{fig:sfrladder} in the main text. In
Figure~\ref{fig:sfr_herschel_W11} we plot the comparison for the 5
CANDELS fields. The typical systematic offset between the two SFR
estimations is always below 0.05~dex, except at $z>3$, and the scatter
is 0.2-0.3~dex. For comparison, the results obtained when comparing
the R13 figures with the SFRs obtained from fits to the MIPS+{\it
  Herschel} data are similar at $z\lesssim2$, 0.04~dex offset and
0.2-0.3~dex scatter, but considerably worse at $z\gtrsim2$, where the
systematic offset is larger than 0.2~dex, reaching 0.5~dex at $z>5$
(in all cases, the SFRs obtained from the fits are larger than those
obtained with the R13 recipe), and the scatter is 0.4~dex.

\slugcomment{Please, plot this figure with the width of two columns}
\placefigure{fig:sfr_herschel_W11}

For the galaxies with no IR detection, we used a UV-based SFR obtained
by applying an attenuation correction, A(UV), to the observed UV
luminosity. Typically, this correction is estimated by measuring the
UV spectral index, $\beta_\mathrm{UV}$, and transforming it to an
attenuation using a recipe such as the one presented in
\citet{meurer99}.  This procedure is based on the fact that young
($\lesssim100$~Myr) starbursts with no dust present a slope of
$\sim-2.2$ (for metallicities as low as 1/20 solar; see, e.g.,
\citealt{leitherer14}) when using $f_\lambda$ (our fiducial
definition; note that this is the same as a flat slope spectrum when
plotting the flux density in terms of $f_\nu$). In the presence of
dust (and also for older stellar populations), and assuming typical
dust properties, the UV slope typically becomes flatter (but not
always, see \citealt{wg00}), offering an opportunity to estimate the
attenuation. However, transforming $\beta_\mathrm{UV}$ into a UV
attenuation depends on the properties of the dust and its relative
location with respect to the stars (\citealt{wg00};
\citealt{popping17}; \citealt{narayanan18}). In addition, as mentioned
above, the UV slope is also mildly dependent on age for young
starbursts ($t<100$~Myr), but the effect of age dominates for more
evolved stellar populations.  Consequently, different types of
galaxies present different relationships between $\beta_\mathrm{UV}$
and A(UV).  For local starbursts, \citet{meurer99} compared UV and
far-IR luminosities, using their ratio (IRX) and its relationship with
$\beta_\mathrm{UV}$ to provide an easy-to-use attenuation recipe.
However, many papers in the literature have shown that the
IRX-$\beta_\mathrm{UV}$ relation presented in \citet{meurer99} is not
universal (among many, \citealt{pettini98}; \citealt{kong04};
\citealt{buat05,buat12}; \citealt{dale09}; \citealt{reddy10,reddy18};
\citealt{overzier11}; \citealt{casey14}, \citealt{popping17};
\citealt{mclure18}; \citealt{narayanan18}). In general, using the
Meurer et al.  IRX-$\beta_\mathrm{UV}$ relationship for all galaxies,
which is common in the literature, will provide acceptable UV
attenuations for active star-forming galaxies similar to those studied
by Meurer and collaborators. However, this procedure will overestimate
the effect of dust for more relaxed systems, which dominate the galaxy
population at low redshifts (see \citealt{dale09}; \citealt{casey14}),
and subestimate the attenuation for heavily obscured galaxies, such as
(U)LIRGs and SMGs at high redshifts (see, e.g., \citealt{overzier10};
\citealt{nordon13}; \citealt{salmon15}; \citealt{bouwens16}).

Based on the results from the literature discussed in the previous
paragraph, in this paper we present an elaborated method to account
for the differences in IRX-$\beta_\mathrm{UV}$ relationships from
galaxy to galaxy. Our empirical recipe is based on the fact that
galaxies present a variety of attenuation laws (which translate to
different IRX-$\beta_\mathrm{UV}$ relationships, see, e.g.,
\citealt{wg00}), but it is also dependent on data depth (in the mid-
and far-IR, as well as in the UV and near-infrared) for the CANDELS
fields. Our method relies on two pillars: the construction of
IRX-$\beta_\mathrm{UV}$ relationships for different types of galaxies
and the actual detection limits of the {\it Spitzer}/MIPS and {\it
  Herschel} surveys, which impose upper limits on the SFR calculations
for IR-undetected galaxies.

\placetable{IRX-beta}
\begin{deluxetable*}{lccccc}
\tabletypesize{\scriptsize}
\vspace{0cm}
\tablecaption{\label{table:IRX-beta}IRX-$\beta_\mathrm{UV}$ relationships as a function of  $\mathrm{SFR}_\mathrm{IR}$}
\tablehead{\colhead{$\mathrm{SFR}_\mathrm{IR}$ range} & \colhead{$T_0$} & \colhead{$T_1$} & \colhead{$T_2$} & \colhead{$T_3$} & \colhead{$T_4$} \\
\colhead{(1)} & \colhead{(2)} & \colhead{(2)} & \colhead{(2)} & \colhead{(2)} & \colhead{(2)}}
\startdata
\multicolumn{6}{c}{160~nm}\\
\hline
$\mathrm{SFR}_\mathrm{IR}>1050$      &+2.920720 & $+2.86435\times10^{-01}$ & $-6.78813\times10^{-03}$  & $+1.88041\times10^{-01}$ & $-5.50707\times10^{-02}$\\
$500<\mathrm{SFR}_\mathrm{IR}<1050$  &+2.741810 & $+6.99146\times10^{-01}$ & $+1.11188\times10^{-01}$  & $-8.58467\times10^{-02}$ & $-1.58543\times10^{-01}$\\
$300<\mathrm{SFR}_\mathrm{IR}<500$   &+2.527450 & $+4.06804\times10^{-01}$ & $-9.03771\times10^{-03}$  & $+1.60448\times10^{-01}$ & $-3.49370\times10^{-02}$\\
$105<\mathrm{SFR}_\mathrm{IR}<300$   &+2.333380 & $+8.51482\times10^{-01}$ & $+8.76543\times10^{-02}$  & $-1.91654\times10^{-01}$ & $-1.66376\times10^{-01}$\\
$75<\mathrm{SFR}_\mathrm{IR}<105$    &+2.084330 & $+8.26576\times10^{-01}$ & $+9.61815\times10^{-02}$  & $-1.43602\times10^{-01}$ & $-1.35397\times10^{-01}$\\
$45<\mathrm{SFR}_\mathrm{IR}<75$     &+1.963200 & $+7.20402\times10^{-01}$ & $+1.69449\times10^{-01}$  & $-5.59624\times10^{-02}$ & $-1.15281\times10^{-01}$\\
$11<\mathrm{SFR}_\mathrm{IR}<45$     &+1.609040 & $+6.55166\times10^{-01}$ & $+1.14031\times10^{-01}$  & $-1.15595\times10^{-01}$ & $-1.19841\times10^{-01}$\\
$7.5<\mathrm{SFR}_\mathrm{IR}<11$    &+1.361420 & $+7.30849\times10^{-01}$ & $+1.04347\times10^{-01}$  & $-2.15491\times10^{-01}$ & $-1.43217\times10^{-01}$\\
$4.5<\mathrm{SFR}_\mathrm{IR}<7.5$   &+1.288050 & $+7.16792\times10^{-01}$ & $+7.21248\times10^{-02}$  & $-2.07808\times10^{-01}$ & $-1.30655\times10^{-01}$\\
$1.1<\mathrm{SFR}_\mathrm{IR}<4.5$   &+1.129310 & $+7.96516\times10^{-01}$ & $+8.45170\times10^{-02}$  & $-2.11392\times10^{-01}$ & $-1.18125\times10^{-01}$\\
$\mathrm{SFR}_\mathrm{IR}<1.1$       &+0.849941 & $+4.50507\times10^{-01}$ & $+8.16457\times10^{-02}$  & $-2.36188\times10^{-02}$ & $-5.08052\times10^{-02}$\\
\hline
\hline
\multicolumn{6}{c}{280~nm}\\
\hline
$\mathrm{SFR}_\mathrm{IR}>1050$      &  $2.52088$ & $+3.49898\times10^{-01}$ & $-8.05563\times10^{-02}$ & $+1.04188\times10^{-01}$ & $-5.78644\times10^{-02}$\\
$500<\mathrm{SFR}_\mathrm{IR}<1050$  &  $2.22664$ & $+4.85975\times10^{-01}$ & $+8.61690\times10^{-02}$ & $-1.04035\times10^{-01}$ & $-1.64763\times10^{-01}$\\
$300<\mathrm{SFR}_\mathrm{IR}<500$   &  $2.00424$ & $+1.30284\times10^{-01}$ & $-1.97642\times10^{-02}$ & $+1.57070\times10^{-01}$ & $-4.37325\times10^{-02}$\\
$105<\mathrm{SFR}_\mathrm{IR}<300$   &  $1.78244$ & $+5.72340\times10^{-01}$ & $+1.02271\times10^{-01}$ & $-1.56439\times10^{-01}$ & $-1.61426\times10^{-01}$\\
$75<\mathrm{SFR}_\mathrm{IR}<105$    &  $1.52895$ & $+5.97391\times10^{-01}$ & $+1.80602\times10^{-01}$ & $-1.29600\times10^{-01}$ & $-1.50702\times10^{-01}$\\
$45<\mathrm{SFR}_\mathrm{IR}<75$     &  $1.46421$ & $+5.65381\times10^{-01}$ & $+1.34942\times10^{-01}$ & $-1.98317\times10^{-01}$ & $-1.73069\times10^{-01}$\\
$11<\mathrm{SFR}_\mathrm{IR}<45$     &  $1.04869$ & $+3.38949\times10^{-01}$ & $+1.47252\times10^{-01}$ & $-4.06656\times10^{-02}$ & $-1.03828\times10^{-01}$\\
$7.5<\mathrm{SFR}_\mathrm{IR}<11$    &  $0.80222$ & $+4.40004\times10^{-01}$ & $+1.42895\times10^{-01}$ & $-1.66367\times10^{-01}$ & $-1.38347\times10^{-01}$\\
$4.5<\mathrm{SFR}_\mathrm{IR}<7.5$   &  $0.72929$ & $+4.08079\times10^{-01}$ & $+1.45188\times10^{-01}$ & $-1.01459\times10^{-01}$ & $-1.07982\times10^{-01}$\\
$1.1<\mathrm{SFR}_\mathrm{IR}<4.5$   &  $0.58280$ & $+4.91613\times10^{-01}$ & $+1.05090\times10^{-01}$ & $-1.42693\times10^{-01}$ & $-1.01624\times10^{-01}$\\
$\mathrm{SFR}_\mathrm{IR}<1.1$       &  $0.31596$ & $+2.53008\times10^{-01}$ & $+6.50686\times10^{-02}$ & $-1.30536\times10^{-01}$ & $-9.86393\times10^{-02}$\\
\enddata
\tablecomments{(1) $\mathrm{SFR}_\mathrm{IR}$ range (in
  $\mathrm{M_\odot\,yr^{-1}}$). (2) Chebyshev polynomial coefficients.}
\end{deluxetable*}

\begin{figure*}
\begin{center}
\includegraphics[angle=-90,width=14cm]{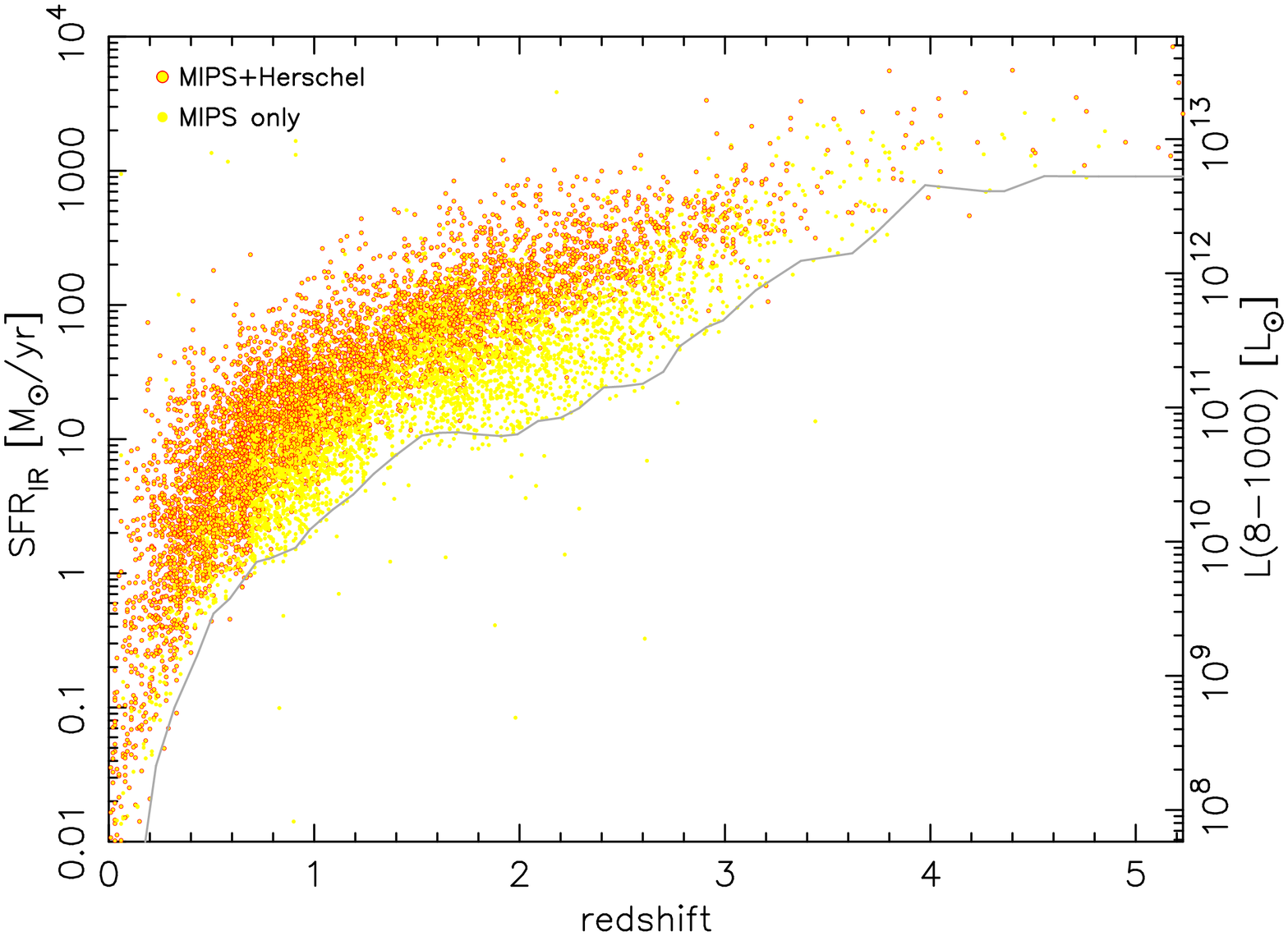}
\figcaption{\label{fig:sfr_redshift}Observational limits of the mid-
  and far-IR surveys carried out by {\it Spitzer} with MIPS and {\it
    Herschel} with PACS and SPIRE in the five CANDELS fields. The
  IR-based SFR, $\mathrm{SFR}_\mathrm{IR}$ (i.e., the unabsorbed part
  of the total SFR has not been taken into account), is plotted as a
  function of redshift. Galaxies with {\it Herschel} detections are
  marked in red (and
  $\mathrm{SFR}_\mathrm{IR}=\mathrm{SFR}_\mathrm{IR}^\mathrm{fit}$,
  see text for details), galaxies with just a MIPS detection are
  plotted in yellow (and
  $\mathrm{SFR}_\mathrm{IR}=\mathrm{SFR}_\mathrm{IR}^\mathrm{W11}$).
  The gray line marks the fluxes above which we can find 90\% of the
  IR sources for GOODS-N and GOODS-S, the two CANDELS fields with the
  deepest MIPS/PACS/SPIRE data. }
\end{center}
\end{figure*}

\slugcomment{Please, plot this figure with the width of two columns}
\placefigure{fig:sfr_redshift}

Figure~\ref{fig:sfr_redshift} shows the $\mathrm{SFR}_\mathrm{IR}$
limits of the MIPS and {\it Herschel} surveys in the CANDELS fields.
The gray line marks the boundary of the IR detections (90\% of
galaxies lie above that line).  We remark that any galaxy presenting a
large enough amount of dust should be detected in the IR and lie above
the gray line. Analogously, for galaxies with no IR detection, the
amount of dust must be limited so the absorbed part of the SFR can
stay below the observational limits. We note that the IR surveys in
each CANDELS field count with different depths. In this plot we show
the $\mathrm{SFR}_\mathrm{IR}$ limits based on the deepest datasets
(in the two GOODS fields), but our method to estimate SFRs from UV
data alone properly scales up the limits for the other CANDELS fields.

\begin{figure*}
\begin{center}
\includegraphics[angle=-90,width=7.5cm]{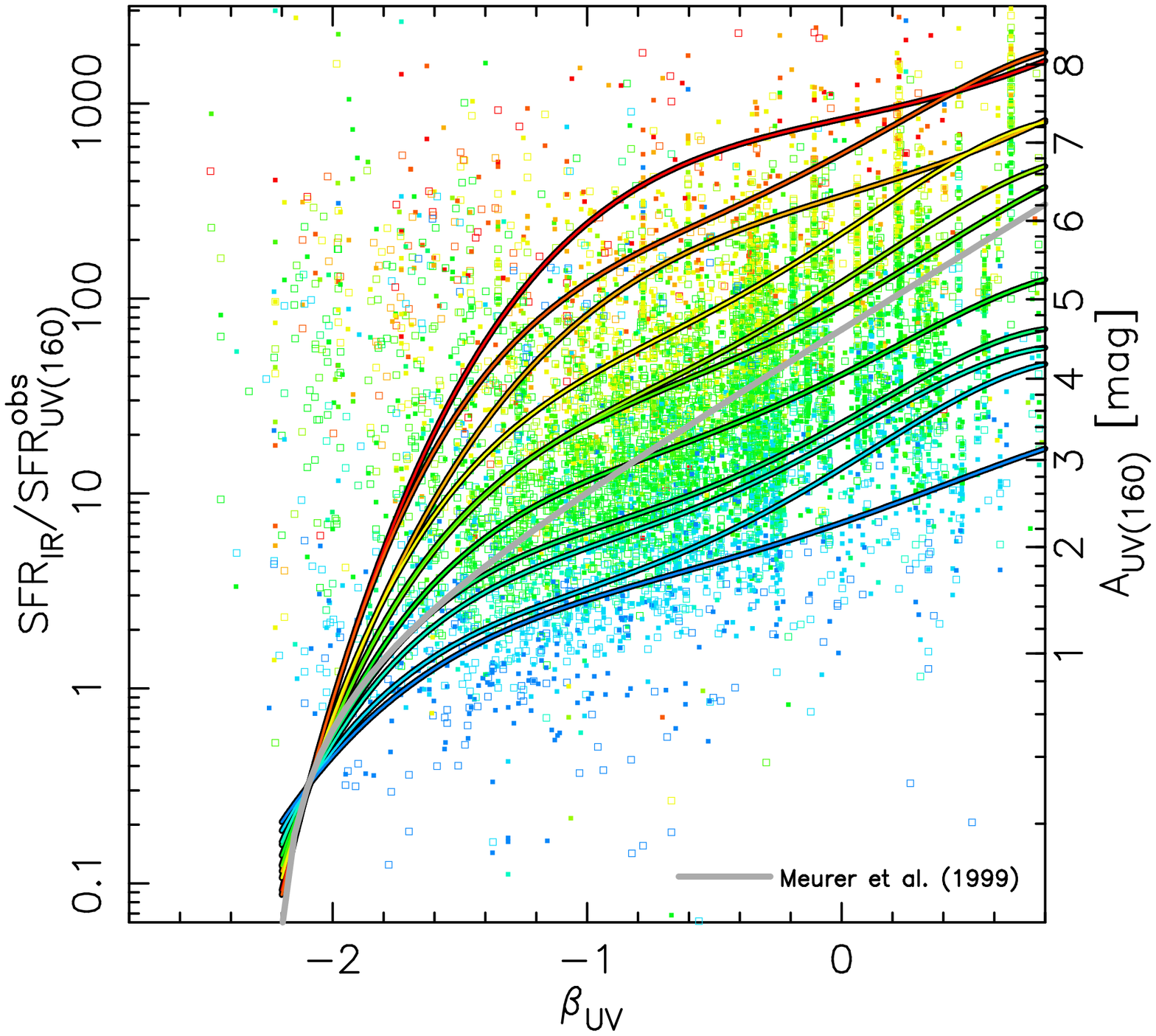}
\includegraphics[angle=-90,width=8.7cm]{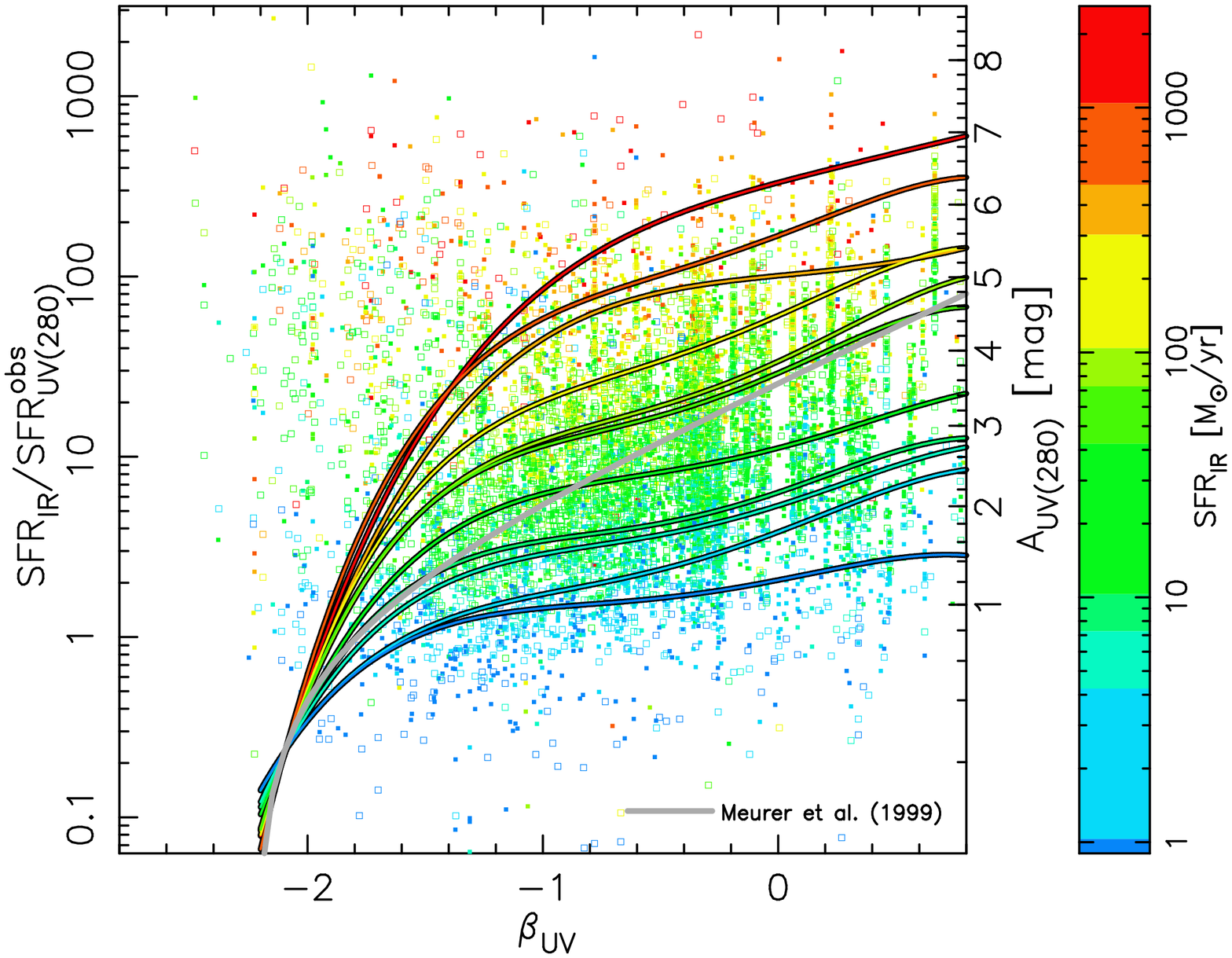}
\figcaption{\label{fig:irx_beta}IRX-$\beta_\mathrm{UV}$ plots for the
  galaxies detected by MIPS/PACS/SPIRE in the five CANDELS fields. On
  the left we show the results for UV-based SFR estimations based on
  the emission at 160~nm rest-frame (Equation D.4, \citealt{ken98}),
  on the right for 280~nm (Equation D.5, \citealt{bell03}). Galaxies
  with {\it Herschel} detections are plotted with filled symbols (and
  $\mathrm{SFR}_\mathrm{IR}=\mathrm{SFR}_\mathrm{IR}^\mathrm{fit}$,
  see text for details), galaxies with just a MIPS detection are
  plotted with open symbols (and
  $\mathrm{SFR}_\mathrm{IR}=\mathrm{SFR}_\mathrm{IR}^\mathrm{W11}$).
  Colors represent different bins of IR-based SFR [or L(8-1000)] as
  shown in the scale plotted on the right. The data for each bin has
  been fitted to a Chebyshev polynomial of order 5, shown with lines
  of the same color. The citet{meurer99} relationship is plotted in
  gray (using the \citealt{calzetti} attenuation law to transform from
  far-UV to near-UV $\mathrm{A_{UV}}$ values). The IRX scale (on the
  left of each panel) is transformed into an attenuation scale (on the
  right of each panel) using Equation~D3.}
\end{center}
\end{figure*}

\slugcomment{Please, plot this figure with the width of two columns}
\placefigure{fig:irx_beta}

Different IRX-$\beta_\mathrm{UV}$ relationships were built for
galaxies as a function of their IR-based SFRs. We divided the total
sample of IR-emitters in eleven bins (roughly equal in linear space,
except at the extremes, where the bins are larger to count with enough
galaxies) of L(8-1000): one bin for HyLIRGs
[$L(8-1000)>10^{13}$~L$_\sun$], three bins for ULIRGs
[$L(8-1000)=10^{12-13}$~L$_\sun$], three for LIRGs
[$L(8-1000)=10^{11-12}$~L$_\sun$], three for starbursts
[$L(8-1000)=10^{10-11}$~L$_\sun$], and one bin for the galaxies with
the lowest luminosities [$L(8-1000)<10^{10}$~L$_\sun$]. For each
subsample we built an IRX-$\beta_\mathrm{UV}$ relationship fitting the
data points to a Chebyshev polynomial of order 5 (given in
Table~\ref{table:IRX-beta}).  Figure~\ref{fig:irx_beta} shows the
results when considering UV-based SFRs using the 160 and 280~nm
estimators. It is readily clear that the galaxies with brighter IR
luminosities present higher IRX values for the same UV slope. We find
that the \citet{meurer99} relationship is a good approximation for
galaxies with
$\mathrm{SFR}_\mathrm{IR}\sim1-100$~$\mathrm{M_\odot\,yr^{-1}}$,
consistent with the SFRs of the local starbursts used in that work
\citep{heckman98}. Galaxies with smaller levels of star formation (in
our sample, this implies $z\lesssim1$, since the IR surveys are not
deep enough to detect such galaxies at high redshift) lie below the
Meurer et al.  relationship. The brightest LIRGs, ULIRGs, and HyLIRGs
present higher attenuations for the same UV slope value compared to
the Meurer et al. curve. Note that the differences in
IRX-$\beta_\mathrm{UV}$ curves imply different attenuation laws
(\citealt{wg00}; \citealt{cf00}).

\begin{deluxetable*}{cclcclcclcclcc}
\tabletypesize{\scriptsize}
\tablewidth{480pt}
\tablecaption{\label{sfr-mass-points}Main sequence data points and scatter}
\tablehead{\colhead{$\log(M)$} & \colhead{$\log(SFR)$} & & \colhead{$\log(M)$} & \colhead{$\log(SFR)$} & & \colhead{$\log(M)$} & \colhead{$\log(SFR)$} & & \colhead{$\log(M)$} & \colhead{$\log(SFR)$} & & \colhead{$\log(M)$} & \colhead{$\log(SFR)$}\\
\colhead{(1)} & \colhead{(2)} & & \colhead{(1)} & \colhead{(2)} & & \colhead{(1)} & \colhead{(2)} & & \colhead{(1)} & \colhead{(2)} & & \colhead{(1)} & \colhead{(2)}}
\startdata
\multicolumn{2}{c}{$0.0<z<0.5$} & & \multicolumn{2}{c}{$0.5<z<1.0$} & & \multicolumn{2}{c}{$1.0<z<1.4$} & & \multicolumn{2}{c}{$1.4<z<1.8$} & & \multicolumn{2}{c}{$1.8<z<2.2$}\\
 8.050 & $-$0.78$\pm$0.21 &  &  8.019 & $-$0.66$\pm$0.21 &  &  8.036 & $-$0.52$\pm$0.30 &  &  8.051 & $-$0.37$\pm$0.29 &  &  8.077 & $-$0.21$\pm$0.30  \\
 8.131 & $-$0.77$\pm$0.23 &  &  8.053 & $-$0.67$\pm$0.25 &  &  8.106 & $-$0.46$\pm$0.28 &  &  8.148 & $-$0.33$\pm$0.30 &  &  8.205 & $-$0.20$\pm$0.33  \\
 8.200 & $-$0.78$\pm$0.22 &  &  8.087 & $-$0.63$\pm$0.25 &  &  8.175 & $-$0.44$\pm$0.29 &  &  8.232 & $-$0.30$\pm$0.30 &  &  8.304 & $-$0.15$\pm$0.34  \\
 8.262 & $-$0.74$\pm$0.23 &  &  8.119 & $-$0.61$\pm$0.26 &  &  8.244 & $-$0.38$\pm$0.29 &  &  8.314 & $-$0.24$\pm$0.30 &  &  8.395 & $-$0.10$\pm$0.33  \\
 8.336 & $-$0.73$\pm$0.22 &  &  8.159 & $-$0.58$\pm$0.27 &  &  8.314 & $-$0.33$\pm$0.27 &  &  8.392 & $-$0.18$\pm$0.30 &  &  8.484 & $-$0.06$\pm$0.34  \\
 8.412 & $-$0.61$\pm$0.24 &  &  8.211 & $-$0.56$\pm$0.27 &  &  8.386 & $-$0.28$\pm$0.29 &  &  8.469 & $-$0.13$\pm$0.28 &  &  8.573 & $+$0.04$\pm$0.35  \\
 8.487 & $-$0.60$\pm$0.26 &  &  8.264 & $-$0.51$\pm$0.28 &  &  8.461 & $-$0.20$\pm$0.28 &  &  8.552 & $-$0.05$\pm$0.31 &  &  8.666 & $+$0.11$\pm$0.34  \\
 8.561 & $-$0.50$\pm$0.25 &  &  8.319 & $-$0.45$\pm$0.27 &  &  8.540 & $-$0.12$\pm$0.29 &  &  8.640 & $+$0.03$\pm$0.31 &  &  8.764 & $+$0.22$\pm$0.33  \\
 8.648 & $-$0.43$\pm$0.29 &  &  8.375 & $-$0.43$\pm$0.29 &  &  8.574 & $-$0.10$\pm$0.31 &  &  8.705 & $+$0.08$\pm$0.29 &  &  8.866 & $+$0.34$\pm$0.31  \\
 8.731 & $-$0.37$\pm$0.28 &  &  8.435 & $-$0.36$\pm$0.30 &  &  8.621 & $-$0.07$\pm$0.28 &  &  8.745 & $+$0.11$\pm$0.30 &  &  8.904 & $+$0.39$\pm$0.31  \\
 8.823 & $-$0.32$\pm$0.29 &  &  8.499 & $-$0.32$\pm$0.30 &  &  8.667 & $-$0.01$\pm$0.28 &  &  8.786 & $+$0.16$\pm$0.30 &  &  8.944 & $+$0.44$\pm$0.35  \\
 8.915 & $-$0.20$\pm$0.32 &  &  8.566 & $-$0.26$\pm$0.31 &  &  8.715 & $+$0.02$\pm$0.30 &  &  8.825 & $+$0.23$\pm$0.31 &  &  8.983 & $+$0.48$\pm$0.31  \\
 9.018 & $-$0.14$\pm$0.35 &  &  8.634 & $-$0.18$\pm$0.31 &  &  8.768 & $+$0.08$\pm$0.27 &  &  8.870 & $+$0.26$\pm$0.31 &  &  9.028 & $+$0.56$\pm$0.31  \\
 9.140 & $-$0.05$\pm$0.34 &  &  8.705 & $-$0.12$\pm$0.29 &  &  8.824 & $+$0.12$\pm$0.29 &  &  8.917 & $+$0.30$\pm$0.32 &  &  9.075 & $+$0.60$\pm$0.30  \\
 9.279 & $+$0.14$\pm$0.35 &  &  8.789 & $-$0.06$\pm$0.31 &  &  8.880 & $+$0.19$\pm$0.30 &  &  8.965 & $+$0.40$\pm$0.31 &  &  9.127 & $+$0.65$\pm$0.31  \\
 9.437 & $+$0.27$\pm$0.38 &  &  8.884 & $+$0.06$\pm$0.31 &  &  8.939 & $+$0.25$\pm$0.31 &  &  9.015 & $+$0.44$\pm$0.34 &  &  9.177 & $+$0.74$\pm$0.28  \\
 9.641 & $+$0.46$\pm$0.35 &  &  8.987 & $+$0.12$\pm$0.28 &  &  9.006 & $+$0.33$\pm$0.26 &  &  9.075 & $+$0.55$\pm$0.29 &  &  9.234 & $+$0.82$\pm$0.28  \\
 9.893 & $+$0.54$\pm$0.37 &  &  9.100 & $+$0.24$\pm$0.30 &  &  9.080 & $+$0.42$\pm$0.29 &  &  9.140 & $+$0.62$\pm$0.31 &  &  9.294 & $+$0.92$\pm$0.27  \\
10.228 & $+$0.70$\pm$0.48 &  &  9.231 & $+$0.33$\pm$0.28 &  &  9.154 & $+$0.48$\pm$0.29 &  &  9.211 & $+$0.71$\pm$0.30 &  &  9.358 & $+$0.98$\pm$0.24  \\
10.754 & $+$0.77$\pm$0.70 &  &  9.379 & $+$0.45$\pm$0.30 &  &  9.236 & $+$0.57$\pm$0.30 &  &  9.287 & $+$0.82$\pm$0.28 &  &  9.430 & $+$1.03$\pm$0.22  \\
       &                  &  &  9.557 & $+$0.60$\pm$0.32 &  &  9.329 & $+$0.70$\pm$0.25 &  &  9.374 & $+$0.93$\pm$0.27 &  &  9.516 & $+$1.12$\pm$0.21  \\
       &                  &  &  9.781 & $+$0.80$\pm$0.35 &  &  9.429 & $+$0.76$\pm$0.26 &  &  9.468 & $+$1.03$\pm$0.24 &  &  9.614 & $+$1.17$\pm$0.19  \\
       &                  &  & 10.083 & $+$1.04$\pm$0.45 &  &  9.542 & $+$0.85$\pm$0.28 &  &  9.581 & $+$1.11$\pm$0.23 &  &  9.723 & $+$1.25$\pm$0.24  \\
       &                  &  & 10.584 & $+$1.21$\pm$0.55 &  &  9.663 & $+$0.95$\pm$0.25 &  &  9.716 & $+$1.18$\pm$0.23 &  &  9.852 & $+$1.31$\pm$0.28  \\
       &                  &  & 11.397 & $+$1.54$\pm$0.28 &  &  9.825 & $+$1.06$\pm$0.28 &  &  9.876 & $+$1.30$\pm$0.30 &  & 10.006 & $+$1.44$\pm$0.37  \\
       &                  &  &        &                  &  & 10.029 & $+$1.18$\pm$0.37 &  & 10.072 & $+$1.42$\pm$0.43 &  & 10.215 & $+$1.67$\pm$0.47  \\
       &                  &  &        &                  &  & 10.296 & $+$1.36$\pm$0.54 &  & 10.341 & $+$1.71$\pm$0.52 &  & 10.476 & $+$1.85$\pm$0.54  \\
       &                  &  &        &                  &  & 10.680 & $+$1.51$\pm$0.56 &  & 10.776 & $+$1.80$\pm$0.62 &  & 10.855 & $+$2.07$\pm$0.55  \\
       &                  &  &        &                  &  & 10.997 & $+$1.70$\pm$0.60 &  & 11.400 & $+$2.17$\pm$0.32 &  & 11.581 & $+$2.76$\pm$0.66  \\
       &                  &  &        &                  &  &        &                  &  &        &                  &  &        &                   \\
\multicolumn{2}{c}{$2.2<z<2.6$} & & \multicolumn{2}{c}{$2.6<z<3.0$} & & \multicolumn{2}{c}{$3.0<z<5.0$} & & \multicolumn{2}{c}{$5.0<z<8.0$} & & \multicolumn{2}{c}{}\\
 8.115 & $-$0.05$\pm$0.31 &  &  8.148 & $+$0.16$\pm$0.30 &  &  8.246 & $+$0.36$\pm$0.33 &  &  8.308 & $+$0.92$\pm$0.43 &  &        &                   \\
 8.278 & $+$0.00$\pm$0.34 &  &  8.348 & $+$0.23$\pm$0.32 &  &  8.512 & $+$0.46$\pm$0.34 &  &  8.630 & $+$0.93$\pm$0.34 &  &        &                   \\
 8.402 & $+$0.05$\pm$0.33 &  &  8.486 & $+$0.28$\pm$0.34 &  &  8.684 & $+$0.54$\pm$0.35 &  &  8.830 & $+$1.10$\pm$0.36 &  &        &                   \\
 8.503 & $+$0.10$\pm$0.33 &  &  8.597 & $+$0.33$\pm$0.37 &  &  8.831 & $+$0.60$\pm$0.36 &  &  8.985 & $+$1.13$\pm$0.33 &  &        &                   \\
 8.601 & $+$0.17$\pm$0.37 &  &  8.710 & $+$0.42$\pm$0.37 &  &  8.961 & $+$0.69$\pm$0.37 &  &  9.124 & $+$1.23$\pm$0.29 &  &        &                   \\
 8.699 & $+$0.23$\pm$0.36 &  &  8.821 & $+$0.52$\pm$0.35 &  &  9.087 & $+$0.83$\pm$0.36 &  &  9.274 & $+$1.27$\pm$0.34 &  &        &                   \\
 8.799 & $+$0.34$\pm$0.35 &  &  8.934 & $+$0.64$\pm$0.37 &  &  9.213 & $+$0.94$\pm$0.36 &  &  9.439 & $+$1.34$\pm$0.32 &  &        &                   \\
 8.899 & $+$0.44$\pm$0.36 &  &  9.052 & $+$0.78$\pm$0.34 &  &  9.351 & $+$1.09$\pm$0.34 &  &  9.598 & $+$1.47$\pm$0.30 &  &        &                   \\
 9.016 & $+$0.61$\pm$0.41 &  &  9.177 & $+$0.94$\pm$0.33 &  &  9.516 & $+$1.24$\pm$0.34 &  &  9.790 & $+$1.62$\pm$0.34 &  &        &                   \\
 9.050 & $+$0.62$\pm$0.33 &  &  9.223 & $+$0.94$\pm$0.33 &  &  9.715 & $+$1.43$\pm$0.34 &  & 10.115 & $+$1.90$\pm$0.33 &  &        &                   \\
 9.085 & $+$0.71$\pm$0.33 &  &  9.252 & $+$1.02$\pm$0.35 &  &  9.738 & $+$1.48$\pm$0.28 &  & 10.270 & $+$2.01$\pm$0.44 &  &        &                   \\
 9.122 & $+$0.69$\pm$0.35 &  &  9.283 & $+$1.04$\pm$0.29 &  &  9.761 & $+$1.51$\pm$0.24 &  & 10.570 & $+$2.20$\pm$0.39 &  &        &                   \\
 9.164 & $+$0.79$\pm$0.31 &  &  9.315 & $+$1.08$\pm$0.31 &  &  9.784 & $+$1.55$\pm$0.30 &  & 11.213 & $+$2.59$\pm$0.47 &  &        &                   \\
 9.209 & $+$0.86$\pm$0.34 &  &  9.351 & $+$1.19$\pm$0.28 &  &  9.808 & $+$1.59$\pm$0.24 &  &        &                  &  &        &                   \\
 9.256 & $+$0.91$\pm$0.29 &  &  9.389 & $+$1.21$\pm$0.29 &  &  9.837 & $+$1.59$\pm$0.31 &  &        &                  &  &        &                   \\
 9.303 & $+$0.98$\pm$0.25 &  &  9.427 & $+$1.20$\pm$0.28 &  &  9.864 & $+$1.63$\pm$0.33 &  &        &                  &  &        &                   \\
 9.356 & $+$1.03$\pm$0.33 &  &  9.471 & $+$1.29$\pm$0.26 &  &  9.894 & $+$1.62$\pm$0.32 &  &        &                  &  &        &                   \\
 9.411 & $+$1.09$\pm$0.30 &  &  9.520 & $+$1.31$\pm$0.30 &  &  9.928 & $+$1.65$\pm$0.29 &  &        &                  &  &        &                   \\
 9.470 & $+$1.15$\pm$0.26 &  &  9.562 & $+$1.36$\pm$0.26 &  &  9.966 & $+$1.67$\pm$0.35 &  &        &                  &  &        &                   \\
 9.529 & $+$1.21$\pm$0.28 &  &  9.613 & $+$1.37$\pm$0.32 &  & 10.007 & $+$1.73$\pm$0.26 &  &        &                  &  &        &                   \\
 9.609 & $+$1.29$\pm$0.21 &  &  9.674 & $+$1.46$\pm$0.27 &  & 10.052 & $+$1.82$\pm$0.37 &  &        &                  &  &        &                   \\
 9.693 & $+$1.36$\pm$0.25 &  &  9.732 & $+$1.53$\pm$0.21 &  & 10.099 & $+$1.83$\pm$0.32 &  &        &                  &  &        &                   \\
 9.783 & $+$1.44$\pm$0.23 &  &  9.799 & $+$1.58$\pm$0.29 &  & 10.157 & $+$1.90$\pm$0.31 &  &        &                  &  &        &                   \\
 9.890 & $+$1.48$\pm$0.20 &  &  9.875 & $+$1.61$\pm$0.29 &  & 10.219 & $+$1.96$\pm$0.32 &  &        &                  &  &        &                   \\
10.022 & $+$1.57$\pm$0.29 &  &  9.969 & $+$1.69$\pm$0.23 &  & 10.289 & $+$1.99$\pm$0.38 &  &        &                  &  &        &                   \\
10.180 & $+$1.67$\pm$0.37 &  & 10.080 & $+$1.77$\pm$0.25 &  & 10.374 & $+$2.03$\pm$0.34 &  &        &                  &  &        &                   \\
10.390 & $+$1.77$\pm$0.50 &  & 10.204 & $+$1.86$\pm$0.39 &  & 10.490 & $+$2.11$\pm$0.40 &  &        &                  &  &        &                   \\
10.773 & $+$2.14$\pm$0.49 &  & 10.384 & $+$2.00$\pm$0.43 &  & 10.675 & $+$2.21$\pm$0.49 &  &        &                  &  &        &                   \\
11.301 & $+$2.56$\pm$0.36 &  & 10.742 & $+$2.24$\pm$0.53 &  & 11.077 & $+$2.64$\pm$0.49 &  &        &                  &  &        &                   \\
       &                  &  & 11.142 & $+$2.48$\pm$0.44 &  & 11.809 & $+$2.94$\pm$0.25 &  &        &                  &  &        &                   \\
\enddata                                      
\tablecomments{(1) Stellar mass in units of $M_\sun$.  (2) Median SFR
  and rms (in $M_\sun\,yr^{-1}$) defining the main sequence.}
\end{deluxetable*}

\begin{figure*}
\begin{center}
\includegraphics[angle=-90,width=18cm]{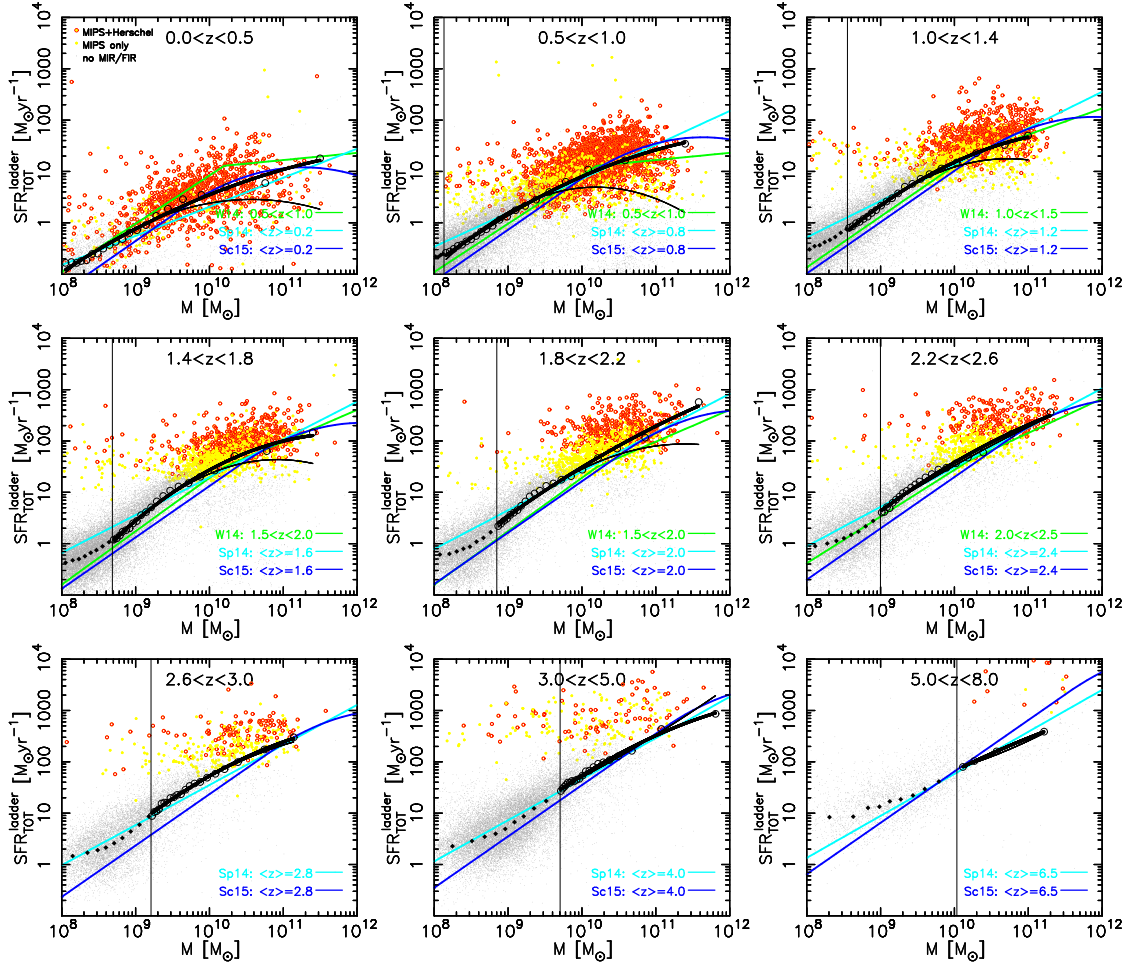}
\figcaption{\label{fig:sfr_mass}Total SFRs vs. stellar masses for all
  cataloged galaxies in the five CANDELS fields, divided in 9 redshift
  bins. The SFRs have been estimated with the ladder approach
  described in this Appendix, and the three layers are marked with
  different colors: MIPS/Herschel sources in red, MIPS-only sources in
  yellow, and UV-only sources in gray. Stellar masses are taken from
  \citet{santini15} for CANDELS fields except GOODS-N, whose masses
  are presented in this paper. For each bin, we plot the main sequence
  curves according to \citet[][green lines]{whitaker14},
  \citet[][cyan]{speagle14} and \citealt[][blue]{schreiber15} for the
  redshifts or Universe ages shown in the legend of each panel. We
  also depict the trends of SFR as a function of mass based on the
  calculations presented in this paper for the sample of star-forming
  galaxies built after removing quiescent systems with $UVJ$
  diagrams. Points plotted in black depict running medians for the
  SFGs: open circles correspond to galaxies with stellar masses larger
  than the stellar mass limits given in \citet[][vertical gray
    lines]{grazian15}, diamonds to galaxies below those limits. The
  former data points are fitted with Chebyshev polynomials (thick
  black lines). The relationship between SFR and stellar mass for all
  galaxies (SFGs jointly with quiescent galaxies) is also shown with
  thin black lines.}
\end{center}
\end{figure*}

Using the results presented in Figures~\ref{fig:sfr_redshift} and
\ref{fig:irx_beta}, we can now estimate attenuation-corrected UV-based
SFRs for sources with no IR detection. For a given galaxy, we have its
redshift and a $\beta_\mathrm{UV}$ estimation obtained by either
fitting directly the UV flux points at wavelengths between 120 and
260~nm, or by measuring slopes in the models used to estimate stellar
masses (see Section~4). We note that both types of estimations provide
very similar results, with an average systematic offset
$\Delta\beta<0.01$. With the redshift we can obtain from
Figure~\ref{fig:sfr_redshift} an upper limit for the SFR based on the
detection limit of the MIPS/{\it Herschel} surveys in the CANDELS
fields.  The upper-limit SFR is used to select an
IRX-$\beta_\mathrm{UV}$ relationship from Figure~\ref{fig:irx_beta},
which allows to translate from $\beta_\mathrm{UV}$ to attenuation in
the UV (in one of the two considered wavelengths).  With this
attenuation we correct the observed
$\mathrm{SFR}_\mathrm{UV-\lambda}^\mathrm{obs}$ to get the
dust-corrected $\mathrm{SFR}_\mathrm{UV}^\mathrm{corr}$, which is
equal to $\mathrm{SFR}_\mathrm{TOT}$. The obtained value of
$\mathrm{SFR}_\mathrm{UV}^\mathrm{corr}$ might be higher than the
detection limit of the IR surveys. In this case, we recalculate the
attenuation using a lower IRX-$\beta_\mathrm{UV}$ relationship until
the estimation is consistent with the non-detection at mid- and far-IR
wavelengths. Two further refinements are included in our method.
First, for each galaxy we calculate a UV-based dust-corrected SFR
using both the 160 and the 280~nm estimator, and we average them to
provide a final value. Typically, the two estimations are consistent
within less than 0.05~dex. Second, we iteratively calculate a main
sequence for different redshift bins (see Figure~\ref{fig:sfr_mass})
and avoid outliers at the bright end that deviate more than $5\sigma$.


Summarizing, we have developed a method to calculate SFRs in a
galaxy-by-galaxy basis following a ladder approach. For the most
extreme galaxies at the highest part of the ladder, which are detected
by both MIPS and {\it Herschel}, total SFRs are calculated with dust
emission models fitting the IR data points and adding the unobscured
star formation. For less extreme cases in the middle of the ladder, we
have measured the obscured star formation directly from MIPS data,
which are highly sensitive to the amount of dust. For sources at the
bottom of the ladder which have not been detected by the mid- and
far-IR surveys, we calculate SFRs taking into account differences in
the attenuation law, linked to different IR-to-UV ratios.

Using this methodology we obtained a total SFR,
$\mathrm{SFR}_\mathrm{TOT}$, for each galaxy.
Figure~\ref{fig:sfr_mass} shows a SFR vs stellar mass plot for the
more than 186,000 galaxies in the CANDELS catalogs, divided in nine
redshift bins and taking into account star formation activity as
inferred from $UVJ$ diagrams. A running median has been run through
the data (with a minimum of 200 points) to obtain the position of the
main sequence (relation between SFR and mass for those galaxies
identified as star-forming based on the $UVJ$ diagram, i.e., after
excluding quiescent galaxies). Medians and rms values for these main
sequences built with the entire CANDELS dataset are given in
Table~\ref{sfr-mass-points}.  Several interesting results can be
extracted from this plot. First, our methodology produces a main
sequence of galaxies where the non-IR detections nicely join the trend
followed by the IR emitters.  Second, the main sequences inferred from
our results are, overall, consistent with the results in the
literature (in particular, we compare with
\citealt{whitaker14},\citealt{speagle14}, and \citealt{schreiber15}).
However, we note that, if we consider the SFR vs mass relation for all
galaxies, including quiescent systems, the SFR-M trend becomes flat at
higher masses, and even reverses the sign of the slope at low
redshifts ($z<1$), where the red sequence is populated by large
numbers of dead galaxies.  Table~\ref{table:sfr-mass} gives the
Chebyshev coefficients of our fits to the main sequence (just
considering SFGs above the mass completeness limit) and general SFR-M
(including all galaxies) relationships shown in
Figure~\ref{fig:sfr_mass}.

\slugcomment{Please, plot this figure with the width of two columns}
\placefigure{fig:sfr_mass}

\placetable{sfr-mass}
\begin{deluxetable*}{lccccc}
\tabletypesize{\scriptsize}
\hspace{-5cm}
\tablecaption{\label{table:sfr-mass}SFR vs. stellar mass and main sequence fits to Chebyshev polynomials}
\tablehead{\colhead{Redshift interval} & \colhead{$M_\mathrm{min}$} & \colhead{Sample} & \colhead{$T_0$} & \colhead{$T_1$} & \colhead{$T_2$} \\
\colhead{(1)} & \colhead{(2)} &  \colhead{(3)} & \colhead{(4)} & \colhead{(4)} & \colhead{(4)}}
\startdata
\hline
$0.0<z<0.5$ &  6.9  & SFG & $-1.39684\times10^{+01}$ &  $+2.32388$ &  $-8.72415\times10^{-02}$ \\
            &       & ALL & $-2.45250\times10^{+01}$ &  $+4.72555$ &  $-2.23490\times10^{-01}$ \\
$0.5<z<1.0$ &  8.1  & SFG & $-1.64662\times10^{+01}$ &  $+2.84365$ &  $-1.10703\times10^{-01}$ \\
            &       & ALL & $-3.47832\times10^{+01}$ &  $+6.96338$ &  $-3.41690\times10^{-01}$ \\
$1.0<z<1.4$ &  8.6  & SFG & $-2.30977\times10^{+01}$ &  $+4.18836$ &  $-1.76068\times10^{-01}$ \\
            &       & ALL & $-3.36487\times10^{+01}$ &  $+6.50548$ &  $-3.03193\times10^{-01}$ \\
$1.4<z<1.8$ &  8.7  & SFG & $-2.75708\times10^{+01}$ &  $+5.02435$ &  $-2.12376\times10^{-01}$ \\
            &       & ALL & $-3.75337\times10^{+01}$ &  $+7.17562$ &  $-3.28619\times10^{-01}$ \\
$1.8<z<2.2$ &  8.9  & SFG & $-1.48351\times10^{+01}$ &  $+2.39121$ &  $-7.58761\times10^{-02}$ \\
            &       & ALL & $-3.08932\times10^{+01}$ &  $+5.76395$ &  $-2.52960\times10^{-01}$ \\
$2.2<z<2.6$ &  9.0  & SFG & $-1.75258\times10^{+01}$ &  $+2.97979$ &  $-1.06905\times10^{-01}$ \\
            &       & ALL & $-1.23610\times10^{+01}$ &  $+1.96159$ &  $-5.73675\times10^{-02}$ \\
$2.6<z<3.0$ &  9.2  & SFG & $-2.11659\times10^{+01}$ &  $+3.76568$ &  $-1.47664\times10^{-01}$ \\
            &       & ALL & $-2.30500\times10^{+01}$ &  $+4.16377$ &  $-1.68796\times10^{-01}$ \\
$3.0<z<5.0$ &  9.7  & SFG & $-2.01471\times10^{-01}$ &  $+3.47129$ &  $-1.28317\times10^{-01}$ \\
            &       & ALL & $+4.48905\times10^{-01}$ &  $-5.13682\times10^{-01}$ &  $+6.38656\times10^{-02}$ \\
$5.0<z<8.0$ & 10.0  & SFG & $-1.16582\times10^{+01}$ &  $+1.98906$ &  $-6.40982\times10^{-02}$ \\
            &       & ALL & $+1.05884\times10^{+01}$ &  $-2.16663$ &  $+1.29366\times10^{-01}$ \\
\enddata
\tablecomments{(1) Redshift interval (same as in Figure~\ref{fig:sfr_mass}). (2) Stellar mass completeness level of the CANDELS catalogs from \citet{grazian15}. (3) Sample of galaxies: star-forming galaxies (SFGs) identified with $UVJ$ diagrams, and entire sample (ALL) adding quiescent galaxies to the previous sample. (4) Chebyshev polynomial coefficients (order 3).}
\end{deluxetable*}

\subsection{Comparison to other IR-based SFR catalogs}
\label{ap:IRcomparison}

In order to further verify the quality the far-IR photometric catalogs
and the SFRs based on them we compare our values to those from two
independent works using similar datasets. These are the MIPS24
photometric and SFR catalog of the 3D-HST survey, presented in
\citet{3dhstgrism} and \citet{whitaker14}, and the {\it
  Spitzer}+Herschel+VLA photometric and SFR catalog of GOODS-N
galaxies, presented in \citet[][hereafter L18]{liu18}.

The 3D-HST catalog is similar to ours in the sense that it provides
far IR photometry to WFC3-selected sources in the 5 CANDELS
fields. Therefore, it is straightforward to identify common sources by
crossmatching celestial coordinates within a 0\farcs3 search radius,
as we have done in \S~\ref{ss:othercats}. The MIPS 24~$\mu$m
photometry in the 3D-HST and CANDELS catalogs is computed following
slightly different procedures. The CANDELS photometry, as described in
appendix~\ref{ap:IRphotometry}, is based on direct detections of
point-like sources whose fluxes are measured using circular apertures
and applying aperture corrections. Then, each MIPS source is
associated to its most likely F160W counterpart using brightness and
proximity criteria. The 3D-HST catalog measures the MIPS 24~$\mu$m
fluxes following the same method as with the all other optical-to-NIR
bands. I.e., they used the convolution software \texttt{MOPHONGO}
(\citealt{labbe13}) on the WFC3 high resolution images to model and
subtract the contributions from neighboring blended sources in the
lower resolution MIPS24 image. Then, they use circular aperture
photometry on the ``clean'' cutout of each source to measure the flux,
and they apply an aperture correction of 20\% to determine the total
magnitude.

In addition to the MIPS24 fluxes, the 3D-HST catalog provides IR and
total SFRs computed following a similar methodology to the one
described in appendix~\ref{ap:sfr}. I.e., they use the MIPS 24~$\mu$m
fluxes and the redshifts of the galaxies to estimate bolometric IR
luminosities using the IR emission templates of
\citealt{wuyts11a}. These luminosities are transformed to SFRs using
the \citet{ken98} calibration, assuming a \citet{chabrier}
IMF. Lastly, they estimate {\it total} SFRs by adding the IR-SFR to a
UV-SFR estimated from the observed UV luminosity computed from the
fitting of the optical-to-NIR SED of each galaxy.

\begin{figure*}[t]
\centering
\includegraphics[scale=0.46, angle=0]{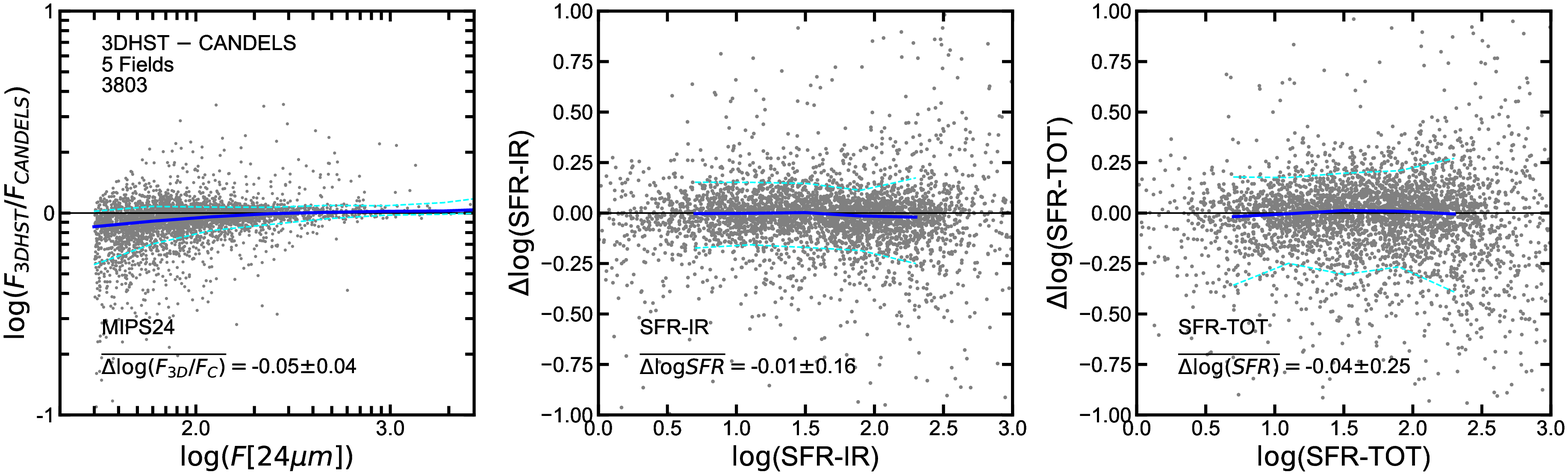}
\caption[]{\label{fig:IR_compare_W14} Comparison of MIPS 24~$\mu$m
  fluxes and SFRs between the CANDELS (this work) and 3D-HST catalogs
  \citep{3dhstgrism} in the 5 fields of the CANDELS survey. From
  left-to-right, comparison of the MIPS24 fluxes, IR-based SFRs and
  total (UV+IR) SFRs. The y-axis shows differences in 3D-HST minus
  CANDELS values. The comparison is restricted to galaxies detected
  above SNR$=$5 in MIPS24 in both catalogs. The blue lines show the
  running median and $\pm$1$\sigma$ scatter of the distribution whose
  average values are indicated in the bottom-left corner. The values
  in each panel have been corrected by a constant offset derived from
  the average running median. Overall, we find a good agreement in the
  fluxes and SFRs between the two catalogs. The CANDELS MIPS24 fluxes
  are only marginally brighter than the 3D-HST by $0.05$~dex. Such
  difference has little impact on the two SFRs which exhibit offsets
  of 0.01 and 0.04~dex, respectively.}
\end{figure*}

Figure~\ref{fig:IR_compare_W14} shows the comparison of the MIPS
24~$\mu$m fluxes and SFRs (IR and total) between the CANDELS and
3D-HST catalogs. The blue and cyan lines indicate the running median
and 1$\sigma$ percentiles. The values in each panel have been {\it
  normalized} by applying a constant offset (indicated in the
bottom-left corner) determined from the average value of the running
median. Overall, the comparison suggests that the values in the
CANDELS and 3D-HST catalogs are quite consistent. The CANDELS MIPS24
fluxes are slightly brighter than the 3D-HST by 0.05~dex. This small
difference is the same in the 5 cosmological fields, which suggest
that the origin could be a small systematic difference in the
photometric zero-points or perhaps in the value of the aperture
correction. Interestingly, despite the small offset in the MIPS24
photometry, the IR-based SFRs are in excellent agreement with a median
difference of just $\Delta$SFR$=-0.01$~dex. The $\sim$0.12~dex scatter
in the comparison of IR-SFRs is also quite small compared for example
to the 0.3~dex scatter in the comparison of stellar masses shown in
\S~\ref{ss:colors_masses}. Such small scatter is probably caused by
small differences in the MIPS flux, redshift or the dust emission
templates. Likewise, the comparison of total (UV+IR) SFRs shows an
excellent agreement with an average offset of $\Delta$SFR$=-0.03$~dex
and a slightly larger scatter of 0.21~dex. The increased scatter
relative to the IR-SFRs comparison is likely caused by additional
random differences in the UV-SFRs which are based on the different
optical-to-NIR SED fitting of each work.

The L18 catalog aims to provide self-consistent far-IR to sub-mm
photometry for galaxies in the GOODS-N region. As such, it is
primarily selected in the far-IR, with the majority of the sources
detected in a merged MIPS 24 + VLA 20cm catalog. Most of these sources
have NIR counterparts, primarily in IRAC and K-band, which are used to
crossmatch to other optical/NIR catalogs, such as the 3D-HST, for the
purpose of obtaining optical-to-NIR photometry and photometric
redshifts for each far-IR source. To first order, the L18 IR
photometric measurements are similar to ours. They are both based on
PSF fitting point-like sources using positional priors from other
bands with higher spatial resolution. As
appendix~\ref{ap:IRphotometry}, these priors help reduce the confusion
effects due to crowding. The primary difference between the L18 and
CANDELS methods is the choice of the detection priors.  For any given
band, we chose priors from the previous, shorter wavelength band with
better resolution (e.g., MIPS24 priors for PACS100), while L18 use the
full SEDs at shorter wavelengths to predict the fluxes of all the
possible counterparts and, based on those, choose only the priors that
are likely to be detected.

\begin{figure*}[t]
\centering
\includegraphics[scale=0.55, angle=0]{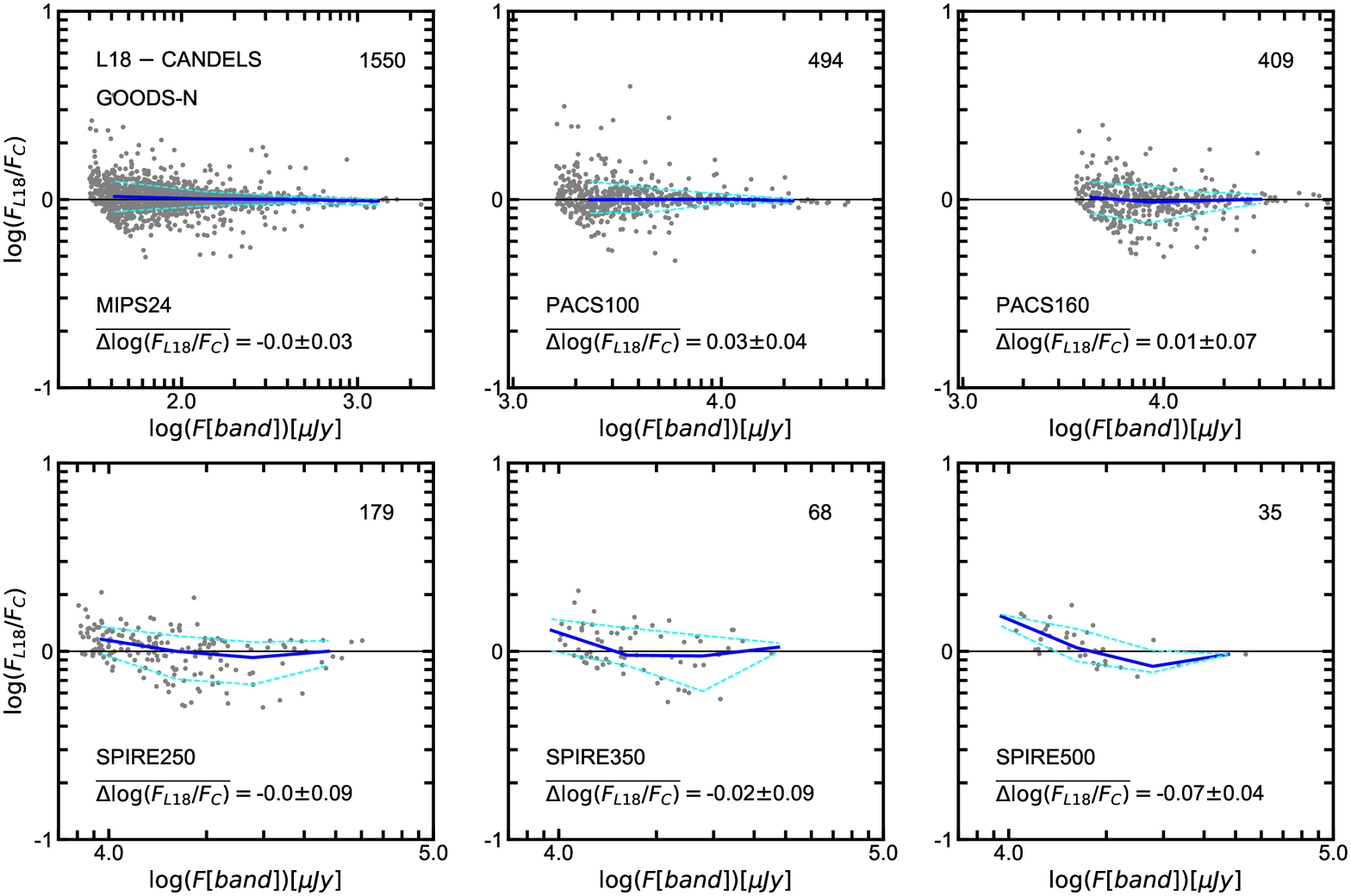}
\caption[]{\label{fig:IR_compare_L18_flux} Comparison of mid-to-far IR
  fluxes between the CANDELS (this work) and the \citet{liu18}
  catalogs in the GOODS-N field. From left-to-right and top-to-bottom,
  the panels show the comparison of the MIPS24, PACS100, PACS160,
  SPIRE250, SPIRE350 and SPIRE500 fluxes. The y-axis shows differences
  in L18 minus CANDELS values. The comparison is restricted to
  galaxies detected above SNR$=$5 in each band in both catalogs (see
  Table~\ref{table:photometry_ir}). The blue lines show the running
  median and $\pm$1$\sigma$ scatter of the distribution whose average
  values are indicated in the bottom-left corner. The values in each
  panel have been corrected by a constant offset derived from the
  average running median. The declining sensitivity and spatial
  resolution with increasing wavelength of the bands leads to fewer
  detections, larger scatter and some outliers likely caused by
  differences in the deblending of particularly crowded
  sources. Nonetheless, we find an overall good agreement in all bands
  with systematic offsets smaller than $\sim0.1$~dex and a scatter
  that is roughly consistent with photometric uncertainties.}
\end{figure*}

This ``informed'' choice of priors can ease the confusion around
particularly crowded sources or can help find sources with peculiar
SEDs (e.g., those detected in SPIRE but not in PACS).  Nonetheless, as
discussed in appendix~\ref{ap:IRphotometry}, the multiplicity in our
sequential crossmatching method is very small (1:1 for MIPS and PACS
and 3 or 2:1 for SPIRE) compared to that of a direct match from F160W
or IRAC to SPIRE, where the multiplicity can be as high as 20 or
30. Thus, we expect that the bulk of the detections and IR fluxes in
both catalogs to be in good agreement. The L18 catalog contains
$\sim$3300 sources over a sligthly larger area of GOODS-N than that
covered in the CANDELS survey. We crossmatch both catalogs within a
0\farcs5 radius and we sucessfully identify all the L18 sources that
already had 3D-HST priors.

Figure~\ref{fig:IR_compare_L18_flux} shows the photometric comparison
between the six mid-to-far IR bands in common between the two
catalogs. Each panel shows only galaxies detected at SNR$>5$ in both
catalogs (see Table~\ref{table:photometry_ir}). As in
Figure~\ref{fig:IR_compare_W14}, the values have been corrected by a
constant average offset determined from the running median (blue
line). Overall, the photometry in the two catalogs exhibits a good
agreement with only small systematic offsets ($\lesssim$0.03~dex in
all bands but SPIRE500) and a typical scatter roughly consistent with
the expected photometric uncertainties. Owing to its higher
sensitivity and spatial resolution, the comparisons in the MIPS and
PACS bands span a broader dynamical range ($\sim$1-1.5~dex) and
include more galaxies, which leads a more homogeneous and smoother
distribution.  In SPIRE, however, both the dynamical range and the
number of detections are smaller. Nonetheless, the bulk of the
galaxies exhibit a good agreement with a scatter smaller than 0.1~dex.

\begin{figure*}[t]
\centering
\includegraphics[scale=0.43, angle=0]{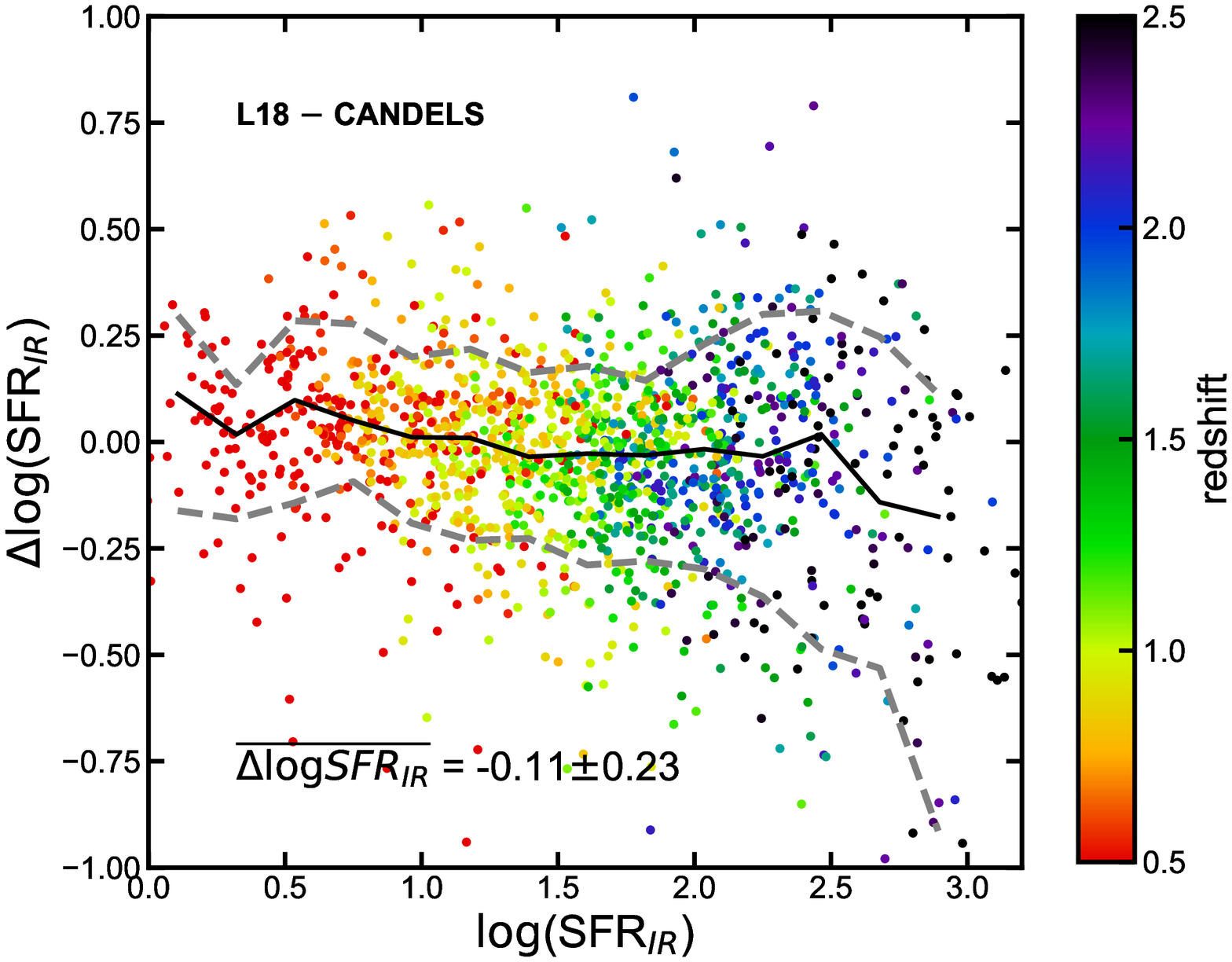}
\hspace*{0.50cm}
\includegraphics[scale=0.43, angle=0]{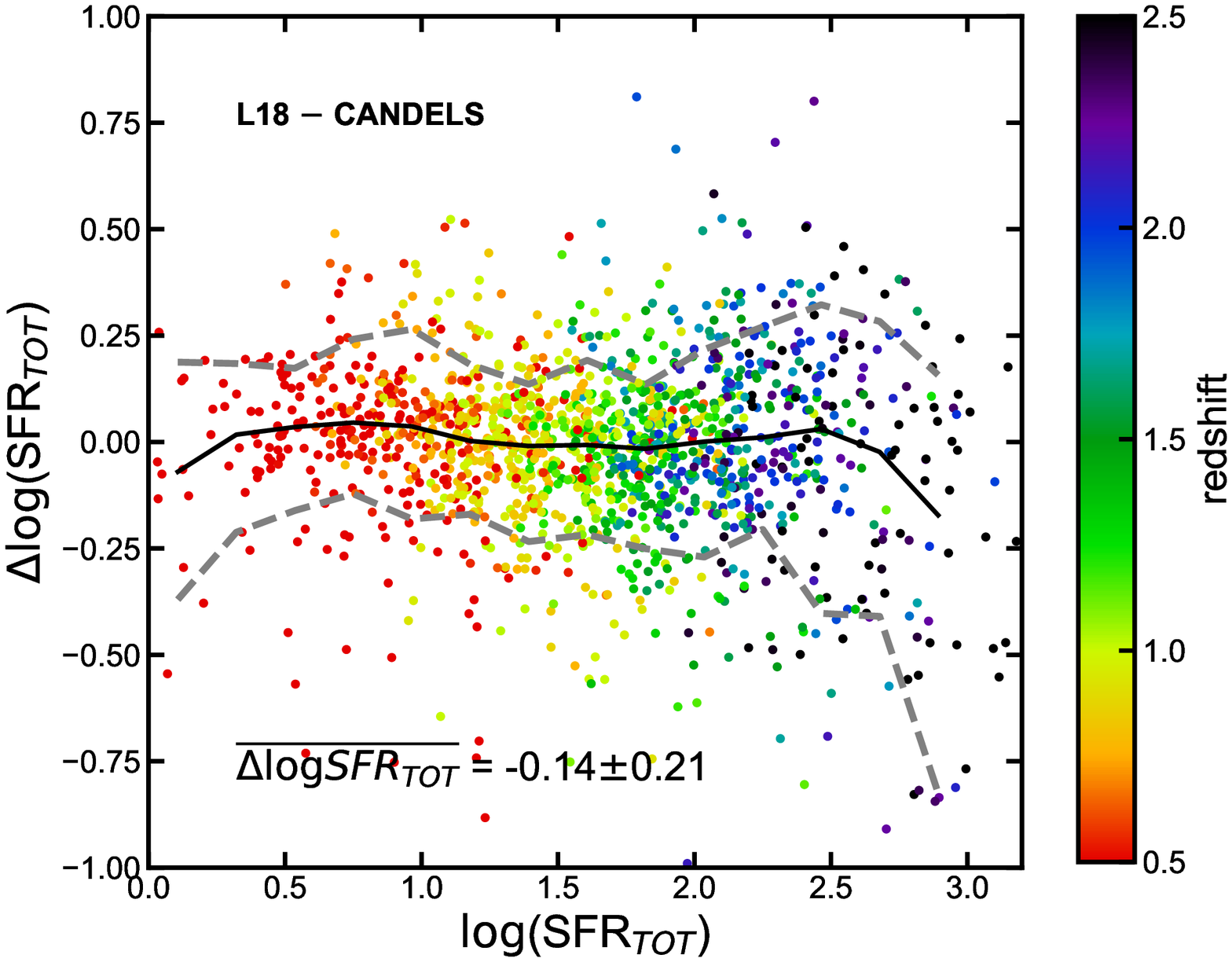}
\caption[]{\label{fig:IR_compare_L18_SFR} Comparison of the IR-based
  SFRs (left) and total UV+IR SFRs (right) in the CANDELS (this work)
  and the \citet{liu18} catalogs in the GOODS-N field, color coded by
  redshift. The y-axis shows differences in L18 minus CANDELS
  values. The comparison is restricted to galaxies detected above
  SNR$=$5 in MIPS24 (the deepest band) in both catalogs. The black and
  grey lines show the running median and $\pm$1$\sigma$ scatter of the
  distribution whose average values are indicated in the bottom-left
  corner. The values in each panel have been corrected by a constant
  offset derived from the average running median. The CANDELS IR- and
  total- SFRs are only marginally larger than those in L18 by
  approximately constant values of 0.1 and 0.14~dex. There is no
  evidence for significant trends with SFR or redshift which appear
  somtimes due to strong features in the dust emission templated used
  in the IR-SED modeling. The scatter in the comparisons,
  $\lesssim0.25$~dex, is consistent or even better than the typical
  0.3~dex. Thus, we conclude that, overall, there is a good agreement
  betwee the CANDELS and L18 SFRs.}
\end{figure*}

In addition to the far-IR photometry, we also compare the IR-based
SFRs derived from the fitting of the full IR SEDs.  This comparison
depends on additional factors such as the number photometric fluxes in
the IR SED, or the set of dust emission templates used for the
fitting. The IR SEDs in L18 include a few more bands in the sub-mm and
radio. However, those detections are restricted to only a few IR
bright objects. Thus, the bulk of the sample has a similar SEDs in
both catalogs. As for the fitting templates, L18 uses the
\citet{magdis12} library plus 2 additional AGN models, while we use a
combination of 3 libraries (see appendix~\ref{ap:sfr}). Despite these
differences, Figure~\ref{fig:IR_compare_L18_SFR} shows an excellent
agreement in the SFRs with only a small average offset of $-0.1$~dex
and a scatter of 0.21~dex, which is consistent with the typical
uncertainties in these modeling-dependent comparisons.  Furthermore,
there is no evidence for any strong systematic offset with
redshift. This is relevant because some dust emission templates
exhibit strong spectral features, such as the PAH emission, that could
bias the fit of some bands at specific redshifts. Lastly, the right
panel of Figure~\ref{fig:IR_compare_L18_SFR} shows the comparison of
the total (UV+IR) SFRs. We find again a good agreement with a similar
average offset and scatter. This suggests that the impact of the
UV-SFRs in the total SFRs of these IR bright galaxies is only minor
and therefore small fluctuations in their values have a negligible
impact in the comparison.

\subsection{SFR catalogs and far-IR photometric flags}
\label{ap:sfrcats}
\begin{table*}
\centering
\caption{Description of the SFR catalog}
\label{table:sfrheader}
\begin{adjustbox}{max width=\textwidth}
\begin{threeparttable}
\begin{tabular}{l l l}
\hline
\hline
Column No. & Column Title  & Description\\
\hline
1& id&			Object identifier \\
2& z &	Photometric redshift used in the IR SED fitting, corresponds to {\tt zbest} in the redshift catalog \\
$3-16$& Flux, Flux\_Err & Flux and flux error in each filter. Filters are included in order:  \\
&                  & MIPS 24 and 70~$\mu$m, PACS 100 and 160$\mu$m and SPIRE 250, 350 and 500~$\mu$m [$\mu$Jy].\\
&                  & Galaxies without IR detections have upper limits in MIPS24 with Flux = 20 - 70 [$\mu$Jy] (see Table~\ref{table:photometry_ir}) and Flux\_Err = 0.\\
&                  & Upper limits in MIPS24 are used to estimate upper limits in SFR-IR, indicated with negative values.\\
17& SFR$_{\rm total}^{\rm ladder}$ & Use as default SFR. Best estimate of the total SFR: either SFR$_{\rm IR}$+SFR$_{\rm UV}^{\rm obs}$ for IR detected sources or SFR$_{\rm UV}^{\rm corr}$ for the rest. \\
18& SFR\_ladder\_type           & Type of SFR indicators used in SFR$_{\rm total}^{\rm ladder}$:\\
  &                             & 1 for SFR$_{\rm total}^{\rm ladder}$ = SFR$_{\rm IR}^{\rm fit}$ + SFR$_{\rm UV}^{\rm obs}$ \\
  &                             & 2 for SFR$_{\rm total}^{\rm ladder}$ = SFR$_{\rm IR}^{\rm W11}$ + SFR$_{\rm UV}^{\rm obs}$ \\
  &                             & 3 for SFR$_{\rm total}^{\rm ladder}$ = SFR$_{\rm UV}^{\rm corr}$ \\
19& SFR$_{\rm UV}^{\rm corr}$     & UV-based star formation rate corrected for extinction using the IRX-$\beta_{\rm UV}$ relations. This value is a weighted average of SFR$_{\rm UV}^{\rm corr}$(160) and SFR$_{\rm UV}^{\rm corr}$(280) \\
20& SFR$_{\rm UV}^{\rm corr}$\_Err     & Uncertainty in the UV-based star formation rate corrected for extinction.\\
21& SFR$_{\rm IR}$                    & IR-based star formation rate. This value is equal to:\\
&                                    &  SFR$_{\rm IR}$ = SFR$_{\rm IR}^{\rm fit}$ when SFR\_ladder\_type=1.\\
&                                    &  SFR$_{\rm IR}$ = SFR$_{\rm IR}^{\rm W11}$ when SFR\_ladder\_type=2.\\
&                                    &  SFR$_{\rm IR}$ = -SFR$_{\rm IR}^{\rm W11}$ when SFR\_ladder\_type=3 (based on upper limits in MIPS24).\\
22& SFR$_{\rm UV}^{\rm obs}(160)$      & UV-based star formation rate not corrected for extinction determined from the UV luminosity at 160~nm.\\
23& SFR$_{\rm UV}^{\rm obs}(280)$      & UV-based star formation rate not corrected for extinction determined from the UV luminosity at 280~nm..\\
24&$\beta_{\rm UV}$              & UV slope\\
25&$A_{\rm UV}(160)$           & UV attenuation derived from the IRX-$\beta_{\rm UV}$ calibration for SFR$_{\rm UV}^{\rm obs}$(160)\\ 
26&$A_{\rm UV}(280)$           & UV attenuation derived from the IRX-$\beta_{\rm UV}$ calibration for SFR$_{\rm UV}^{\rm obs}$(280)\\
27&$A$(V)                     & Optical attenuation in the V-band derived from $A_{\rm UV}$ by assuming a Calzetti et al. (2000) attenuation law.\\

28& SFR$_{\rm IR}^{\rm W11}$      & IR-based star formation rate derived from the MIPS 24$\mu$m flux following \citep{wuyts11b}.\\
29& SFR$_{\rm UV}^{\rm corr}(160)$      & UV-based star formation rate corrected for extinction using the IRX-$\beta_{\rm UV}$ relations determined from the UV luminosity at 160~nm.\\
30& SFR$_{\rm UV}^{\rm corr}(280)$      & UV-based star formation rate corrected for extinction using the IRX-$\beta_{\rm UV}$ relations determined from the UV luminosity at 280~nm.\\
\hline
\end{tabular}
\end{threeparttable}
\end{adjustbox}
\end{table*}

\begin{table*}
\centering
\caption{Description of the SFR Flag catalogs}
\label{table:sfr_flags}
\begin{adjustbox}{max width=\textwidth}
\begin{threeparttable}
\begin{tabular}{l l l}
\hline
\hline
Column No. & Column Title & Description \\
\hline
1     & id                       & Object identifier in the F160W catalog\\
2     & MIPS\_ID\_order          & ID of the MIPS24 counterpart in the catalog from P\'erez-Gon\'alez et al. (2018)\\
      &                          & Multiple F160W counterparts can be associated with this source the order of likelihood is indicated with \_1,\_2, etc.\\
3     & MIPS\_discriminator      & Criteria used to determine the likelihood order of the F160W counterparts to a MIPS source: mips24, irac3.6, irac8.0, dist\\
      &                          & There is only one counterpart within 2\farcs5 (mips24), or the primary counterpart in the brightest this IRAC band or it is the closest in coordinates (dist) \\
4--9  & Flux, Flux\_Err          & Flux and flux error in the MIPS24, IRAC80, IRAC36 filters. In units of $\mu$Jy.\\
10    & MIPS\_distance           & Distance between the F160W source and the closest MIPS source in arcsec.\\
11    & MIPS\_order              & Likelihood of being the F160W source being true counterpart of the MIPS source. From 1 to N, with 1 being the highest.\\
12    & MIPS\_n\_counterparts    & Number of F160W counterparts candidates for the closest MIPS source within 2\farcs5.\\
13    & MIPS24\_snr\_cuts        & Flag regarding the SNR cuts applied in MIPS24: 0 no-flux, 1 flux $>$ SNR limit, -1 flux $<$ SNR limit. Only sources with flag > 0 are included in table~\ref{table:sfrheader}.\\
14--18& N\_F160W\_MIPS24\_PSF    & Number of F160W counterparts within 1, 0.5, 0.25, 2, 3 times the size of the MIPS24 PSF (2\farcs0) around the MIPS24 primary.\\
19--23& N\_F160W\_MIPS24\_WCS    & Number of F160W counterparts within 1, 0.5, 0.25, 2, 3 times the size of the WCS accuracy of the MIPS24 mosaic (2\farcs0)around the MIPS24 primary.\\
24--28& N\_MIPS24\_MIPS24\_PSF   & Number of MIPS24 counterparts within 1, 0.5, 0.25, 2, 3 times the size of the MIPS24 PSF (2\farcs0) around the MIPS24 primary.\\
29--33& N\_MIPS24\_MIPS24\_WCS   & Number of MIPS24 counterparts within 1, 0.5, 0.25, 2, 3 times the size of the WCS accuracy of the MIPS24 mosaic (2\farcs0) around the MIPS24 primary.\\
34--38& N\_IRAC36\_MIPS24\_PSF   & Number of IRAC36 counterparts within 1, 0.5, 0.25, 2, 3 times the size of the MIPS24 PSF (2\farcs0) around the MIPS24 primary.\\
39--43& N\_IRAC36\_MIPS24\_WCS   & Number of IRAC36 counterparts within 1, 0.5, 0.25, 2, 3 times the size of the WCS accuracy of the MIPS24 mosaic (2\farcs0) around the MIPS24 primary.\\
\hline
\hline
1     & id                       & Object identifier in the F160W catalog\\
2     & PACS\_ID\_order          & ID of the PACS counterpart in the catalog from P\'erez-Gon\'alez et al. (2008,2010)\\
      &                          & Multiple F160W counterparts can be associated with this source the order of likelihood is indicated with \_1,\_2, etc.\\
3     & PACS\_discriminator      & Criteria used to determine the likelihood order of the F160W counterparts to a PACS source: pacs160, pacs100, mips24, irac3.6, irac8.0, dist\\
      &                          & There is only one counterpart within 3\farcs0 (pacs160/pac100), or the primary counterpart in the brightest this MIPS/IRAC band or it is the closest in coordinates (dist)\\
4--13 & Flux, Flux\_Err          & Flux and flux error in the PACS160, PACS100, MIPS24, IRAC80, IRAC36 filters. In units of $\mu$Jy.\\
14    & PACS\_distance           & Distance between the F160W source and the closest PACS source within 3\farcs0.\\
15    & PACS\_order              & Likelihood of being the F160W source being true counterpart of the PACS source. From 1 to N, with 1 being the highest.\\
16    & PACS\_n\_counterparts    & Number of F160W counterparts candidates for the closest PACS source (IN WHICH RADIUS).\\
17    & PACS100\_snr\_cuts       & Flag regarding the SNR cuts applied in PACS100: 0 no-flux, 1 flux $>$ SNR limit, -1 flux $<$ SNR limit. Only sources with flag > 0 are included in table~\ref{table:sfrheader}.\\
18    & PACS160\_snr\_cuts       & Flag regarding the SNR cuts applied in PACS160: 0 no-flux, 1 flux $>$ SNR limit, -1 flux $<$ SNR limit. Only sources with flag > 0 are included in table~\ref{table:sfrheader}.\\
19--28& N\_F160W\_PACS\_PSF    & Number of F160W counterparts within 1, 0.5, 0.25, 2, 3 times the size of the PACS100 and PACS160 PSF (4\farcs5/7\farcs0) around the PACS primary.\\
29--38& N\_F160W\_PACS\_WCS    & Number of F160W counterparts within 1, 0.5, 0.25, 2, 3 times the size of the WCS accuracy of the PACS100 and PACS160 mosaics (2\farcs0/2\farcs5) around the PACS primary.\\
39--48& N\_PACS\_PACS\_PSF    & Number of PACS counterparts within 1, 0.5, 0.25, 2, 3 times the size of the PACS100 and PACS160 PSF (4\farcs5/7\farcs0) around the PACS primary.\\
49--58& N\_PACS\_PACS\_WCS    & Number of PACS counterparts within 1, 0.5, 0.25, 2, 3 times the size of the WCS accuracy of the PACS100 and PACS160 mosaics (2\farcs0/2\farcs5) around the PACS primary.\\
59--68& N\_MIPS24\_PACS\_PSF    & Number of MIPS24 counterparts within 1, 0.5, 0.25, 2, 3 times the size of the PACS100 and PACS160 PSF (4\farcs5/7\farcs0) around the PACS primary.\\
69--78& N\_MIPS24\_PACS\_WCS    & Number of MIPS24 counterparts within 1, 0.5, 0.25, 2, 3 times the size of the WCS accuracy of the PACS100 and PACS160 mosaics (2\farcs0/2\farcs5) around the PACS primary.\\
79--88& N\_IRAC36\_PACS\_PSF    & Number of IRAC36 counterparts within 1, 0.5, 0.25, 2, 3 times the size of the PACS100 and PACS160 PSF (4\farcs5/7\farcs0) around the PACS primary.\\
89--98& N\_IRAC36\_PACS\_WCS    & Number of IRAC36 counterparts within 1, 0.5, 0.25, 2, 3 times the size of the WCS accuracy of the PACS100 and PACS160 mosaics (2\farcs0/2\farcs5) around the PACS primary.\\
\hline
\hline
1     & id                       & Object identifier in the F160W catalog\\
2     & SPIRE\_ID\_order          & ID of the SPIRE counterpart in the catalog from P\'erez-Gon\'alez et al. (2008,2010)\\
      &                          & Multiple F160W counterparts can be associated with this source the order of likelihood is indicated with \_1,\_2, etc.\\
3     & SPIRE\_discriminator     & Criteria used to determine the likelihood order of the F160W counterparts to a SPIRE source: spire500, spire350, spire250, pacs160, pacs100, mips24, irac3.6, irac8.0, dist\\
      &                          & There is only one counterpart within 9\farcs0 (spire500,350,250), or the primary counterpart in the brightest this PACS/MIPS/IRAC band or it is the closest in coordinates (dist)\\
4--19 & Flux, Flux\_Err          & Flux and flux error in the PACS160, PACS100, MIPS24, IRAC80, IRAC36 filters. In units of $\mu$Jy.\\
20    & SPIRE\_distance           & Distance between the F160W source and the closest SPIRE source in arcsec.\\
21    & SPIRE\_order              & Likelihood of being the F160W source being true counterpart of the SPIRE source. From 1 to N, with 1 being the highest.\\
22    & SPIRE\_n\_counterparts    & Number of F160W counterparts candidates for the closest SPIRE source within 9\farcs0.\\
23    & SPIRE250\_snr\_cuts       & Flag regarding the SNR cuts applied in SPIRE250: 0 no-flux, 1 flux $>$ SNR limit, -1 flux $<$ SNR limit. Only sources with flag > 0 are included in table~\ref{table:sfrheader}.\\
24    & SPIRE350\_snr\_cuts       & Flag regarding the SNR cuts applied in SPIRE350: 0 no-flux, 1 flux $>$ SNR limit, -1 flux $<$ SNR limit. Only sources with flag > 0 are included in table~\ref{table:sfrheader}.\\
25    & SPIRE500\_snr\_cuts       & Flag regarding the SNR cuts applied in SPIRE500: 0 no-flux, 1 flux $>$ SNR limit, -1 flux $<$ SNR limit. Only sources with flag > 0 are included in table~\ref{table:sfrheader}.\\
26--30& N\_F160W\_SPIRE\_PSF    & Number of F160W counterparts within 1, 0.5, 0.25, 2, 3 times the size of the SPIRE250, SPIRE 350 and SPIRE500 PSF (11\farcs0/11\farcs0/17\farcs0) around the SPIRE primary.\\
41--55& N\_F160W\_SPIRE\_WCS    & Number of F160W counterparts within 1, 0.5, 0.25, 2, 3 times the size of the WCS accuracy of the SPIRE250, SPIRE 350 and SPIRE500 mosaics (9\farcs0/9\farcs0/15\farcs0) around the SPIRE primary.\\
56--70& N\_SPIRE\_SPIRE\_PSF    & Number of SPIRE counterparts within 1, 0.5, 0.25, 2, 3 times the size of the SPIRE250, SPIRE 350 and SPIRE500 PSF (11\farcs0/11\farcs0/17\farcs0) around the SPIRE primary.\\
71--85& N\_SPIRE\_SPIRE\_WCS    & Number of SPIRE counterparts within 1, 0.5, 0.25, 2, 3 times the size of the WCS accuracy of the SPIRE250, SPIRE 350 and SPIRE500 mosaics (9\farcs0/9\farcs0/15\farcs0) around the SPIRE primary.\\
86--100 & N\_PACS\_SPIRE\_PSF    & Number of PACS counterparts within 1, 0.5, 0.25, 2, 3 times the size of the SPIRE250, SPIRE 350 and SPIRE500 PSF (11\farcs0/11\farcs0/17\farcs0) around the SPIRE primary.\\
101--115& N\_PACS\_SPIRE\_WCS    & Number of PACS counterparts within 1, 0.5, 0.25, 2, 3 times the size of the WCS accuracy of the SPIRE250, SPIRE 350 and SPIRE500 mosaics (9\farcs0/9\farcs0/15\farcs0) around the SPIRE primary.\\
116--130& N\_MIPS\_SPIRE\_PSF    & Number of MIPS counterparts within 1, 0.5, 0.25, 2, 3 times the size of the SPIRE250, SPIRE 350 and SPIRE500 PSF (11\farcs0/11\farcs0/17\farcs0) around the SPIRE primary.\\
131--145& N\_MIPS\_SPIRE\_WCS    & Number of MIPS counterparts within 1, 0.5, 0.25, 2, 3 times the size of the WCS accuracy of the SPIRE250, SPIRE 350 and SPIRE500 mosaics (9\farcs0/9\farcs0/15\farcs0) around the SPIRE primary.\\ 
146--160& N\_IRAC36\_SPIRE\_PSF  & Number of IRAC36 counterparts within 1, 0.5, 0.25, 2, 3 times the size of the SPIRE250, SPIRE 350 and SPIRE500 PSF (11\farcs0/11\farcs0/17\farcs0) around the SPIRE primary.\\
161--175& N\_IRAC36\_SPIRE\_WCS  & Number of IRAC36 counterparts within 1, 0.5, 0.25, 2, 3 times the size of the WCS accuracy of the SPIRE250, SPIRE 350 and SPIRE500 mosaics (9\farcs0/9\farcs0/15\farcs0) around the SPIRE primary.\\
\hline
\end{tabular}
\end{threeparttable}
\end{adjustbox}
\end{table*}

This section describes the content of the FIR photometry and SFR
catalogs for all sources in the 5 CANDELS
fields. Table~\ref{table:sfrheader} contains the FIR fluxes, dust
attenuations and different SFR tracers computed following the methods
described in \S~\ref{ss:sfrs} and appendix~\ref{ap:IRphotometry} . In
addition Table~\ref{table:sfr_flags}, provides FIR photometric and
proximity flags that can be used to clean the catalog or to apply more
restrictive conditions on the sources with FIR detections.

Note that, while in Table~\ref{table:sfrheader} there is only one
possible F160W counterpart to each detection in a FIR band, the
catalog described in Table~\ref{table:sfr_flags} list all the possible
F160W counterparts to a given FIR source (MIPS, PACS and SPIRE)
indicating their likelihood (from 1 to N) of being the primary
counterpart of the IR detection. Only the F160W sources with the
maximum likelihood (e.g., MIPS\_order=1) have IR fluxes in
Table~\ref{table:sfrheader}. The total number of F160W counterparts to
a given IR source as well as their distances such source and their
respective IRAC fluxes are also indicated.

In addition to the F160W multiplicity for a given FIR source, the
catalog lists the multiplicity of that source in all other mid-to-far
IR bands up to itself. These multiplicities are computed for several
cross-match radius relative to the typical spatial resolution of the
FIR band (e.g., radius of 0.5, 1, 2 or 3$\times$ the FWHM of the
PSF). For the example, the MIPS flag catalog includes the
multiplicities of F160W, IRAC and MIPS sources within different
radius. The PACS flag catalog includes multiplicities of F160W, IRAC,
MIPS and PACS sources, etc. The values in the flag catalogs provide a
quick a simple way to find relative isolated FIR sources, or to
identify FIR sources in crowded environments which might lead to
some contamination in the photometry.

\end{appendix}
\end{document}